\documentclass[twocolumn,         
               showpacs,longbibliography,            
               showkeys,preprintnumbers,     
               aps,                 
               prd,          	    
               letterpaper,             
               superscriptaddress,      
               nofootinbib,         
               tightenlines,        
               floats,floatfix      
               ]{revtex4-1}
\usepackage{physics}
\usepackage{epsf}
\usepackage{graphicx,color}
\usepackage{subfigure}
\usepackage{latexsym}
\usepackage{amsmath,amssymb}        
\usepackage[colorlinks=true,linkcolor=blue,citecolor=blue]{hyperref}
\usepackage{mathrsfs}
\usepackage{comment}
\usepackage{soul}
\usepackage{cancel}
\definecolor{purple}{rgb}{0.58,0.0,0.83}
\definecolor{orange}{rgb}{1,0.5,0}
\DeclareSymbolFontAlphabet{\mathrsfs}{rsfs}
\DeclareMathAlphabet{\mathcal}{OMS}{cmsy}{m}{n}

\begin{document}


\title{The Effect of Boundary Conditions on Structure Formation in Fuzzy Dark Matter}


\author{Iv\'an  \'Alvarez-Rios}
\email{ivan.alvarez@umich.mx}
\affiliation{Instituto de F\'{\i}sica y Matem\'{a}ticas, Universidad
              Michoacana de San Nicol\'as de Hidalgo. Edificio C-3, Cd.
              Universitaria, 58040 Morelia, Michoac\'{a}n,
              M\'{e}xico.}

\author{Francisco S. Guzm\'an}
\email{francisco.s.guzman@umich.mx}
\affiliation{Instituto de F\'{\i}sica y Matem\'{a}ticas, Universidad
              Michoacana de San Nicol\'as de Hidalgo. Edificio C-3, Cd.
              Universitaria, 58040 Morelia, Michoac\'{a}n,
              M\'{e}xico.}  

\author{Paul R. Shapiro}
\email{shapiro@astro.as.utexas.edu}
\affiliation{Department of Astronomy and Texas Center for Cosmology and Astroparticle Physics, The University of Texas at Austin, Austin, TX 78712-1083, USA.}             


\date{\today}


\begin{abstract}
 We illustrate the effect of boundary conditions on the evolution of structure 
  in Fuzzy Dark Matter. Scenarios explored include the evolution of single, 
  ground-state equilibrium solutions of the Schr\"odinger-Poisson system,
  relaxation of a Gaussian density fluctuation, mergers of two equilibrium configurations, 
  and the random merger of many solitons. For comparison, each scenario is evolved twice, 
  with isolation vs. periodic boundary conditions, the two commonly 
  used to simulate isolated systems and structure formation, respectively. Replacing isolation 
  boundary conditions (implemented by an absorbing sponge at large radius) by periodic 
  boundary conditions changes the domain topology and dynamics of each scenario, 
  by affecting the outcome of gravitational cooling. With periodic boundary conditions, 
  the initial ground-state equilibrium solution and Gaussian fluctuation each evolve toward 
  the single equilibrium solitonic core of the isolated case, but surrounded by an envelope, 
  or tail, in which additional mass is distributed nearly uniformly, unlike the isolated versions. 
  The case of head-on, binary merger introduces additional effects caused by the pull on
  the system due to the infinite network of periodic images along each axis of 
  the domain. Adding angular momentum to this binary merger results in a tail with polynomial 
  profile when using a periodic domain. Finally, the 3-D merger of many, randomly-placed solitonic 
  cores of different mass makes a solitonic core surrounded by a tail with power-law-like profile, 
  for periodic boundary conditions, while producing a core with a much sharper fall-off in the 
  isolated case. This suggests the conclusion of earlier work that the ground-state equilibrium 
  solution is an attractor for the asymptotic state is true even in 3-D and for more general 
  circumstances than previously considered, but only if gravitational cooling is able to carry 
  mass and energy off to infinity, which isolation boundary conditions allow, but periodic ones do not.   
\end{abstract}


\keywords{self-gravitating systems -- dark matter -- Bose condensates}


\maketitle

\section{Introduction}
\label{sec:intro}

The Fuzzy Dark Matter (FDM) model \cite{Hu:2000} is based on the idea that the dark matter particle is an ultralight spin-less boson (e.g. \citep{Matos-Urena:2000,Sahni:2000,Hui:2016}), which has shown to have potential at solving some of the traditional problems of the standard Cold Dark Matter (CDM) model, namely the too-big-to-fail and the cusp-core problems \citep{Hui:2021tkt,Niemeyer_2020,ElisaFerreiraAApRev2021,Lee:2017,Hui:2016,Suarez:2013}. The properties of the model are ordinarily tested with simulations of structure formation as well as other local astrophysical scenarios, which are based on the numerical solution of the Schr\"odinger-Poisson (SP) system of equations, which rules the dynamics of this type of matter.  In view of its importance as a potentially observable discriminant of FDM over CDM, we shall focus here on the numerically-derived internal structure of virialized objects -- haloes-- that form.  As described below, our goal is to shed light on the important role that the treatment of outer boundary conditions plays in shaping their final mass distribution.   

In phenomenological or theoretical analyses, the solution of the SP system of equations is essential within this dark matter model. It happens that no exact solutions are known of this system, from the simplest stationary scenarios \cite{Ruffini:1969,GuzmanUrena2004}, interaction between a few structures (e. g. \cite{BernalGuzman2006a,Schwabe:2016,ShapiroCreTail,PhysRevD.97.063507}) until structure formation simulations with very elaborate dynamics (e.g. \cite{Schive:2014dra,Mocz:2017wlg,mocz19,mocz19b,Niemeyer_2020,Gotinga2022}), the solutions constructed are numerical, and therefore the results are subject to numerical methods and conditions imposed by the restrictions of each problem.

The SP system is solved on a finite size numerical domain and boundary conditions are used to implement the desired effects on the system under study. Two types of boundary conditions are used that distinguish two physical scenarios. The first one corresponds to isolated systems, oriented to the study of the phenomenology of isolated structures, mostly cores under different circumstances, like perturbations and mergers. The second scenario is that of structure formation, where the assumptions of homogeneity and isotropy at large scales is not consistent with isolation conditions, and periodic boundary conditions are used.

Isolated systems are expected to radiate the excess of kinetic energy toward infinity, and relax through the gravitational cooling process \cite{SeidelSuenCooling,GuzmanUrena2006}. Instead of attempting the implementation of well posed absorbing boundary conditions for Schr\"odinger equation, a sponge is used to absorb the modes approaching the boundary of the numerical domain, which acts as a sink of particles independently of the wave front orientation. The resulting configurations in a number of configurations are isolated blobs with a solitonic density profile at the center and an exponential decay outside (e.g. \cite{SeidelSuenCooling,GuzmanUrena2004,Schwabe:2016}).

Structure formation simulations on the other hand, change the topology of the domain from a piece of $R^3$ to the 3-torus ${\cal T}^3$ through the implementation of periodic boundary conditions to the SP system of equations. A consequence is that the matter does not leave the domain and it redistributes across the domain instead. The relaxation process of some density fluctuations consists in the formation of cored clumps with the profile of isolated equilibrium configurations surrounded by restless outskirts with a polynomial density profile in average, as shown in fits from structure formation simulations \cite{Schive:2014dra,Mocz:2017wlg,Schwabe:2016}, that eventually can match the Navarro-Frenk-White (NFW) density profile \cite{NFW}.

Here we solve the SP system with isolation and periodic boundary conditions in order to study the effect of the topology on the dynamical properties of the interaction among structures.

The paper is organized as follows. In Section \ref{sec:Background}, we trace the background and motivation underlying our effort here to reconcile the apparent contrast between earlier developments on the equilibrium structure of isolated objects with recent work on halos from structure formation in the FDM cosmology.
 In Section \ref{sec:NM} we describe the numerical methods implemented for the study of the effects of the two types of boundary conditions. In Section \ref{sec:results} we compare the results for different scenarios, and in Section \ref{sec:conclusions} we draw some conclusions.


\section{Background and Motivation}
\label{sec:Background}
The last several years have witnessed an explosion of interest in the possibility that cosmic dark matter is a scalar field of ultralight (e.g. $m\gtrsim10^{-23}$ eV) bosons for which quantum coherent effects cause structure that forms from primordial density fluctuations to be smoothed-out on small scales, perhaps even on scales as large as galaxies (e.g.  \citep{Marsh2016,Niemeyer_2020, Hui:2021tkt, ElisaFerreiraAApRev2021} and references therein). For the free-field limit (i.e. no self-interaction), sometimes referred to as Fuzzy Dark Matter (``FDM''), the filtering scale is related to the de Broglie wavelength, which acts like a quantum Jeans length.   There is just one parameter that determines the outcome of structure formation in FDM from a given spectrum of primordial density perturbations in a given background universe, namely, particle mass $m$.  The smaller the mass $m$, the larger is the associated Jeans length.  In the linear regime, the transfer of primordial perturbations after horizon entry introduces a short-wavelength cut-off in the power-spectrum on this scale, which filters-out structure formation below this cut-off, during construction of the cosmic web.   In the nonlinear regime, the internal structure of gravitationally-bound objects (i.e. dark matter haloes) is smoothed-out on scales smaller than the local de Broglie wavelength determined by the particle mass and the local virial velocity inside the halo, i.e. $\lambda_{\rm deB}=h/mv$, for which halos of mass $M$ and size $R$ have virial or circular velocity $v\simeq(GM/R)^{1/2}$ . 
This means haloes for which  $R\lesssim\lambda_{\rm deB}$, or, equivalently, particle mass 
$m\lesssim(\hbar/R^2)(\pi G\bar{\rho})^{-1/2}\approx10^{-22}(R/1~\rm{kpc})^{-1/2}(M/10^8~M_\odot)^{-1/2}~\rm{eV}$, where
$\bar\rho$ is the mean mass density of the halo, are not expected to form, while those for which $R\gtrsim\lambda_{\rm deB}$ will be smoothed on scales smaller than $\lambda_{\rm deB}$. 

This nonlinear smoothing property was a strong motivation for considering FDM as an alternative to standard WIMP CDM, since the latter produces density profiles for virialized dark matter haloes which N-body simulations show are well-fit by an empirical formula, known as the NFW profile, which diverges in a cusp toward the center, while galaxy observations suggest that a flattened core may be preferred in a range of systems.  For FDM, by tuning the particle mass to be sufficiently small ($m \lesssim 10^{-22}$ eV$/c^2$), the size of the de Broglie smoothing scale can approach that of the Kpc-sized cores observed in some galaxies.    On the other hand, the linear-regime smoothing that leads to a cut-off in the transfer function is also of interest, since it can reduce the abundance of haloes at the small-mass end during the galaxy formation era, as well as the amplitude of density fluctuations in the intergalactic medium in the quasi-linear regime. The former effect was initially thought to be beneficial as a way to reconcile the paucity of observed satellite dwarf galaxies in the Local Group with the over-abundance predicted by N-body simulations of standard CDM. However, further study suggests that these reductions of small-mass halo abundance and intergalactic density fluctuations may go too far if $m$ is small enough to yield dwarf- galaxy cores as large as $\sim1$ Kpc during the nonlinear phase of structure formation, as compared with observations of Local Group satellites and fluctuations in the Ly-$\alpha$ forest of quasar absorption-line spectra.  This is sometimes referred to as the ``catch-22 problem'' for FDM.  

Our interest here is in the nonlinear regime, in general, and the internal structure of gravitationally-bound objects comprised of FDM, in particular. Study of the latter has a long history that predates much of the modern treatment of FDM as an alternative to standard CDM, including the early description of compact objects supported against gravitational collapse by quantum pressure as ``boson stars'' (for the latter, see, e.g. \cite{Jetzer1992} and references therein).
For FDM structure formation on scales that are well within the Hubble volume and for objects that are not dense enough to be subject to relativistic gravitational instability, the fully-relativistic description of the scalar field by coupled Klein-Gordon and Einstein equations reduces to that of the coupled Nonlinear Schr\"odinger and Poisson equations, in the Newtonian approximation. That is our regime of interest here.

Early literature on boson stars, including a paper by Membrado, Pacheco and Sanudo \cite{Membrado1989}, found that the equilibrium solution of the coupled Schr\"odinger-Poisson equations in 1D spherical symmetry yielded a centrally-flattened density profile which, at large radius, as $r\rightarrow\infty$, drops off as $r^{-4}$, enclosing a finite mass but extending to infinity.  That purely numerical solution yields a radius which encloses $99\%$ of the finite mass to infinity given by $R_{99} = 9.9 (\hbar^{2}/(G M m^2)$, where $M$ is the total mass and $m$ is the boson mass.   This is comparable to the de Broglie wavelength for particles of mass $m$ and velocity given by the virial velocity for a body of total mass $M$ inside a radius $R_{99}$.  The time-dependent problem of the formation of such equilibrium objects was not addressed by this solution, however.

Subsequently, \cite{SeidelSuenCooling} showed numerically that a self-gravitating scalar field of bosons would be able to form solitonic boson stars as equilibrium objects, which virialized upon their collapse, by ``gravitational cooling'', a term invented to describe the emission of mass and energy to infinity.  In further literature on boson stars,  
e.g. \cite{GuzmanUrena2004,GuzmanUrena2006}, the time-dependent S-P equations in spherical symmetry were solved numerically, finding that the same equations yielded a static equilibrium solution for isolated spherical cores which is the ``ground state'' eigenfunction if one assumes the wavefunction has a harmonic time-dependence.   Furthermore, they showed, a ``multistate'' initial condition which includes excited states, as well, decays quickly to this ground state, by radiating mass and energy to infinity (i.e. ``gravitational cooling'').  So, one could expect that the ground state solution was, in a sense, the ``attractor''.   This work established this numerical solution of the ground-state eigenfunction, therefore, as the expectation for gravitationally-bound objects that might form out of ultralight bosons, with its flattened central profile characterized by a size comparable to the de Broglie wavelength computed from $m$ and the virial velocity, surrounded by a very sharply decreasing density profile.  Subsequent work, in fact, has sometimes approximated this ground-state numerical profile by a Gaussian, for analytical convenience.  

More recently, numerical solutions of the S-P equations in 3-D 
\cite{Schive:2014hza,Schwabe:2016,Mocz:2017wlg} were reported that modeled the formation of gravitationally-bound objects by gravitationally merging pre-existing, 1D, spherical, solitonic cores, each of which was initially in the self-gravitating, static equilibrium appropriate for isolated cores.  The outcome was a final object with a single core surrounded by an envelope, in virial equilibrium.  The core had a radius comparable to the deBroglie wavelength of the final object, and a centrally-flattened density profile like that of the isolated solitons described above.  Beyond that radius, however, the extended envelope, while also in virial equilibrium, had a density profile that dropped off much more gently.  The envelope profile, in fact, resembled the power-low drop-off at large radius of the empirical NFW fit to the halo profiles in CDM N-body simulations
\cite{NFW}
, i.e.  $r^{-n}$, with $n=3$.   The mass of the central solitonic core of this final object, however, was not the same as that of the original cores that were merged to make it. 

A similar core-envelope structure was also found for the virialized halos that formed in 3-D \emph{cosmological} simulations of FDM starting from cosmological, Gaussian-random, linear-perturbation density fluctuations
\cite{Schive:2014hza,Schwabe:2016,Mocz:2017wlg}.
Like the cores formed from the idealized mergers of isolated cores described above, the solitonic cores formed from 3-D perturbation growth from the linear to the nonlinear stage in cosmological simulations of FDM were also of size comparable to 
$\lambda_{\rm deB}$ evaluated with a velocity equal to the virial velocity of the core-envelope halo.   

That a core-envelope structure would result for the halos in virial equilibrium formed in the nonlinear stage of cosmological perturbation growth, with mass and radius larger than that of their solitonic cores, was not entirely unanticipated.  While this core-envelope structure of the final objects appeared to be a surprise when first reported, in view of the wide-spread assumption that the natural equilibrium structure to expect for an isolated object was that of the solitonic cores described above, some literature had already noted that those simple density profiles should be interpreted as halo cores, rather than entire halos 
(\cite{RindlerDallerShapiro2014}).
Otherwise, they could not easily explain the mass-radius relation of halos in our observed universe. According to the Membrado et al. solution described above, for instance, 
$M R_{99} = $ constant, which is not what we observe nor what the standard CDM model predicts. As
\cite{RindlerDallerShapiro2014}
pointed out,
however, 
halo formation in cosmology proceeds by mergers and infall during hierarchical clustering,
which causes the quantum fluid to make waves, and waves can interfere, so random internal wave motions can provide an effective kinetic pressure support equivalent to the random particle orbits that balance gravity when halos of CDM virialize. They concluded that, for scalar field dark matter to resemble CDM-like structure formation on larger scales than $\lambda_{\rm deB}$, with halos that obey a CDM-like mass-radius relation, this wave-like behavior must ensure that, when mass infalls and merges to make an object of mass and size \emph{larger} than the core, it virializes on scales well beyond the core.  In that case, a virialized object can form with the same total mass and size as it does in CDM, even if its central region has a non-CDM-like solitonic core of much smaller size.  The 3D simulations of merging cores and cosmological structure formation in FDM described above were, therefore, a verification of this hypothesis.  

As recently discussed by 
\cite{DawoodbhoyShapiroRindlerDaller2021,ShapiroDawoodbhoyRindlerDaller2022}, 
the CDM-like structure of FDM halos \emph{outside} their solitonic cores (i.e. on scales larger than $\lambda_{\rm deB}$) can be understood as an inevitable consequence of solving the coupled NLSE and Poisson equations in the presence of cosmological infall boundary conditions.  The dynamics of CDM as a nonrelativistic gas of non-interacting, massive particles, coupled only by gravity, in the collisionless limit (i.e. in which the 2-body gravitational relaxation time is infinitely long) is described, instead, by the collisionless Boltzmann equation (``CBE'') coupled to the Poisson equation.   There is a correspondence, however, between the NLSE and the CBE (on scales larger than $\lambda_\text{deB}$) that was first discussed in the context of CDM by
\cite{WidrowKaiser1993}.
They pointed out that CDM structure formation can be modeled by solving the NLSE in the free-field limit, as a computational alternative to simulating gravitational N-body dynamics of CDM particles -- a kind of quantum analog for modelling CDM, in which the analog particle mass is tuned to make its $\lambda_\text{deB}$ much smaller than any scale of interest (i.e. equivalent to taking the limit of $\hbar\rightarrow0$). More recently, \cite{MoczFialkovBecerraChavanis2018} followed this idea in the opposite direction, of solving the coupled CBE and Poisson equations (e.g. by well-known N-body methods used to model CDM) as an approximation to describe large-scale structure formation in FDM on scales larger than $\lambda_\text{deB}$.   

The general correspondence between the NLSE and the CBE, with additional terms in the latter that encode the effects of wave mechanics, goes back much further, however, and makes clearer how the wave-mechanical behavior of the scalar field creates an effective ``pressure'' support against gravity in the virialized halo even outside the solitonic core (see, e.g., \cite{DawoodbhoyShapiroRindlerDaller2021}  for a review and additional references). It is often noted in recent FDM literature that, by writing the field in polar form, in terms of a real amplitude and phase,

    \begin{equation}
        \psi = |\psi|e^{iS},
        \label{eq:polar}
    \end{equation}

\noindent (referred to as ``the Madelung transformation'', after \cite{Madelung1927}) 
and substituting this into the NLSE, it is possible to derive continuity and momentum equations similar to those of  classical hydrodynamics, as an exact alternative to the NLSE, sometimes referred to as the equations of ``quantum hydrodynamics'' (QHD).  The momentum equation that results resembles the Euler equation of classical fluid mechanics, except that the usual term involving the gradient of gas pressure is replaced by the gradient of the ``quantum potential'', $Q$ \citep[sometimes also referred to as the ``Bohm potential'', after][]{Bohm1952a,Bohm1952b}, given in terms of the mass density $ \rho = |\psi|^2$, by

 \begin{equation}
Q = -\frac{\hbar^2}{2m^2} \frac{\nabla^2 \sqrt{\rho}}{\sqrt \rho} \label{eq:Q}.
\end{equation}

\noindent It is the gradient of $Q$ that must balance the gravitational acceleration in this Euler-like equation, if FDM halos are in a static equilibrium (i.e. for which the ``bulk velocity'' $v$ in the Euler equation is essentially zero), even at radii well-beyond that of their solitonic cores.   Unfortunately, neither the NLSE nor these equivalent QHD equations give much insight into how to construct $Q$ to achieve this balance nor into how this $Q$ relates to the effective kinetic pressure we described above. [Some conditions for the numerical integration of the equations, limitations, and equivalence of the equations in the NLSE and QHD frames are discussed in \cite{SPvsMadelung}].

To provide that insight, \citep{DawoodbhoyShapiroRindlerDaller2021} appealed to a second approach to obtaining these QHD equations, the \emph{phase-space formulation}, pioneered by
\citet{Takabayasi1954}.  
This approach starts from a phase-space distribution function constructed from $\psi$, known as the Wigner function \citep{Wigner1932}, and derives its equation of motion, known as the Wigner-Moyal equation, by taking the partial time derivative of the Wigner function and substituting-in the NLSE. This Wigner-Moyal equation resembles the CBE but with additional terms that encode the effects of wave mechanics.  Momentum moments of this equation are then taken to derive the continuity and momentum equations of QHD.   In this version of the QHD momentum equation, the usual term in the classical Euler equation involving the gradient of gas pressure is replaced by a new, pressure-like term, called  the ``quantum pressure'' tensor, $\bm{\Pi}$, which is sourced by the \emph{velocity dispersion tensor}:
\begin{equation}
\Pi_{ij} = \rho \sigma^2_{ij}
        \label{eq:PiRhoSigma}
    \end{equation}
The divergence of this quantum pressure tensor in the Euler equation of QHD accounts for the transport of momentum associated with the kinetic energy term in the NLSE; it is a momentum flux density tensor.   It is possible to re-express this $\bm{\Pi}$, however, in terms of the density $\rho$ and its spatial derivatives, alone, to reassure us that the final QHD equations that result from this phase-space formulation have the same content as those from the Madelung-Bohm formulation above. But we gain additional insight by comparing the two formulations to show that the quantum potential and quantum pressure tensor are related according to:
\begin{equation}
        \frac{\partial Q}{\partial x_i} = \frac{1}{\rho} \frac{\partial \Pi_{ij}}{\partial x_j}.
        \label{eq:QeqivPi}
    \end{equation}

This shows that the acceleration associated with the quantum potential term in the Madelung-Bohm formulation corresponds to the effective kinetic pressure in the momentum flux density tensor, associated with the internal spread of momentum in the phase space derivation.  Both terms arise from the kinetic term in the NLSE and are what is responsible for providing support against gravity inside FDM halos.   Inside solitonic cores, this support is provided on the scale of $\lambda_\text{deB}$, but, as shown by \citep{DawoodbhoyShapiroRindlerDaller2021}, when FDM halos form by gravitational instability and collapse, the coupling of gravitational and quantum dynamics leads to a large-enough velocity dispersion that support against gravity is also possible on much larger scales, as well.

To demonstrate this explicitly, \cite{DawoodbhoyShapiroRindlerDaller2021,ShapiroDawoodbhoyRindlerDaller2022} 
reduced the complexity of the system required to create such conditions, to 1-D, spherical symmetry (as considered previously to solve the NLSE for isolated solitonic cores), to show that, even in this case, halo formation from gravitational instability and collapse are sufficient to produce a CDM-like halo structure outside the core.   First, they followed \cite*{Skodje1989}, \cite{WidrowKaiser1993}, and \cite{MoczFialkovBecerraChavanis2018}, who adopted a \textit{smoothed} phase space representation of $\psi$, known as the Husimi representation \citep{Husimi1940}, deriving its equation of motion by steps similar to that which led to the Wigner-Moyal equation, then setting the smoothing scale to be much larger than $\lambda_\text{deB}$, to find that the equation of motion reduces to the CBE, with no additional terms.    Second, they adopted the fluid approximation for solving the CBE derived by \citet{AhnShapiro2005} to describe CDM dynamics, by taking moments of the CBE and closing the moment hierarchy in spherical symmetry by assuming the velocity distribution is symmetric about its mean (i.e. skewless) and isotropic in the frame of bulk motion.
For CDM, the latter is a good approximation inside virialized regions that form during structure formation (with no loss of accuracy outside those regions, since infall is supersonic during halo formation and motion is ballistic). This reduced the CBE to the familiar fluid equations of conservation of mass, momentum and energy for an ideal, compressible gas with adiabatic index $\gamma = 5/3$.  As \citep{DawoodbhoyShapiroRindlerDaller2021} recognized, these same steps apply to the CBE for FDM, as well, when smoothed on scales larger than $\lambda_\text{deB}$.   This meant that one could now understand the role of quantum pressure in the dynamical formation of FDM halos, smoothed on the de Broglie scale, by its correspondence to the gas pressure associated with random, thermal motions of a $\gamma = 5/3$ ideal gas, in the same way \cite{AhnShapiro2005} had previously used this approach for CDM, to understand the effective kinetic pressure support provided by random orbital motions of the collisionless particles.   The ``temperature'' associated with this ideal gas pressure corresponds directly to the velocity dispersion ($\sigma^2$) obtained from moments of the phase space distribution function.
  
Finally, \citep{DawoodbhoyShapiroRindlerDaller2021, ShapiroDawoodbhoyRindlerDaller2022} adopted initial conditions suitable for forming a halo by gravitational instability of an initial, 1-D, spherically-symmetric linear density perturbation in a ``cold'' gas (i.e. which is  Jeans unstable on wavelengths larger than the deBroglie wavelength of the final object), and solved these fluid approximation equations numerically, by a Lagrangian hydrodynamics method.   

The result of the nonlinear outcome of this gravitational collapse was the formation of a virialized central region, bounded by a strong accretion shock, inside of which was ``post-shoc'' gas, in hydrostatic equilibrium, at high temperature and pressure, with a density profile for this post-shock interior region just like that for CDM halos observed in 3-D N-body simulations. 
This showed that the origin of the quantum pressure support that allowed a halo of FDM to be in virial equilibrium over a region much larger and more massive than its solitonic core was the ``thermalization'' of the kinetic energy of infall by passage through this accretion shock.     
Furthermore, this work showed, the NFW-like shape of the FDM envelope, i.e. the profile shape beyond the scale of the inner de Broglie wavelength -- and the evolution over time of the halo virial radius, mass, and concentration parameter during 3-D cosmological structure formation -- are the inevitable consequence of the shape of the initial linear perturbation surrounding the density peak that made the halo, which controls the rate of mass infall in the nonlinear regime.
\citet{AlvarezAhnShapiro2003}, \citet{Shapiroetal2004}, and \citet{Shapiro2006} 
had previously demonstrated this for CDM.  In particular, they showed that, when the initial, spherically-symmetric, linear-perturbation profile is shaped to yield the average mass assembly history (MAH) of a halo in the nonlinear stage, as found by CDM N-body simulations, the shock-bounded sphere in virial equilibrium that results reproduces the NFW profile and all its evolutionary properties. 
Now, by applying the same fluid approximation and initial conditions to describe FDM halo formation -- an approximation that is suitable to describe FDM dynamics averaged over scales that are larger than de Broglie but without losing the ``unresolved'' effects of quantum pressure -- this explains why FDM halos must generically have CDM-like envelopes beyond their solitonic cores.
\footnote{In \citep{Chavanis22}, the dynamics of ultralight bosonic dark matter haloes was modelled, instead, by replacing the NLSE by a wave equation with heuristic terms added to account for the effects of violent relaxation, gravitational cooling and dissipation. That approach also identified an additional, effective thermal pressure, distinct from the quantum pressure, associated with an effective temperature.  In 1-D, spherical symmetry, static equilibrium solutions of this heuristic equation, coupled to the Poisson equation, then showed solitonic cores surrounded by isothermal atmospheres. By contrast, in \citep{DawoodbhoyShapiroRindlerDaller2021, ShapiroDawoodbhoyRindlerDaller2022}, there was no appeal to the addition of heuristic terms to the original NLSE. Rather, their assumption of skewless and isotropic velocity distribution (in the frame of bulk motion),  consistent with N-body simulations of halo formation from Gaussian random density fluctuations, is all they required to show haloes have CDM-like envelopes on scales larger than their de Broglie-wavelength-sized solitonic cores, as a consequence of forming by gravitational infall.}

According to this, the key difference between the solution of the isolated solitonic cores calculated from the 1-D, time-dependent NLSE and Poisson equations -- for which gravitational cooling enabled them to relax to the ground-state eigenfunction -- and the core-envelope structure of virialized objects that formed dynamically from the growth of density perturbations or the mergers of isolated cores, must be the outer boundary conditions adopted in each case.   The purpose of this paper will be to demonstrate this explicitly, by comparing the results of the time-dependent formation of gravitationally-bound objects in FDM, as simulated by solving the NLSE in 3-D for different boundary conditions, \emph{without} smoothing over the de Broglie wavelength scale.  In particular, we will show that, by replacing the absorbing ``sponge'' boundary conditions of the previous 1-D calculations of isolated solitonic cores, with the periodic boundary conditions common to 3-D cosmological simulation, we can alter the outer shape of the density field, outside the core, to make it decline more gently.

\section{Numerical methods}
\label{sec:NM}

The SP system is a constrained evolution Initial Value Problem, where Schr\"odinger equation rules the dynamics of the wave function $\Psi$, whereas Poisson equation for the gravitational potential $V$ is a constraint that has to be satisfied during the evolution. We define the problem on the domain $D\times [0,t_f]=[x_{min},x_{max}]\times[y_{min},y_{max}]\times[z_{min},z_{max}]\times[0,t_f]$ described with Cartesian coordinates $x,y,z,t$ provided initial conditions for $\Psi$ and $V$. 

This Initial Value Problem is solved numerically on the discrete domain with $D_d=\{(x_i,y_j,z_k)\}$ with 
$x_i= x_{min}+i\Delta x$, 
$y_i= y_{min}+j\Delta y$, 
$z_i= z_{min}+k\Delta z$, where $i=0,...,N_x$, $j=0,...,N_y$, $k=0,...,N_z$ label cells along each direction, 
$\Delta x = (x_{max}-x_{min})/N_x$, 
$\Delta y = (y_{max}-y_{min})/N_y$, 
$\Delta z = (z_{max}-z_{min})/N_z$,  are the spatial resolutions. Time is discretized with labels $t^n = n\Delta t$, where 
$\Delta t =C \min(\Delta x,\Delta y, \Delta z)^2$ is the time resolution and $C$ is a CFL factor. In the simulations of this paper we use $\Delta x=\Delta y=\Delta z$ in a cubic box centered at the origin where $x_{min}=y_{min}=z_{min}=-x_{max}=-y_{max}=-z_{max}$. 

We denote the wave function and gravitational potential at the arbitrary point $(x_i,y_j,z_k)\in D_d$ and time $t^n$ by $\Psi^{n}_{i,j,k}$ and $V^n_{i,j,k}$ respectively.

\subsection{Methods for the isolated domain}

The Schr\"odinger-Poisson system of equations in the continuum is written as

\begin{eqnarray}
{\rm i}\frac{\partial \Psi}{\partial t} & = & -\dfrac{1}{2}\nabla^2\Psi + V\Psi, \label{eq:SchroIso}\\
\nabla^2 V &=& |\Psi|^2,\label{eq:PoiIso}
\label{eq:SPiso}
\end{eqnarray}

\noindent where Planck constant and the boson mass have been absorbed with an appropriate rescaling of coordinates and variables. 

Provided initial conditions for $\Psi$ and a consistent gravitational potential $V$, we integrate in time the semi-discrete version of Schr\"odinger equation for $\Psi$ from time $t^n$ to $t^{n+1}$, using the Method of Lines, and a Finite Differences approximation of the right hand side of Eq. (\ref{eq:SchroIso}) using fourth order finite difference stencils

\begin{eqnarray}
\frac{\partial \Psi}{\partial t} &=& \frac{{\rm i}}{2} \left(\delta^2_x[\Psi^{n}_{i,j,k}] + \delta^2_y[\Psi^{n}_{i,j,k}]+\delta^2_z[\Psi^{n}_{i,j,k}]\right)
-{\rm i} V^{n}_{i,j,k}\Psi^n_{i,j,k}\nonumber
\end{eqnarray}

\noindent where $\delta^2_x,\delta^2_y,\delta^2_z$ are the second order derivative operators with fourth order accuracy. The evolution is carried out using a third order accurate Runge-Kutta integrator.

Near the boundary of $D_d$ at all times, for $\Psi$ we impose a sponge, which is implemented by adding an imaginary potential to the gravitational potential such that $V = V+{\rm i}V_{im}$. The effect is that the continuity equation for $\rho=|\Psi|^2$ becomes $\frac{\partial \rho}{\partial t} + \nabla\cdot[\frac{\rm i}{2} (\Psi\nabla\Psi^* - \Psi^* \nabla\Psi)]=2V_{im}|\Psi|^2$. This means that when $V_{im}<0$ the density $\rho$ enters a sink of particles that we implement only in the region near the boundary of $D_d$ using the function $V_{im} = -\frac{V_0}{2}[2+\tanh(r-r_s)/\delta - \tanh(r_s/\delta)]$, a smooth version of a step function along the radial direction with $r=\sqrt{x^2+y^2+z^2}$ \cite{GuzmanUrena2004}. Our simulations are carried out in a cubic domain centered at the origin, which allows the implementation of this recipe straightforwardly. We use $V_0=1$, $r_s=0.8 x_{max}$ and $\delta=4\Delta x$.

Finally, Poisson equation is solved using a monopolar boundary condition $V(\partial D_d)=-M/(4\pi r_{\partial D_d})$, where $r_{\partial D_d}$ is the distance from the origin to each point of the boundary of $D_d$, and implement a multigrid solver with a two-level  V-cycle.

\subsection{Methods for the periodic domain}

In this case the SP system is written differently: 

\begin{eqnarray}
{\rm i}\frac{\partial \Psi}{\partial t} & = & -\dfrac{1}{2}\nabla^2\Psi + V\Psi, \\
\nabla^2 V &=&|\Psi|^2-\langle|\Psi|^2\rangle,
\label{eq:SP}
\end{eqnarray}

\noindent where $\langle|\Psi|^2\rangle$ is the average density over the domain, introduced in order to satisfy that the integral on the right hand side of Poisson equation vanishes and also to allow the periodicity of the potential.\\

We solve both, Schr\"odinger and Poisson equations using the Fourier Transform (FT) because it is convenient to the  implementation of periodic boundary conditions on $\Psi$ and $V$.  
Poisson equation in the Fourier space reads

\begin{equation}
-p^2\mathcal{F}(V) = \mathcal{F}\left(|\Psi|^2-\langle|\Psi|^2\rangle\right),
\end{equation}

\noindent where the FT is approximated by a Fast Fourier Transform. Notice that for the mode $p:=|\vec{p}|=0$ the identity $0=0$ holds, which allows $\mathcal{F}(V)(\vec{p}=0)$ to take any value, particularly useful to choose a value at the boundary. We choose the condition $\mathcal{F}(V)(\vec{p}=0)=0$. Finally, the solution to Poisson equation is given by the inverse FT:

\begin{equation}
V = \mathcal{F}^{-1}\left(
\dfrac{-\mathcal{F}\left(|\Psi|^2-\langle|\Psi|^2\rangle\right)}{p^2}
\right).\label{eq:IFTPoisson}
\end{equation}

\noindent On the other hand, Schr\"odinger equation is discretized using the implicit Crank-Nicolson average, so that the evolution from time $t^n$ to $t^{n+1}$ is formally written as

\begin{equation}
\left(1+\dfrac{1}{2}{\rm i}\Delta t\hat{H}^{n+1}\right)\Psi^{n+1} = \left(1-\dfrac{1}{2}{\rm i}\Delta t\hat{H}^{n}\right)\Psi^{n}.
\label{eq: CN}
\end{equation}

\noindent The integration from time $t^n$ to $t^{n+1}$ uses a three step splitting

\begin{eqnarray}
\left( 1+\frac{\rm i}{4} \Delta t ~V^n\right) \Psi^{n+\alpha} &=& 
\left( 1-\frac{\rm i}{4} \Delta t ~V^n\right) \Psi^{n}\label{eq:split1}\\
\left( 1-\frac{\rm i}{4} \Delta t \nabla^2 \right) \Psi^{n+\beta} &=& 
\left( 1+\frac{\rm i}{4} \Delta t \nabla^2 \right) \Psi^{n+\alpha}\label{eq:split2}\\
\left( 1+\frac{\rm i}{4} \Delta t ~V^n\right) \Psi^{n+1} &=& 
\left( 1-\frac{\rm i}{4} \Delta t ~V^n\right) \Psi^{n+\beta}\label{eq:split3}
\end{eqnarray}

\noindent where $V^n$ is the gravitational potential at time $t^n$. The first step (\ref{eq:split1}) can be solved easily

\begin{equation}
\Psi^{n+\alpha} = \dfrac{1-\frac{1}{4}{\rm i}\Delta t V^n}{1+\frac{1}{4}{\rm i}\Delta t V^n}\Psi^{n},
\label{eq: first step CN}
\end{equation}

\noindent whereas the second step (\ref{eq:split2}) is solved using the Fourier Transform as follows:

\begin{equation}
\Psi^{n+\beta} = \mathcal{F}^{-1}\left\lbrace
\dfrac{1-\frac{1}{4}{\rm i}\Delta t ~ p^2}{1+\frac{1}{4}{\rm i}\Delta t ~ p^2}\mathcal{F}\left(\Psi^{n+\alpha}\right)
\right\rbrace,
\label{eq: second step CN}
\end{equation}

\noindent and finally the third step, equation (\ref{eq:split3}), is solved:

\begin{equation}
\Psi^{n+1} = \dfrac{1-\frac{1}{4}{\rm i}\Delta t V^n}{1+\frac{1}{4}{\rm i}\Delta t V^n}\Psi^{n+\beta}. 
\label{eq: third step CN}
\end{equation}

\noindent Notice a very important subtlety. In equations (\ref{eq:split1}) and (\ref{eq:split3}) we use the gravitational potential evaluated at time $t^n$, whereas the Crank-Nicholson method requires the average of the Hamiltonian in time,  between times $t^n$ and $t^{n+1}$. Thus we use the so far calculated $\Psi^{n+1}$ in (\ref{eq: third step CN}), integrate Poisson equation and obtain the potential $V^{n+1}$ with (\ref{eq:IFTPoisson}). We then define the average potential $V^{n+1/2} = \frac{1}{2}(V^n+V^{n+1})$ that we use to implement again the steps (\ref{eq:split1})-(\ref{eq:split3}). The result will be $\Psi^{n+1}$ with the appropriate average potential.

We have tested the code in the case of isolation conditions (e.g. \cite{Nkode3d,AlvarezGuzman2022}). However, for the implementation of periodic boundary conditions we do not have previous tests, that is why in the Appendix we add an essential testbed related to the evolution of a boosted equilibrium configuration.

\subsection{Diagnostics}

The evolution is monitored using expectation values of the variables calculated within the spatial numerical domain $D_d$. These quantities can be the mass $M$, kinetic energy $K$, gravitational energy $W$, momentum $\vec{P}$ and angular momentum $\vec{L}$:

\begin{equation}
\begin{array}{rcl}
M_{\Omega} & = & \int_{\Omega} |\Psi|^2 d^3x, \\
\\
K_{\Omega} & = & -\frac{1}{2}\int_{\Omega} \Psi^* \nabla^2 \Psi d^3x, \\
\\
W_{\Omega} & = & \frac{1}{2}\int_{\Omega} V |\Psi|^2 d^3x, \\
\\
\vec{P}_{\Omega} & = & -{\rm i} \int_{\Omega} \Psi^* \grad\Psi d^3x, \\
\\
\vec{L}_{\Omega} & = & -{\rm i}\int_{\Omega} \Psi^* \vec{x}\times\nabla\Psi d^3x.
\end{array}
\end{equation}

\noindent In our analysis below, the domain $\Omega$ will be two regions. The region $\Omega=core$ is a sphere of radius $r_c$, the core of a fluctuation that admits a fitting with the solitonic profile \cite{Schive:2014dra,Mocz:2017wlg}:

\begin{equation}
\rho_{core} = \rho_{0,core}\left(1+0.091\left(\dfrac{r}{r_c}\right)^{2}\right)^{-8}.
\label{eq:coreprofile}
\end{equation}

\noindent The second region $\Omega=tail$ is what we call the tail, which is the region within $D_d$ but outside of the $core$. Other additional useful quantities are the total energy $E_{\Omega}=K_{\Omega} + W_{\Omega}$, which helps measuring the dissipation of the methods during the evolution and $2K_{\Omega} + W_{\Omega}$ that helps monitoring when the system is near a virialized state. 

Of importance to our analysis of mergers is the calculation of $\vec{P}$ on semi-domains, in that case we estimate the momentum along the $x-$direction in the domains $\Omega$ such that $x<0$ and $x \ge 0$.

\section{Comparison in various scenarios}
\label{sec:results}

We perform the comparison between the isolated and periodic domain using a set of scenarios. These include the evolution of a ground state equilibrium configuration, the collapse of a fluctuation with a Gaussian profile and the collision of two equilibrium configurations. In each case we monitor specially the density of matter and dynamical variables in order to measure the effects of boundary conditions. The simulations are carried out in a convergence regime with spatial resolution $\Delta x=\Delta y = \Delta z = 0.62$, time resolution $\Delta t=0.05$ and various specified domains sizes.

\subsection{Evolution of a ground state configuration}

Ground state configurations were originally constructed assuming isolation boundary conditions \cite{Ruffini:1969,GuzmanUrena2004}. When this configuration is evolved assuming isolation boundary conditions, specifically, if a sponge is implemented for the wave function $\Psi$ to be absorbed near the boundary, and the gravitational potential is constructed with isolation boundary conditions, this configuration remains stationary in the continuum limit as shown with convergence tests in \cite{GuzmanUrena2006}.

\begin{figure}
\includegraphics[width=4.15cm]{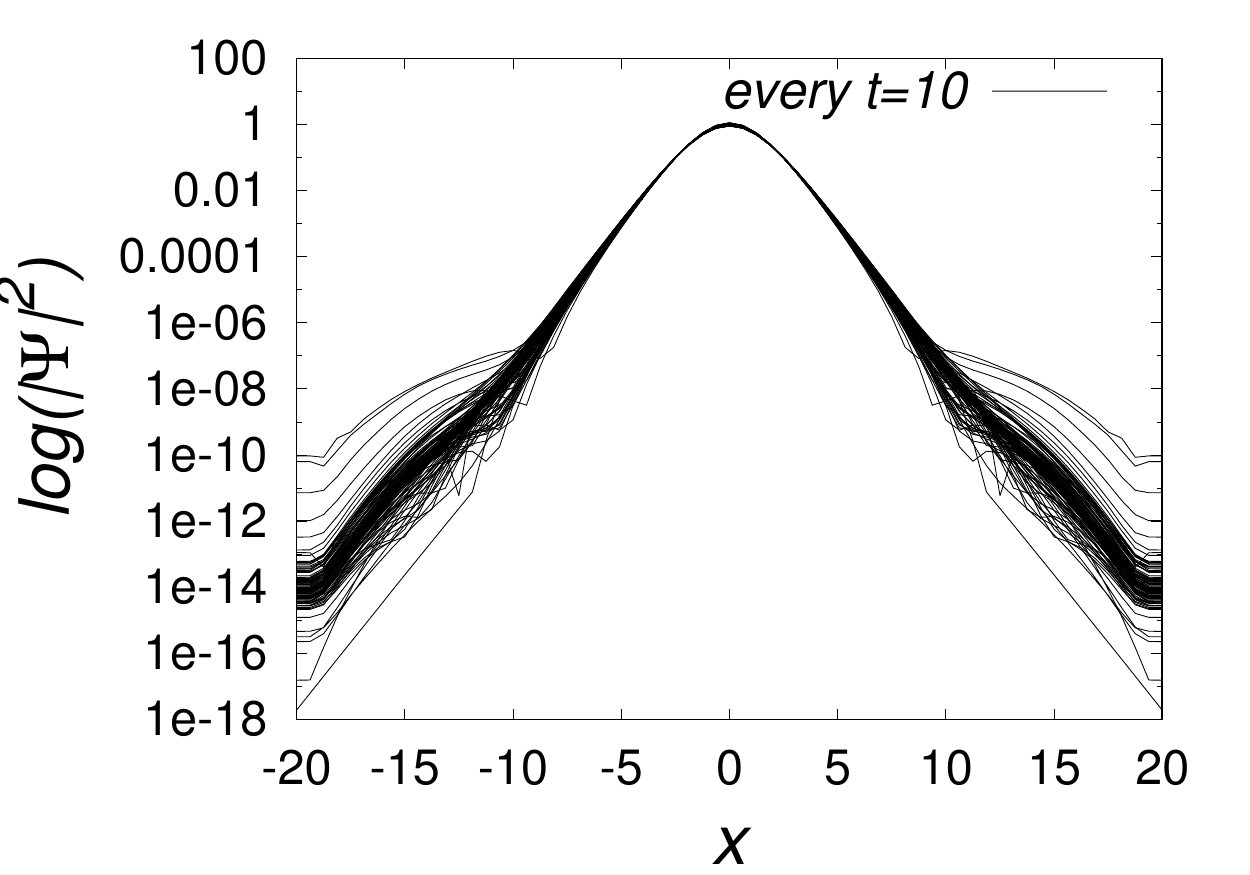}
\includegraphics[width=4.15cm]{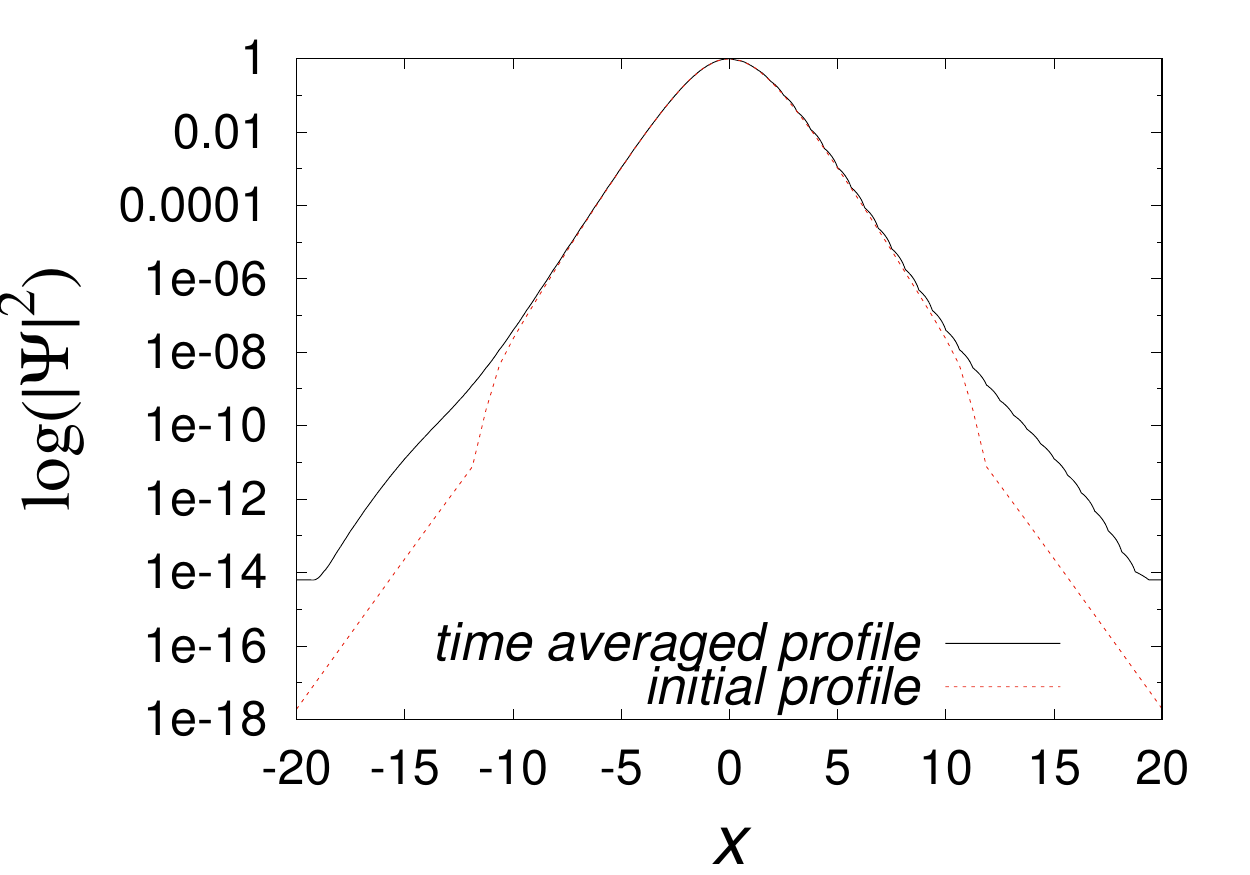}
\includegraphics[width=4.15cm]{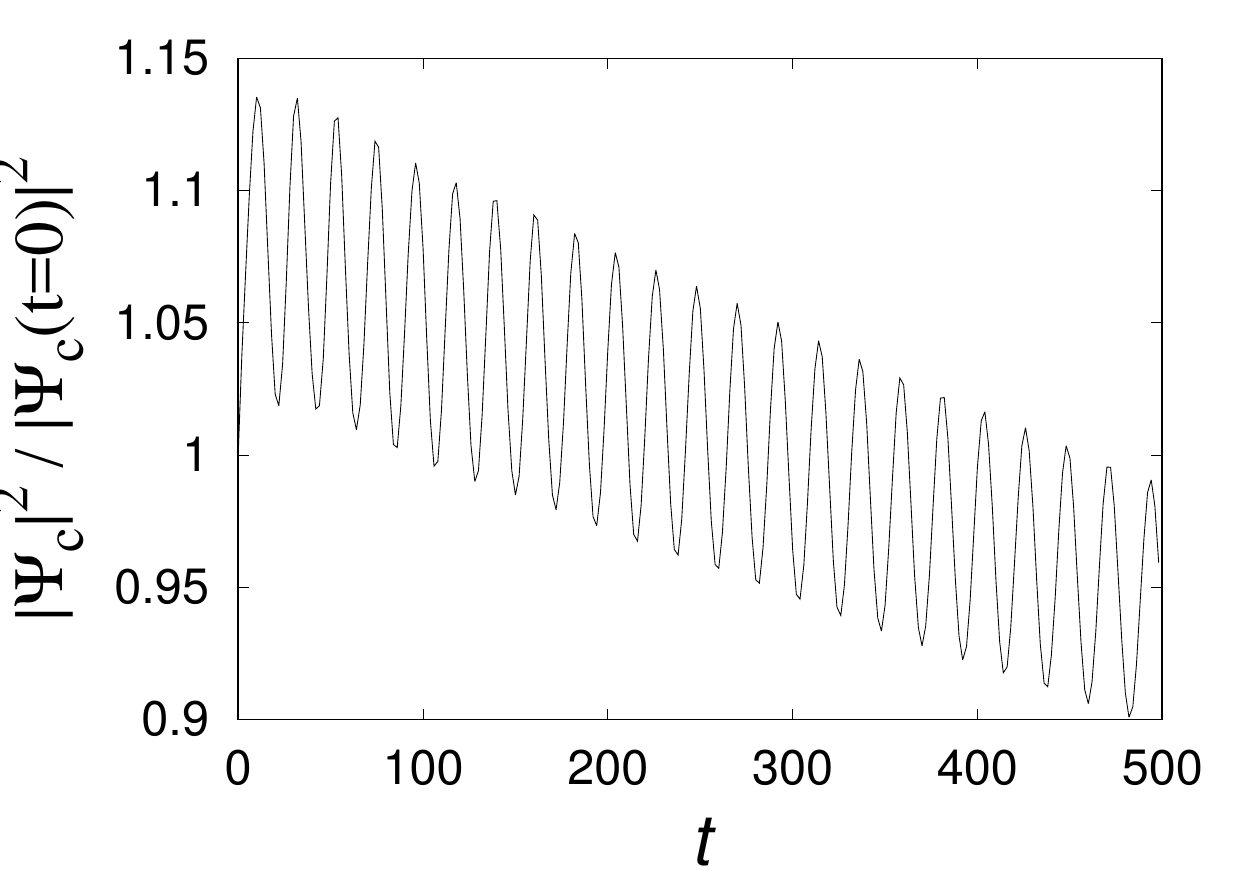}
\includegraphics[width=4.15cm]{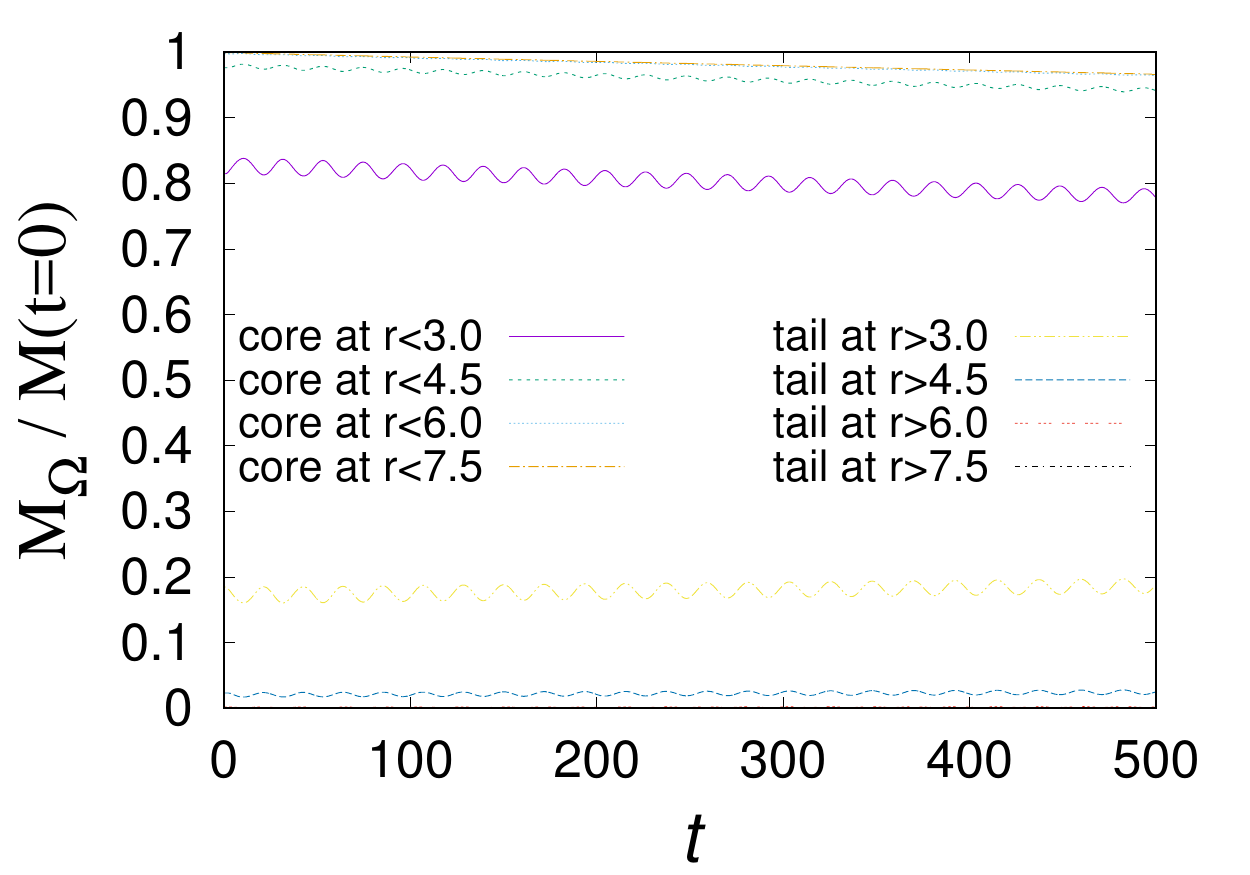}
\caption{\label{fig:isolated equilibrium20} 
Evolution of a ground state equilibrium configuration on the domain $[-20,20]^3$ using isolated boundary conditions. 
(Top) Snapshots of the density $|\Psi|^2$ at various times that illustrates the core and the restless behavior of the density outside of the core. Also shown is the time average of the density profile from $t$=250 to $t$=500 together with the initial profile.
(Bottom) Central density as function of time that shows the oscillations of the core and the mass calculated using different core radius $r_c=3, 4.5, 6.0$ and $7.0$.}
\end{figure}

\begin{figure}
\includegraphics[width=4.15cm]{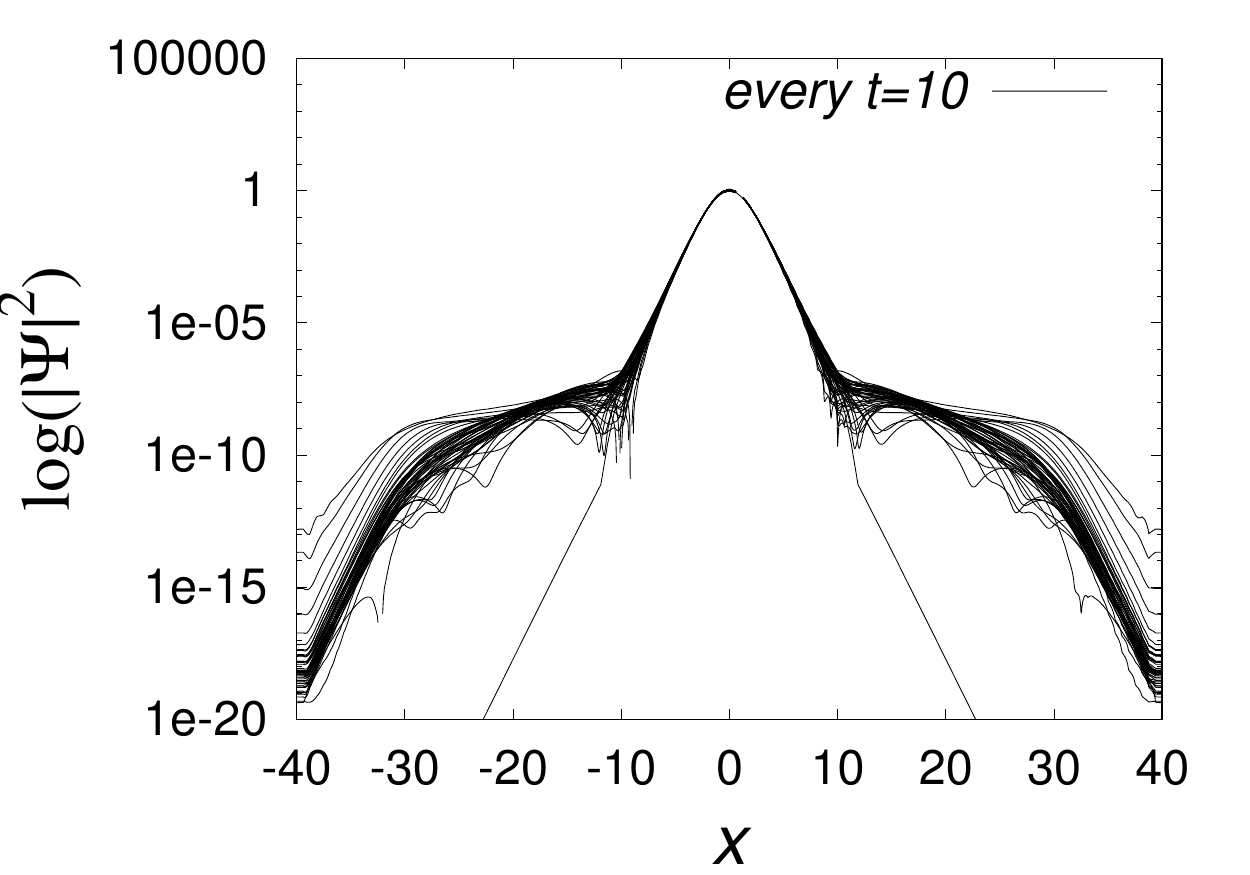}
\includegraphics[width=4.15cm]{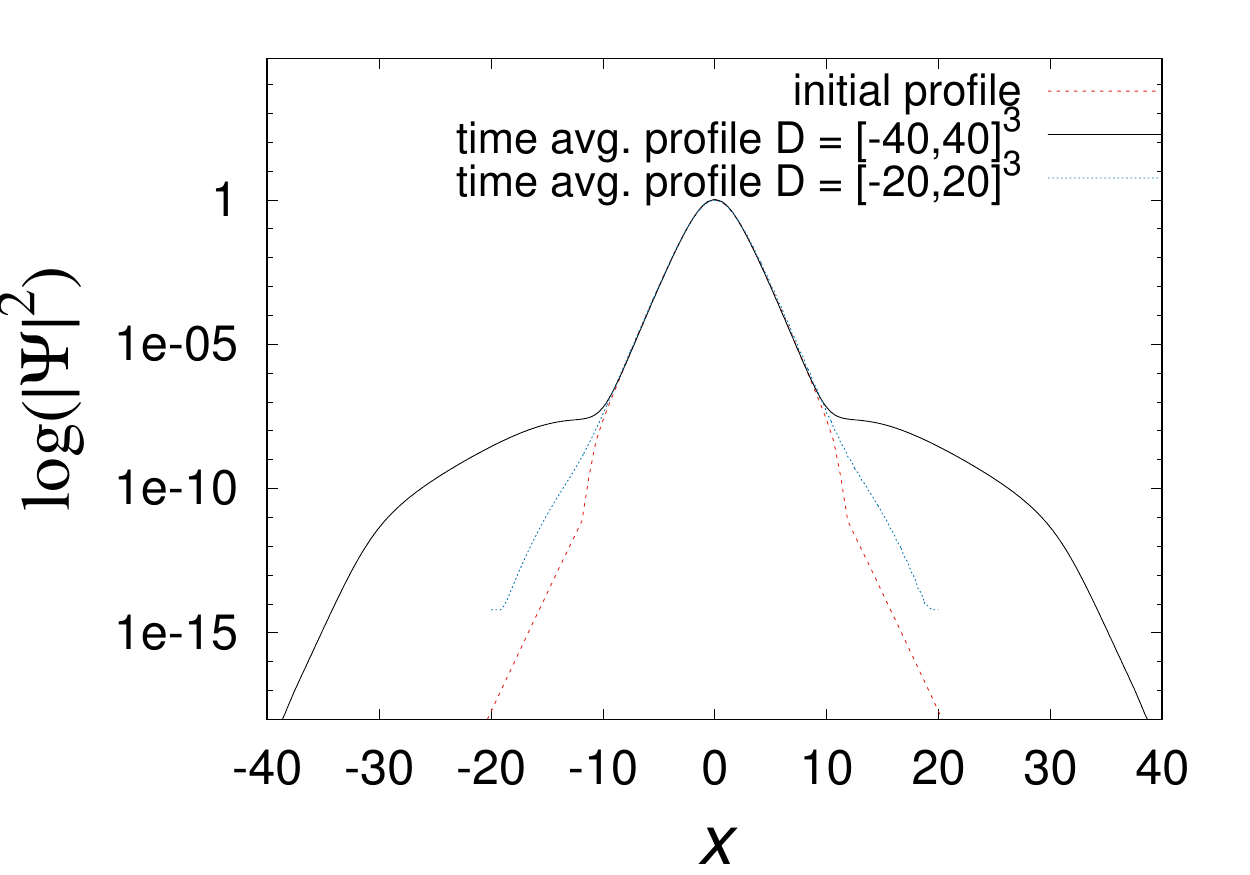}
\includegraphics[width=4.15cm]{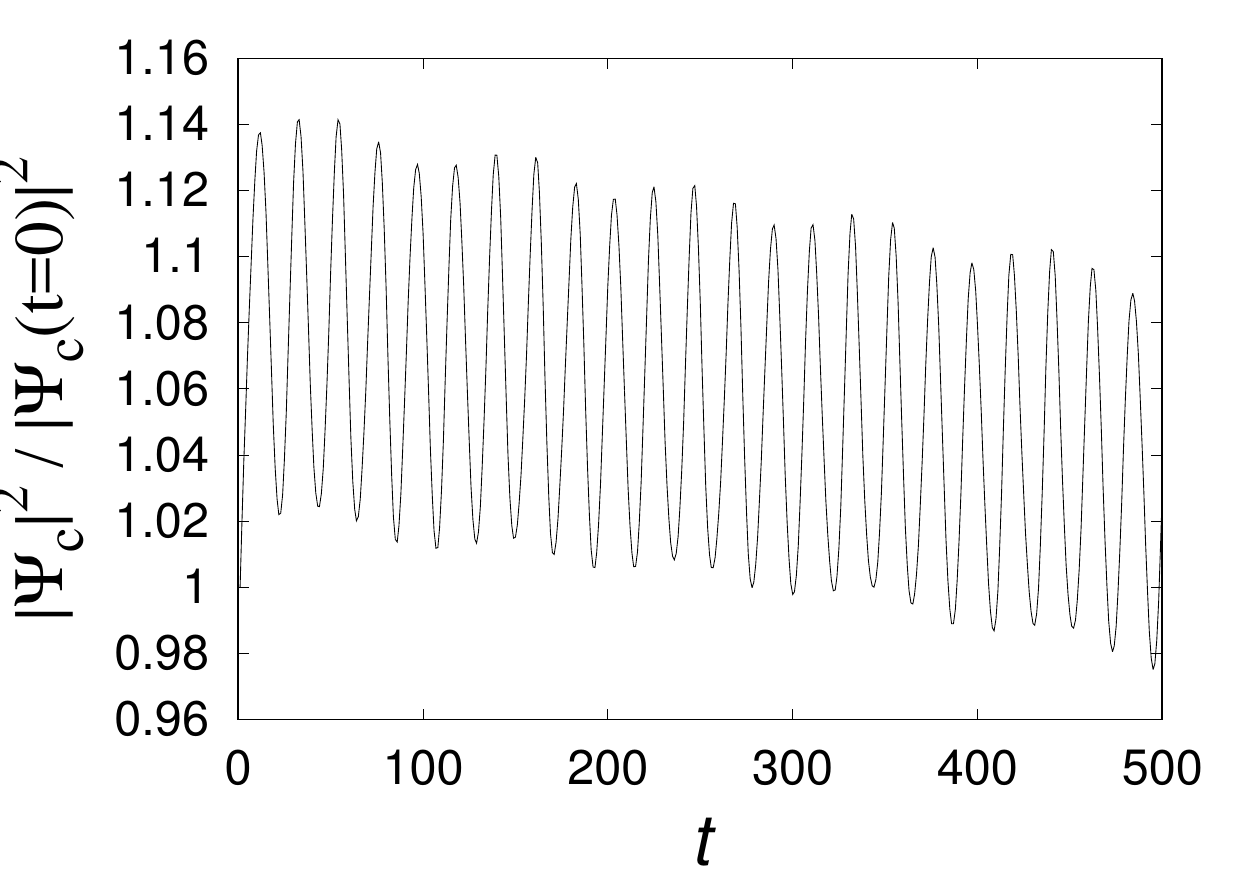}
\includegraphics[width=4.15cm]{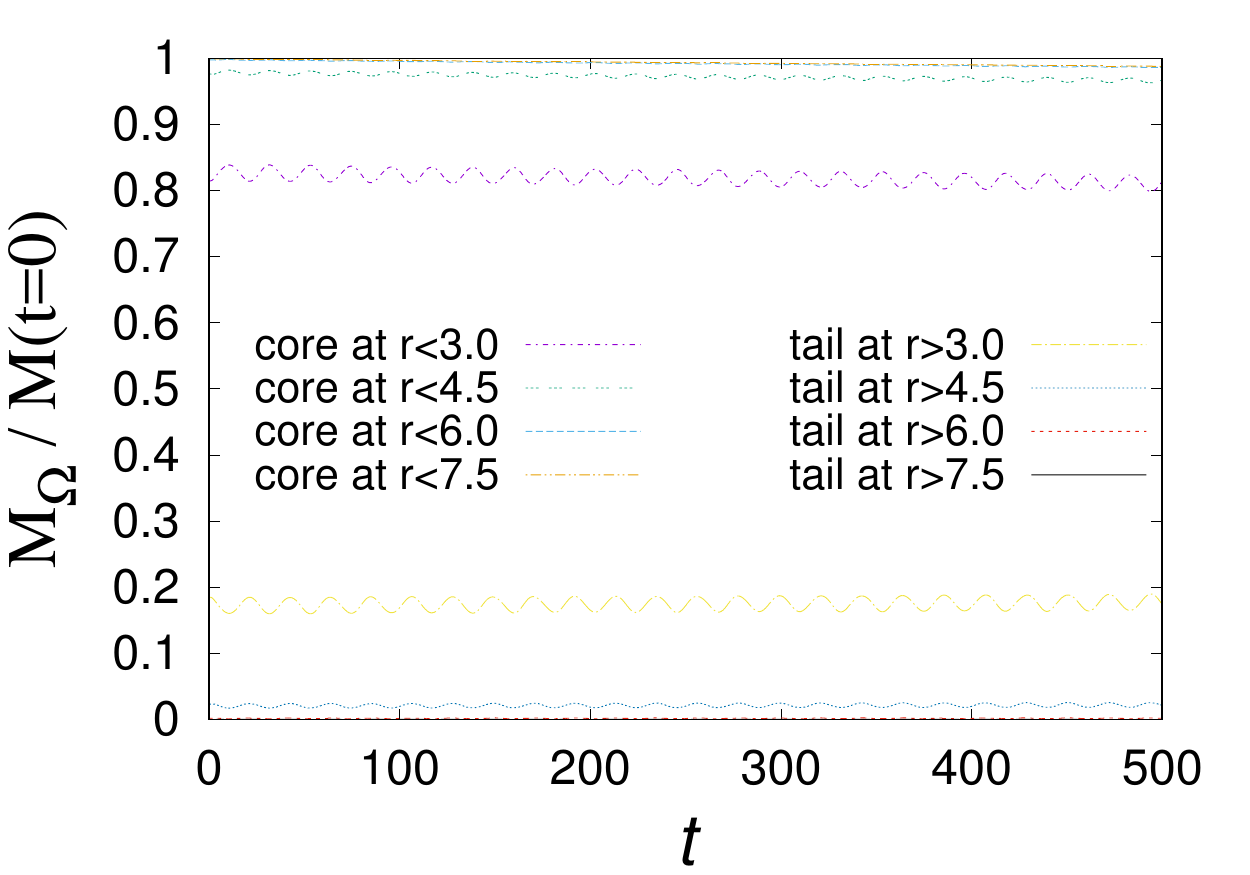}
\caption{\label{fig:isolated equilibrium40} 
Evolution of a ground state equilibrium configuration on the domain $[-40,40]^3$ using isolated boundary conditions. 
(Top) Snapshots of the density $|\Psi|^2$ at various times that illustrates the core and the restless behavior of the density outside of the core. Also shown is the time average of the density profile from $t$=250 to $t$=500 together with  the initial profile.
(Bottom) Central density as function of time that shows the oscillations of the core and the mass calculated using different core radius $r_c=3, 4.5, 6.0$ and $7.0$.}
\end{figure}

We evolve this configuration first using {\it isolation} boundary conditions in two different numerical domain sizes, a small one $[-20,20]^3$ and a big one $[-40,40]^3$, that illustrate the effects of the boundary. The results are in Figures \ref{fig:isolated equilibrium20} and \ref{fig:isolated equilibrium40} that we comment together. In the first plot we show various snapshots of the density $|\Psi|^2$ where a nearly stationary central core and an outer region where the density oscillates can be seen. A time average of density is also shown that illustrates the fall-off of the density profile with distance. The comparison of the time-average of density shows the effects of the sponge, which absorbes the density outside a sphere of radius $r_c=$18 in the small domain and $r_c=$36 in the big domain. The results in these figures illustrate that isolation conditions do not necessarily imply transparent boundary conditions. 
In the second row of results in these two Figures, we show the central density that shows the oscillations of the core and total mass as functions of time. In the big domain the effects of the sponge, and therefore of the domain size, are smaller and the mass is better preserved.

\begin{figure}
\includegraphics[width=4.15cm]{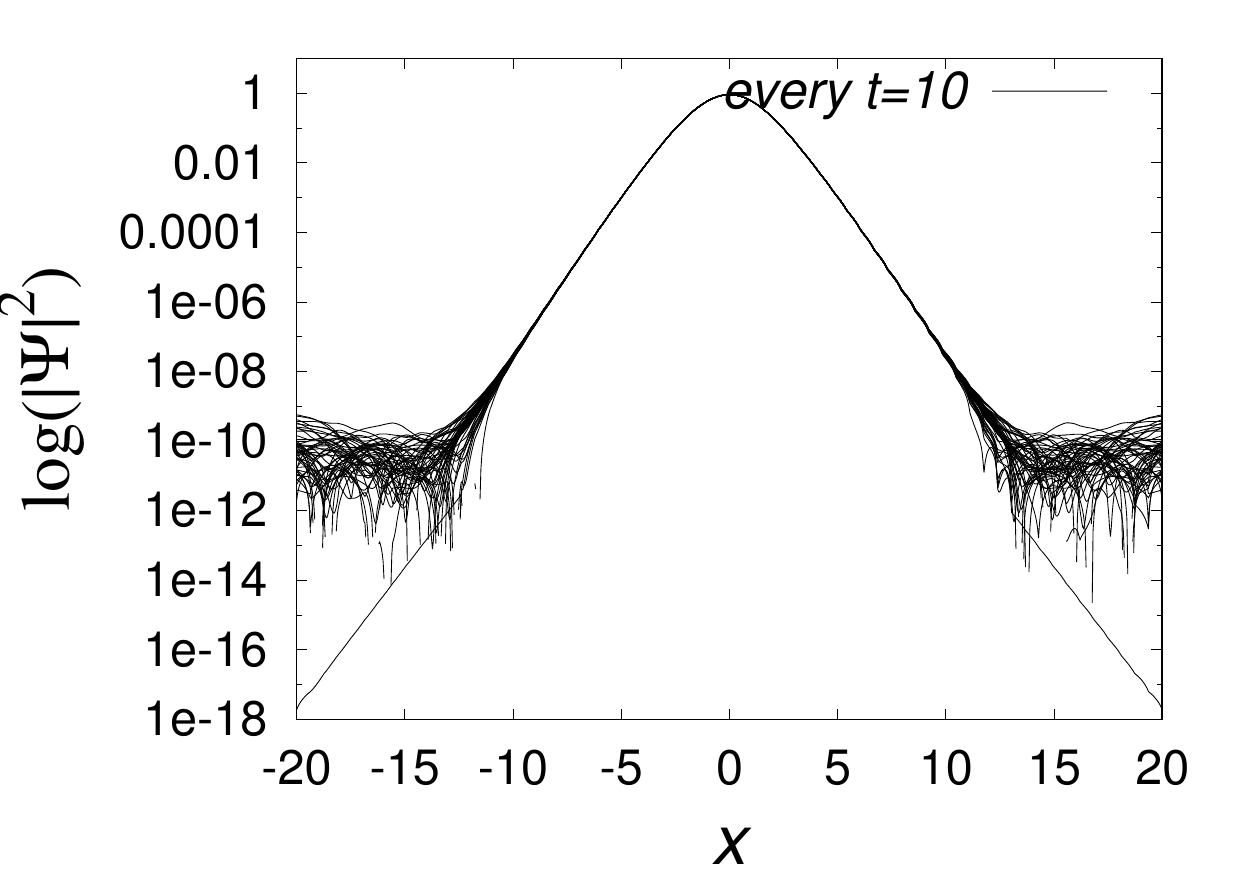}
\includegraphics[width=4.15cm]{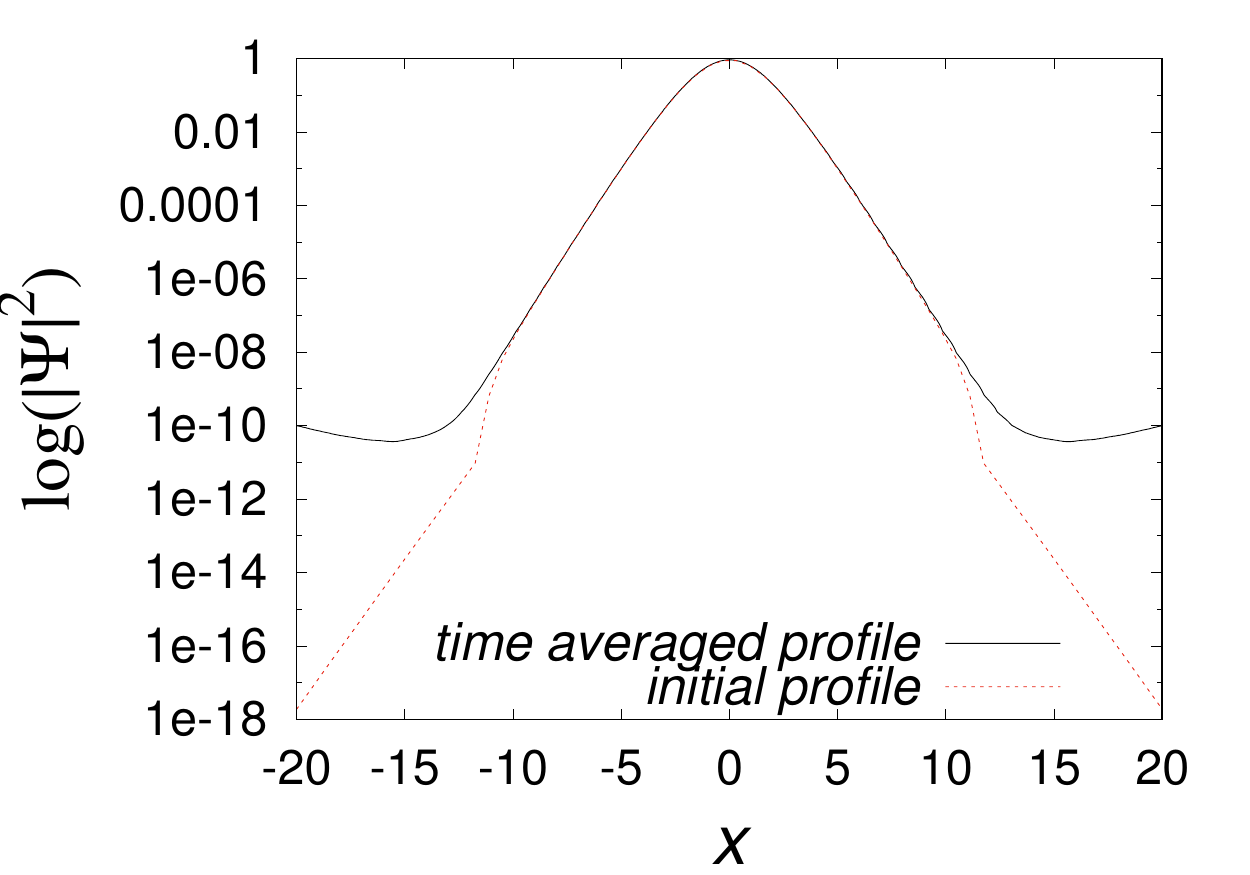}
\includegraphics[width=4.15cm]{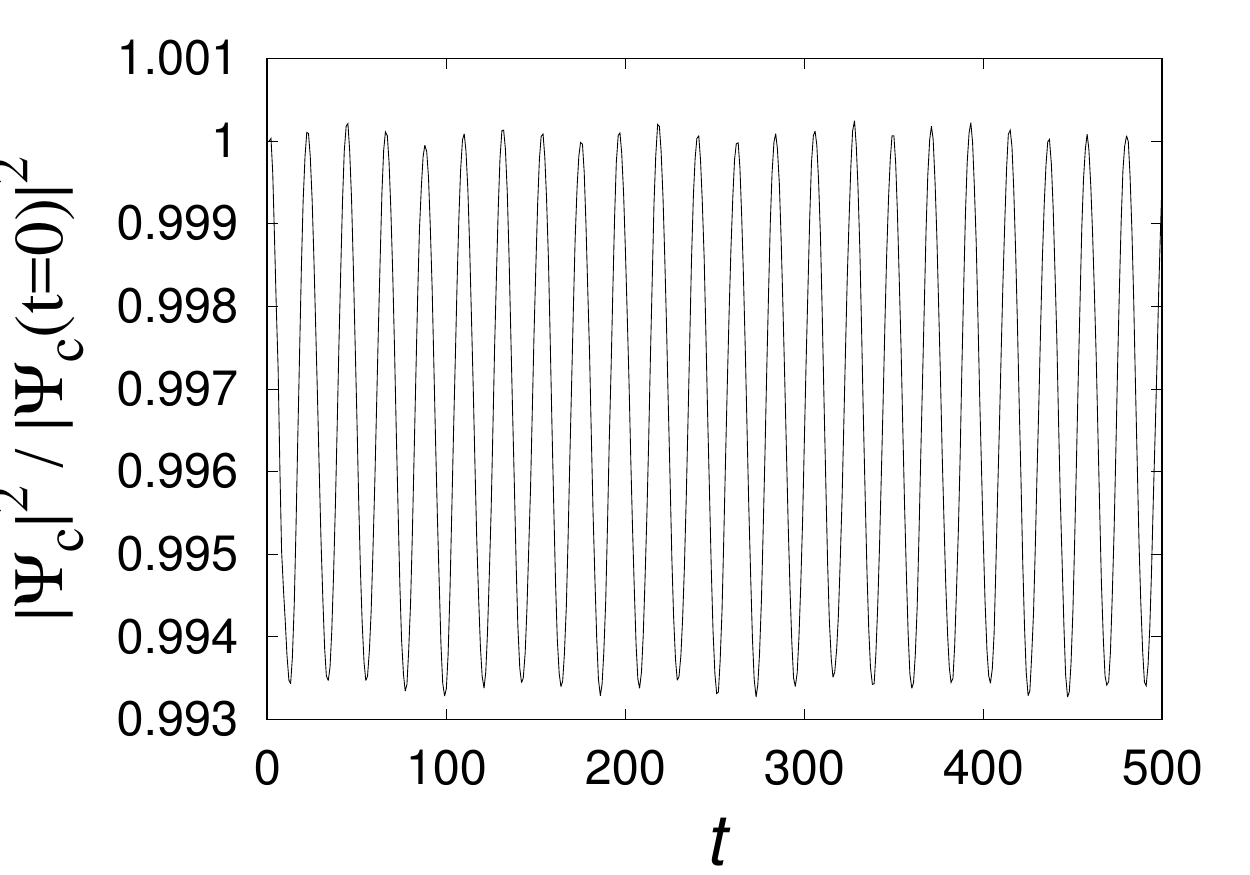}
\includegraphics[width=4.15cm]{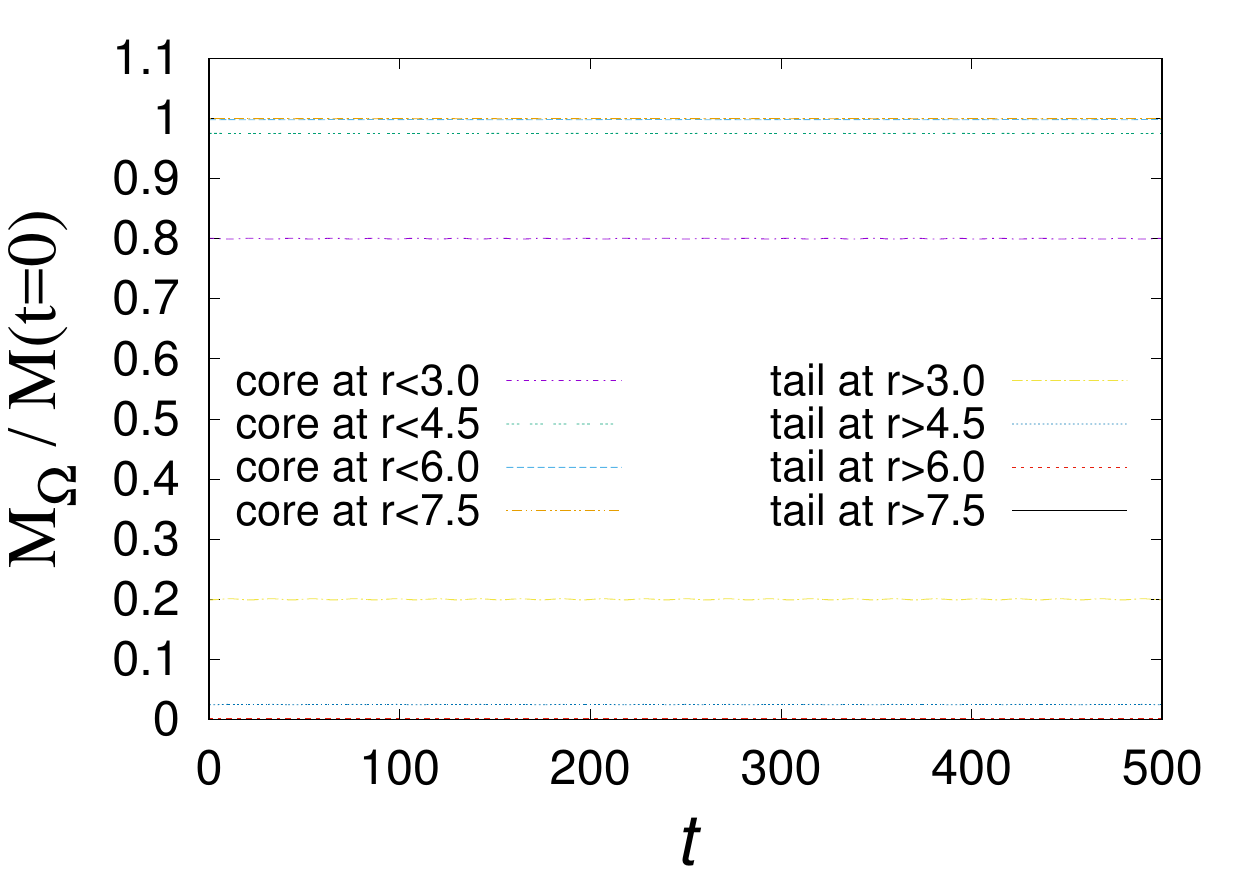}
\caption{\label{fig:periodic equilibrium20} 
Evolution of a ground state equilibrium configuration on the domain $[-20,20]^3$ using periodic boundary conditions. 
(Top) Snapshots of the density $|\Psi|^2$ at various times that illustrates the core and the restless behavior of the density outside of the core. Also shown is the time average of the density profile from $t$=250 to $t$=500 together with  the initial profile.
(Bottom) On the left the central density as function of time that shows the oscillations of the core and on the right the 
mass of core and tail, with respect to their initial value using different core radius $r_c=3, 4.5, 6.0$ and $7.0$.}
\end{figure}

\begin{figure}
\includegraphics[width=4.15cm]{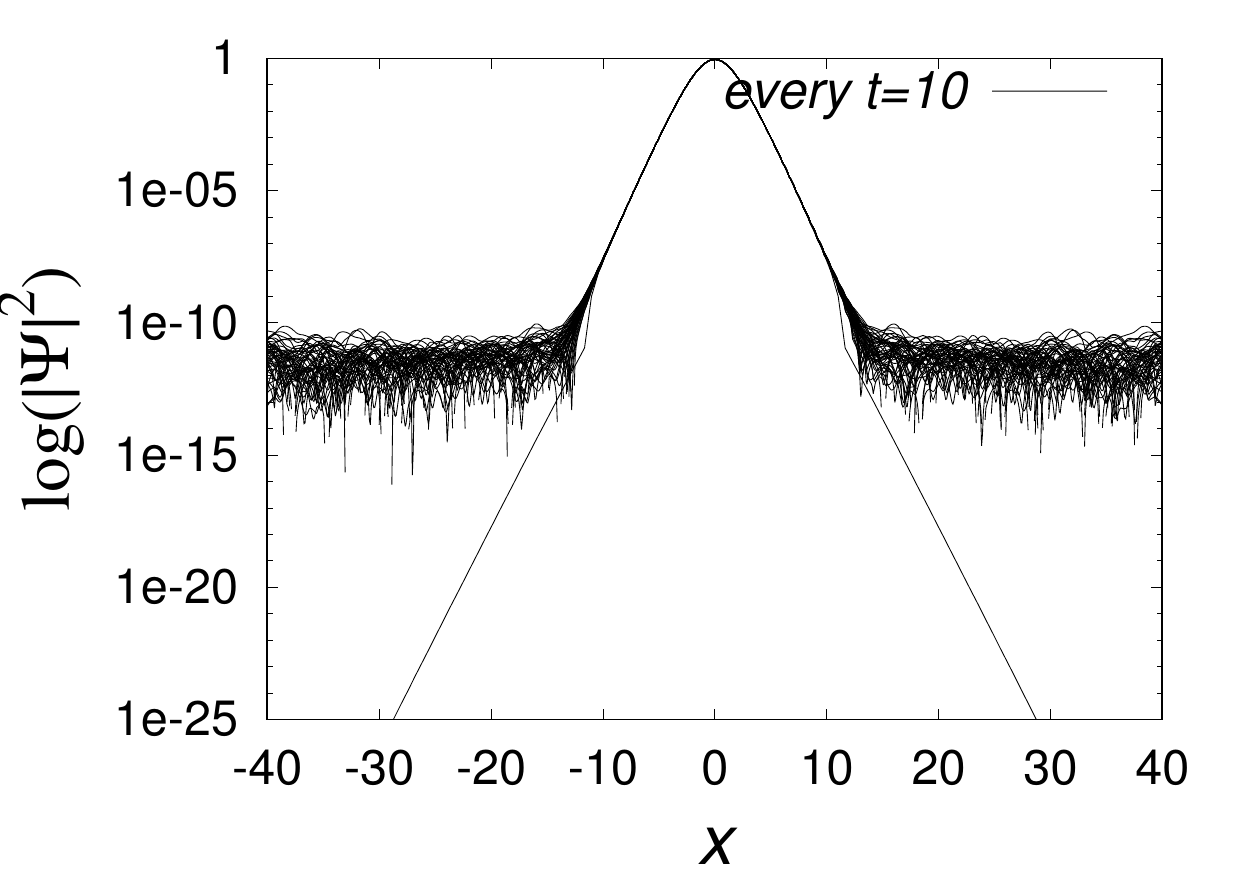}
\includegraphics[width=4.15cm]{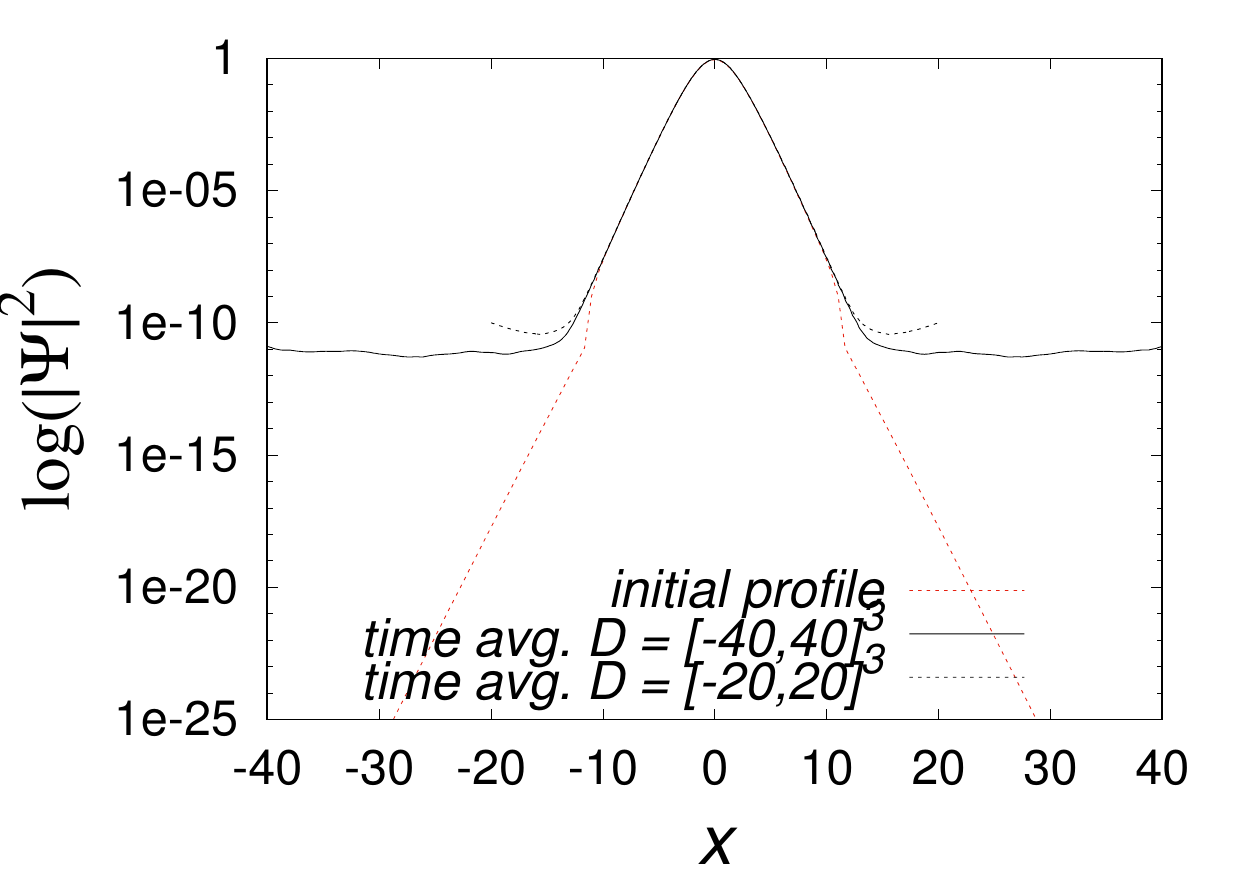}
\includegraphics[width=4.15cm]{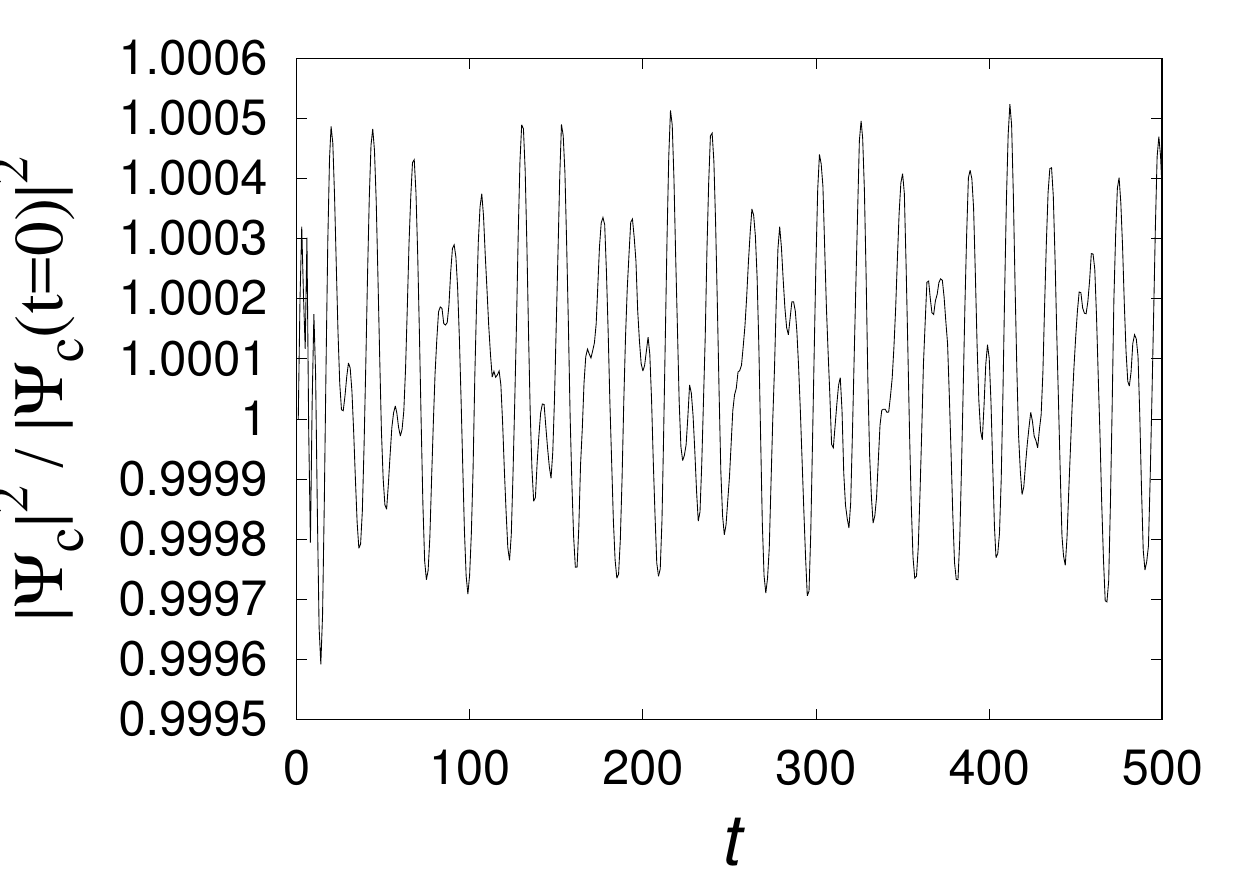}
\includegraphics[width=4.15cm]{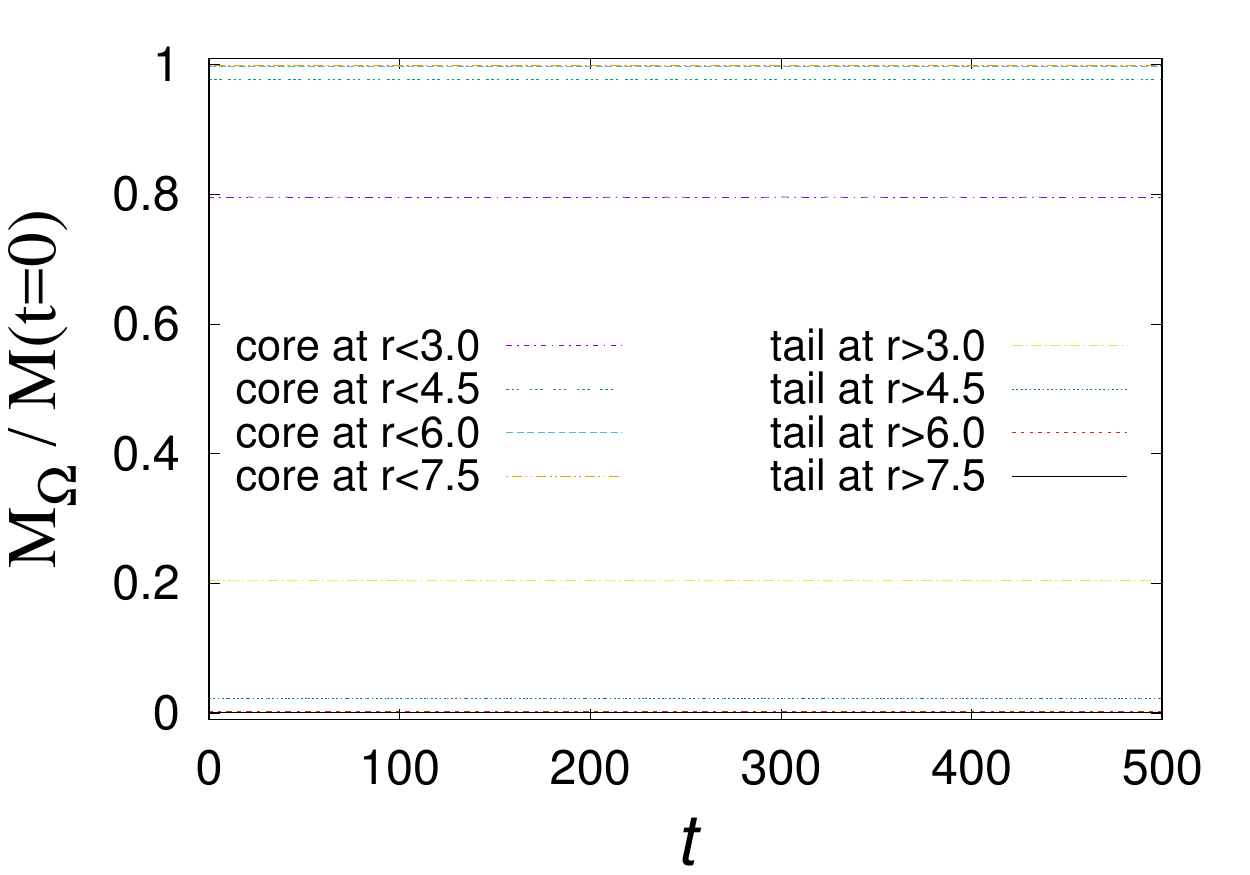}
\caption{\label{fig:periodic equilibrium40} 
Evolution of a ground state equilibrium configuration on the domain $[-40,40]^3$ using periodic boundary conditions. 
(Top) Snapshots of the density $|\Psi|^2$ at various times that illustrates the core and the restless behavior of the density outside of the core. Also shown is the time average of the density profile from $t$=250 to $t$=500, together with the time average solution on the domain $[-20,20]^3$ and the initial density.
(Bottom) On the left the central density as function of time that shows the oscillations of the core and on the right the  mass of core and tail with respect to their initial time using various values of the core radius $r_c=3, 4.5, 6.0$ and $7.0$.}
\end{figure}

We now evolve the same configuration using {\it periodic} boundary conditions. In this case the gravitational potential does not correspond to an isolated system and the boundary conditions at the faces of the cubic domain allow the potential to interact with itself, since the domain is a 3-Torus. The wave function is allowed to leak out from the gravitational potential well and redistribute across the domain, unlike the isolated case, where the sponge absorbs the density near the faces of the box. The evolution of the configuration is shown in Figure \ref{fig:periodic equilibrium20} for the small domain $[-20,20]^3$ and in Figure \ref{fig:periodic equilibrium40} for the big domain $[-40,40]^3$ that we describe together. 

The first row of plots shows a set of snapshots of the density $|\Psi|^2$ starting from the initial configuration. Likewise in the isolated case above, a nearly stationary central core with the initial profile prevails, whereas the density outside the core has a very dynamical behavior and redistributes differently from the isolated case. The time average is also shown and indicates how the tail acquires a slow fall-off with distance that becomes nearly constant in the case of a big domain. The results of this distribution is only due to the periodicity of the domain. Notice that the nearly constant density of the tail is different for the small and big domains. If we consider the core radius $r_c = 7.5$, we find that the mass of the tail redistributes with a factor of $\sim$8.5, which is the factor between the approximately constant tail densities in Figures \ref{fig:periodic equilibrium20} and \ref{fig:periodic equilibrium40}.

The second row of results contains the central density as function of time that illustrates the oscillations of the core density, with a frequency that can be associated to the fundamental quasi normal modes, in this case for a soliton on top of a background density, unlike in vacuum \cite{Guzman2019}. The simulation in the big domain develops a superposed mode that is not seen in the results in the small domain. It is also shown the conservation of mass of the core and tail using different values of the core radius. Notice that unlike the isolated domain, these quantities are better preserved during the evolution since there is no sponge where part of the density would sink.

A summary of results is that the time-average core density profile is independent of the domain size, although it oscillates with modes that are excited differently for different domain size.
The tail has a different profile, because the mass distributes in a bigger volume for the domain $[-40,40]^3$ than in the case $[-20,20]^3$. 

\begin{figure}
\includegraphics[width=7.5cm]{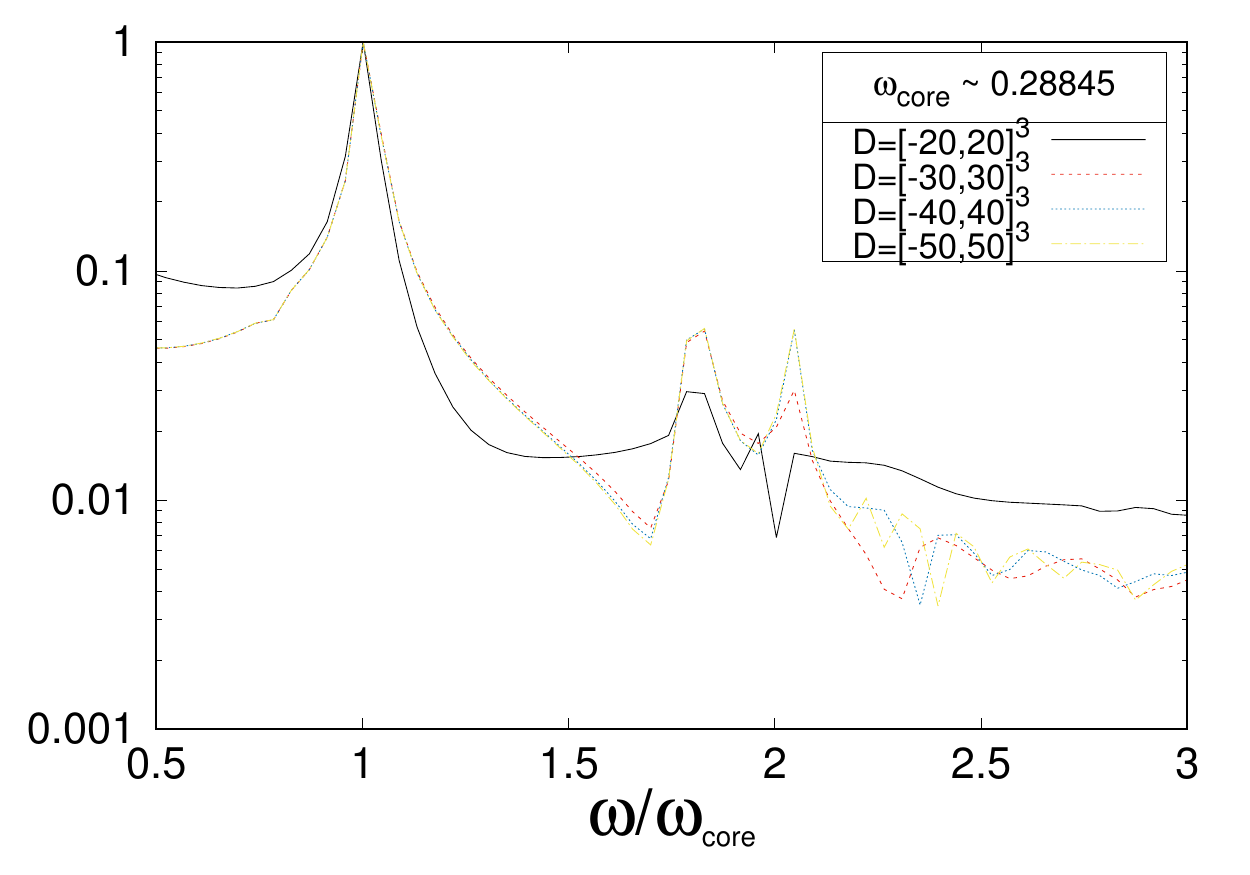}
\includegraphics[width=7.5cm]{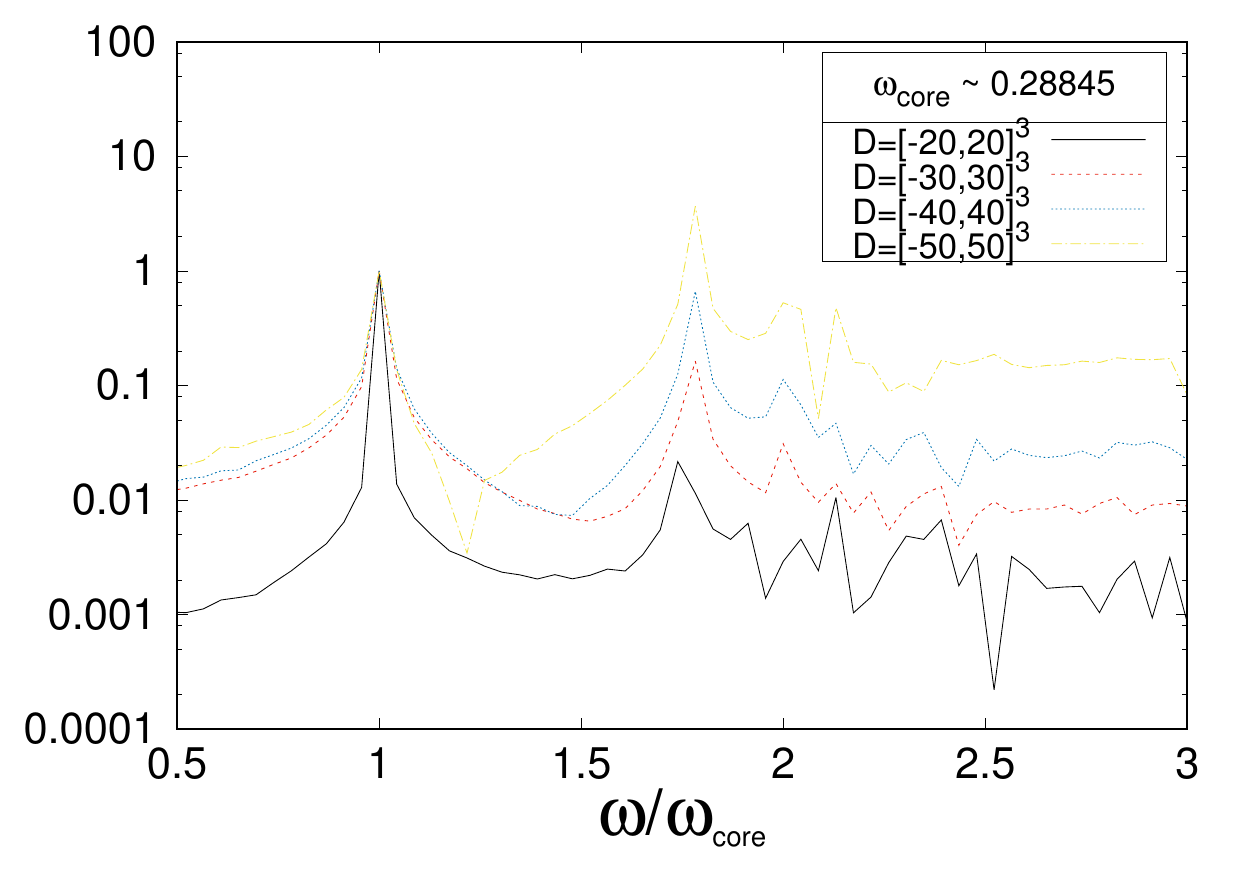}
\caption{\label{fig:frequency equilibrium}
Fourier Transform of the central density as function of time, resulting from the evolution of a ground state equilibrium configuration. At the top the case of an isolated domain, where the two first modes resulting from a spherical perturbation with peaks at 1 and at $\sim 1.768$, as found in \cite{Guzman2019}. At the bottom we show the result when using periodic boundary conditions with two peaks at frequencies 1 and $\sim 1.78$. There is a second peak with high power attributed to the distortion seen in Figure \ref{fig:periodic equilibrium40}. This second peak gains power with the domain size, but it  appears at the same location of the second mode of ground state equilibrium solutions as discussed in \cite{Guzman2019}.
}
\end{figure}

In order to investigate whether periodic boundary conditions trigger the extra oscillation mode seen in the solution with the big domain, we calculate the Fourier Transform of the central density as function of time from Figures \ref{fig:periodic equilibrium20} and \ref{fig:periodic equilibrium40} and two other simulations using spatial domains $[-30,30]^3$ and $[-50,50]^3$ to see the effects of domain size better. The results appear in Figure \ref{fig:frequency equilibrium}. The FT of the evolution using isolation conditions shows two main peaks at frequencies at 1 and at $\sim 1.768$, which coincide with the two first modes of oscillation when a ground state configuration is perturbed with a spherical perturbation \cite{Guzman2019}. 

When using periodic boundary conditions, the use of a bigger domain adds power to the second mode, which is reflected in the height of the second peak in the figure. Based on this observation, the mode superposed on the time-series of the central density of Figure \ref{fig:periodic equilibrium40} can be associated to the second mode of ground state equilibrium configurations, and is not a new oscillation mode, however excited by the reentrance of matter into the domain.

\subsection{Relaxation of a Near-Equilibrium Gaussian}
\label{subsec:GaussianCollapse}

A second problem is the dynamical relaxation of a spherically-symmetric object in near equilibrium, with a Gaussian density profile, as sometimes used to approximate the numerical ground-state equilibrium solution analytically.  In this case, we adopt 
the initial conditions $\Psi({\bf x},0)=A e^{-r^2/(2\sigma^2)}$,  $A=\sqrt{8} (\frac{K^3}{27\pi^3 M})^{1/4} $ and $\sigma=\sqrt{\frac{3 M}{4 K}}$ with approximately the same mass $M$ and kinetic energy $K$ as that of the equilibrium configuration above. The evolution is carried out again in domains $[-20,20]^3$ and $[-40,40]^3$, covered with the same resolution as in the equilibrium configuration case.
\footnote{This relaxation of a spherical object with a Gaussian density profile should not be confused with the Jeans-unstable gravitational collapse and infall calculated by \citep{DawoodbhoyShapiroRindlerDaller2021, ShapiroDawoodbhoyRindlerDaller2022}, described in our introduction.  In the latter case, the initial condition was so ``cold'' that
the ``Jeans length'' in the initial condition was much smaller than the initial Gaussian width, so it was highly
unstable gravitationally. Mass shells in that latter case fall inward, unopposed by pressure, destined to reach the origin in sequence, according to their initial radius (i.e. with shells at initially smaller radii reaching the center first). Before they reach the origin, however, shells are halted by a strong accretion shock that forms at finite radius.  It is this post-shock region that appears as the envelope, outside the solitonic core, in the core-envelope structures identified with FDM halos that form from cosmological initial conditions.    
In the case of the Gaussian presented here, however, the initial condition is not Jeans unstable; it is intended, instead, to be similar to the ground-state profile of the
isolated solution, so does not fit this description.}

\begin{figure}
\includegraphics[width=4.15cm]{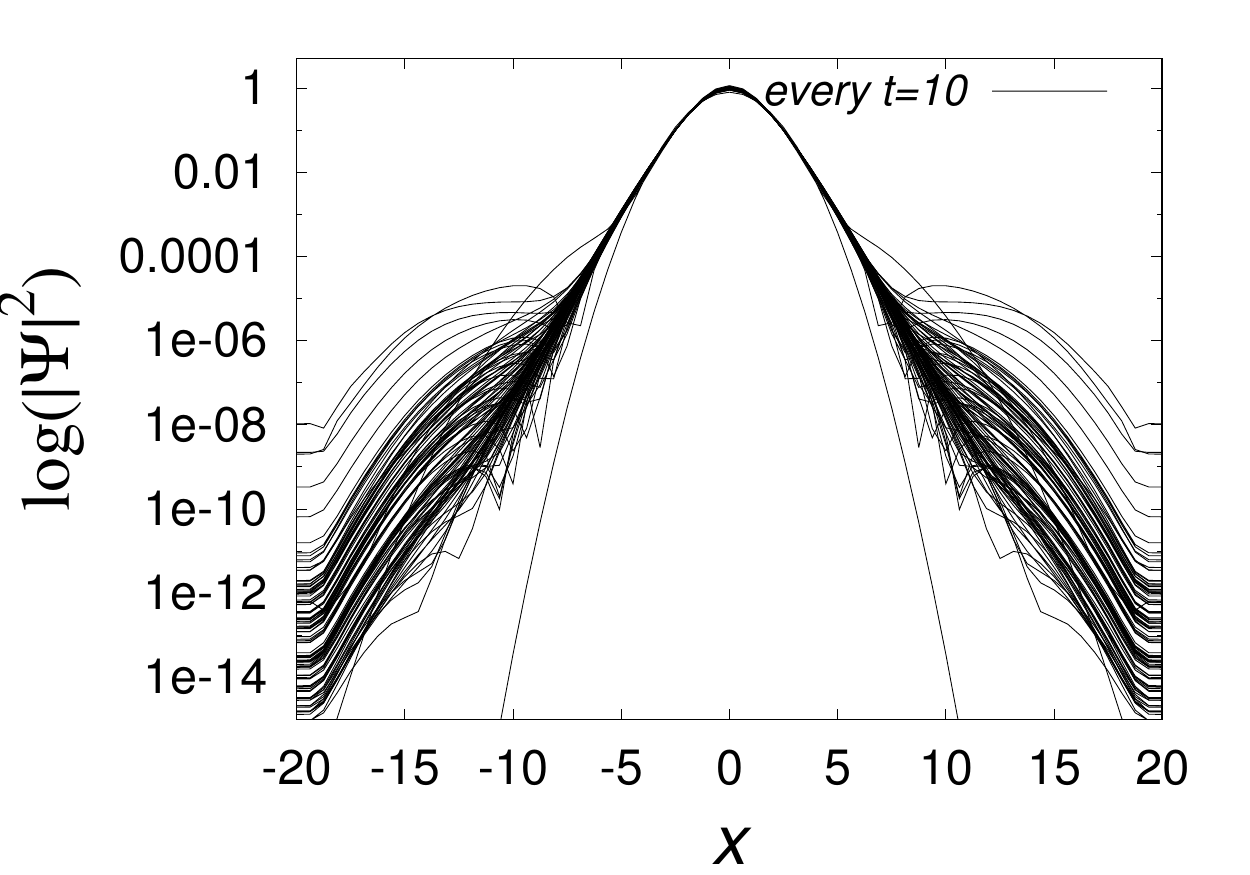}
\includegraphics[width=4.15cm]{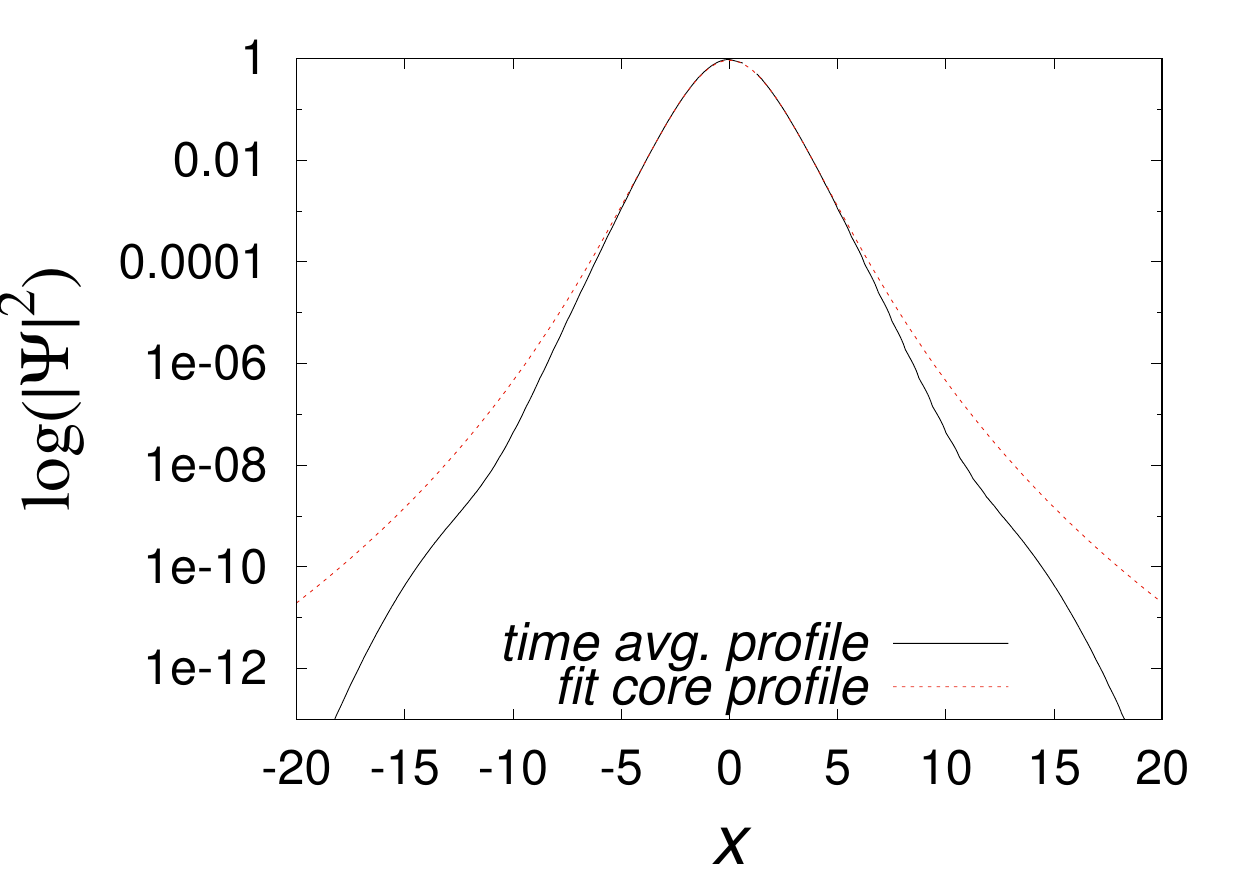}
\includegraphics[width=4.15cm]{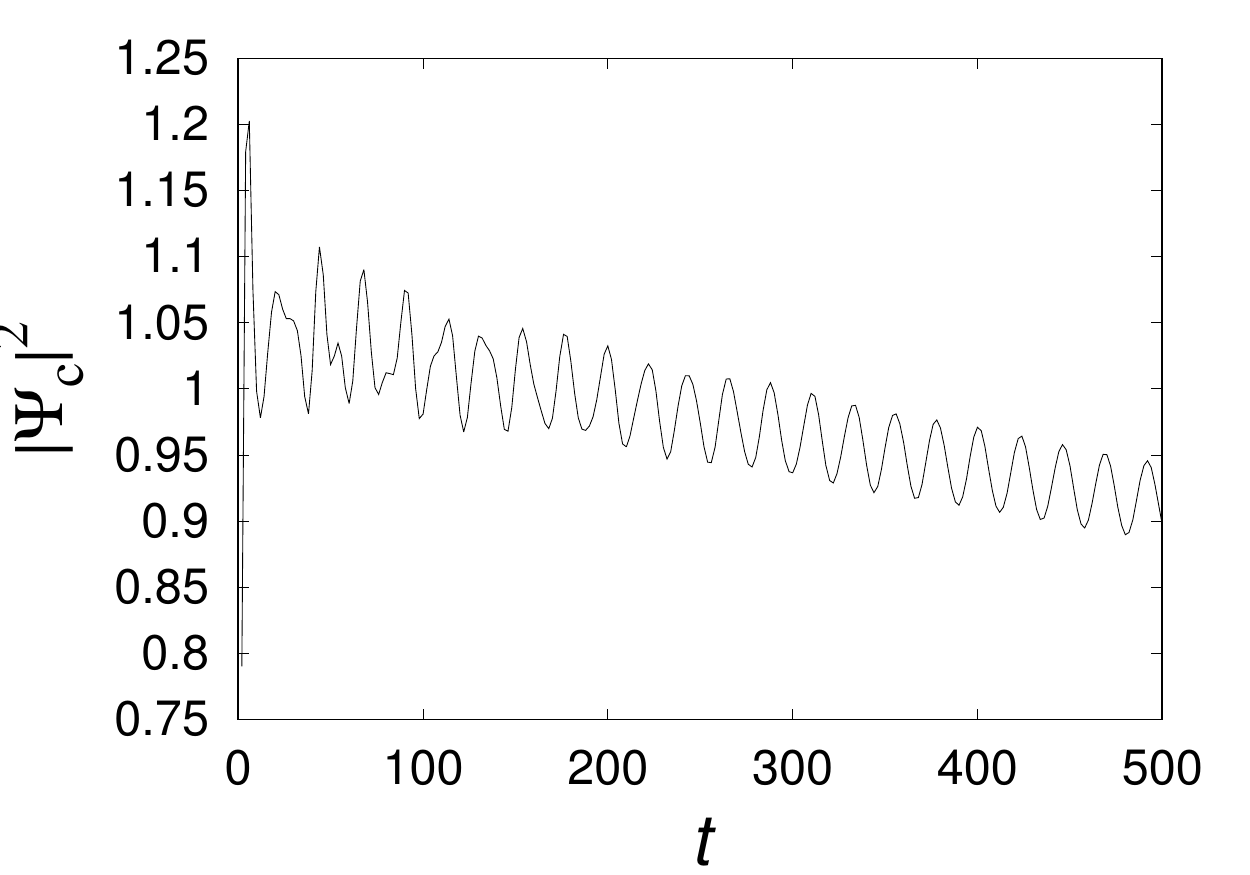}
\includegraphics[width=4.15cm]{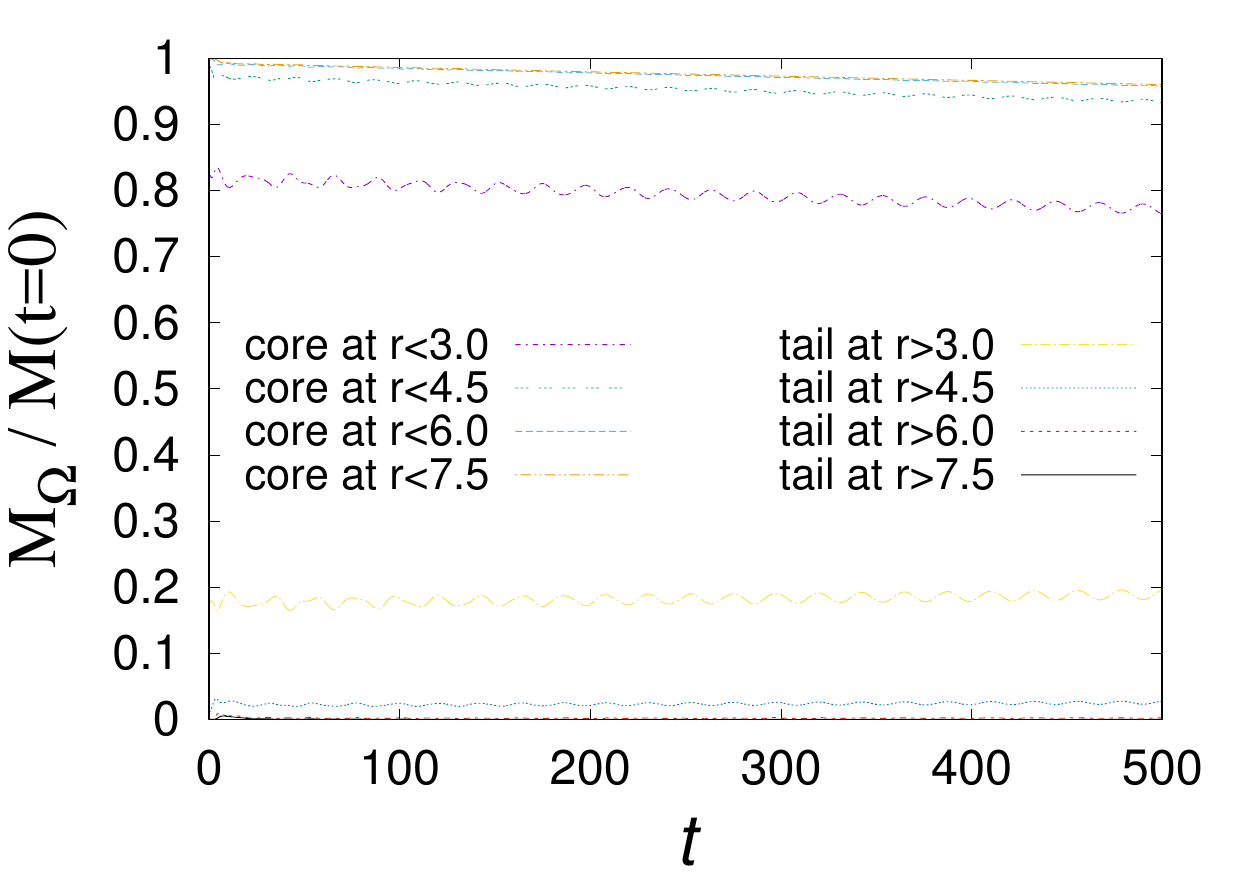}
\caption{\label{fig:isolated gaussian20} 
Evolution of a Gaussian in the domain $[-20,20]^3$ using isolated boundary conditions. 
(Top) Snapshots of the density $|\Psi|^2$ at various times that illustrates the core formation and the restless behavior of the density outside of the core. Also shown is the time average of the density profile from $t$=250 to $t$=500 together with the  solitonic fitting profile.
(Bottom) On the left the central density as function of time that shows the oscillations of the core and on the right the mass of core and tail with respect to their initial value using various values of the core radius $r_c=3, 4.5, 6.0$ and $7.0$.}
\end{figure}

\begin{figure}
\includegraphics[width=4.15cm]{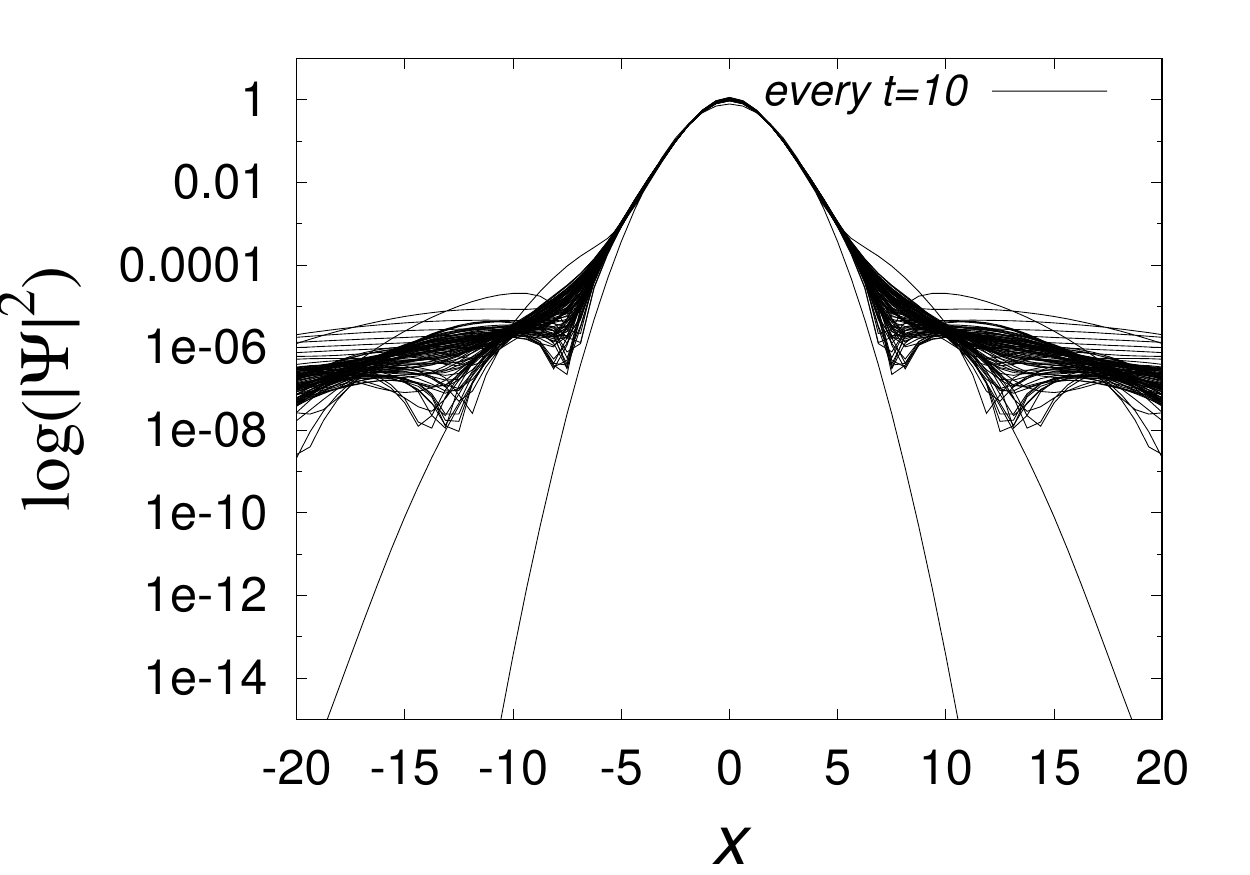}
\includegraphics[width=4.15cm]{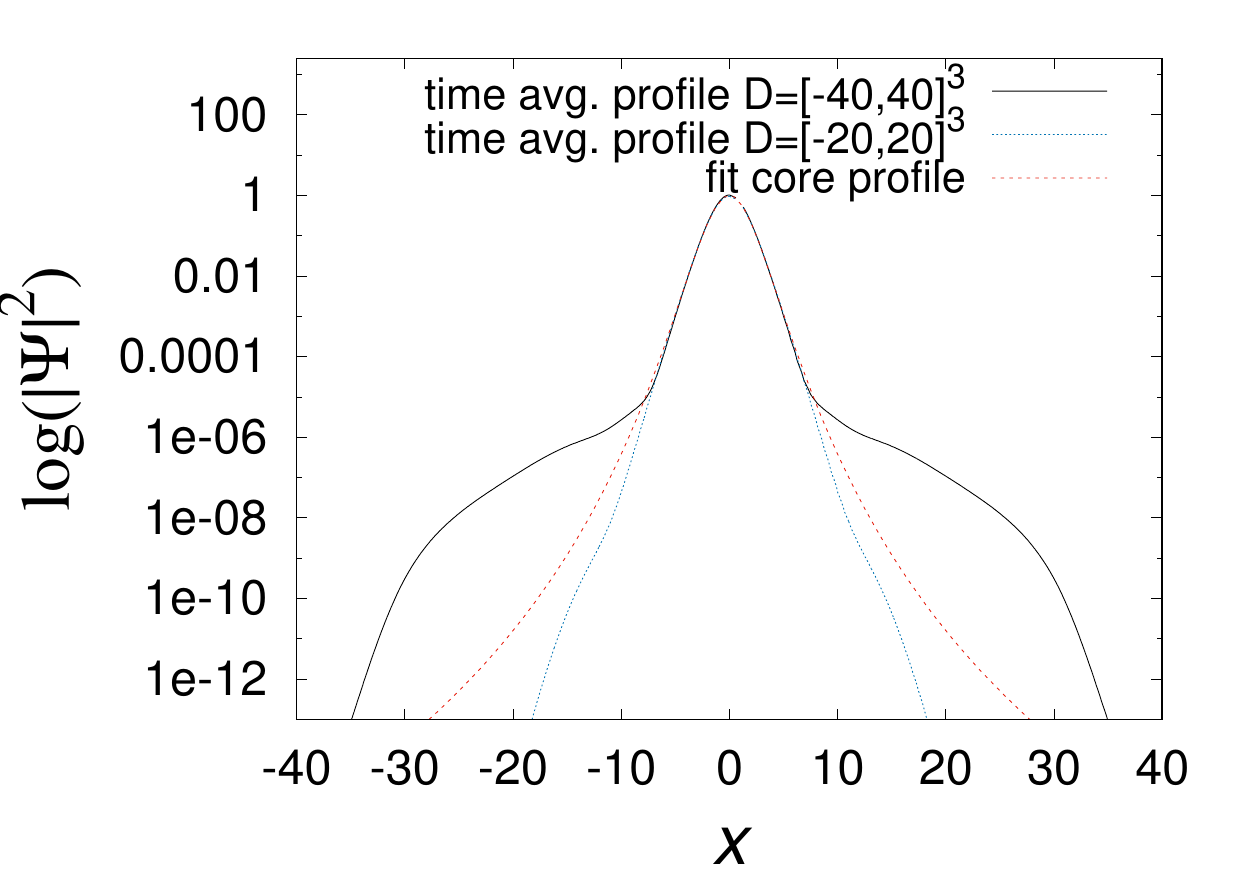}
\includegraphics[width=4.15cm]{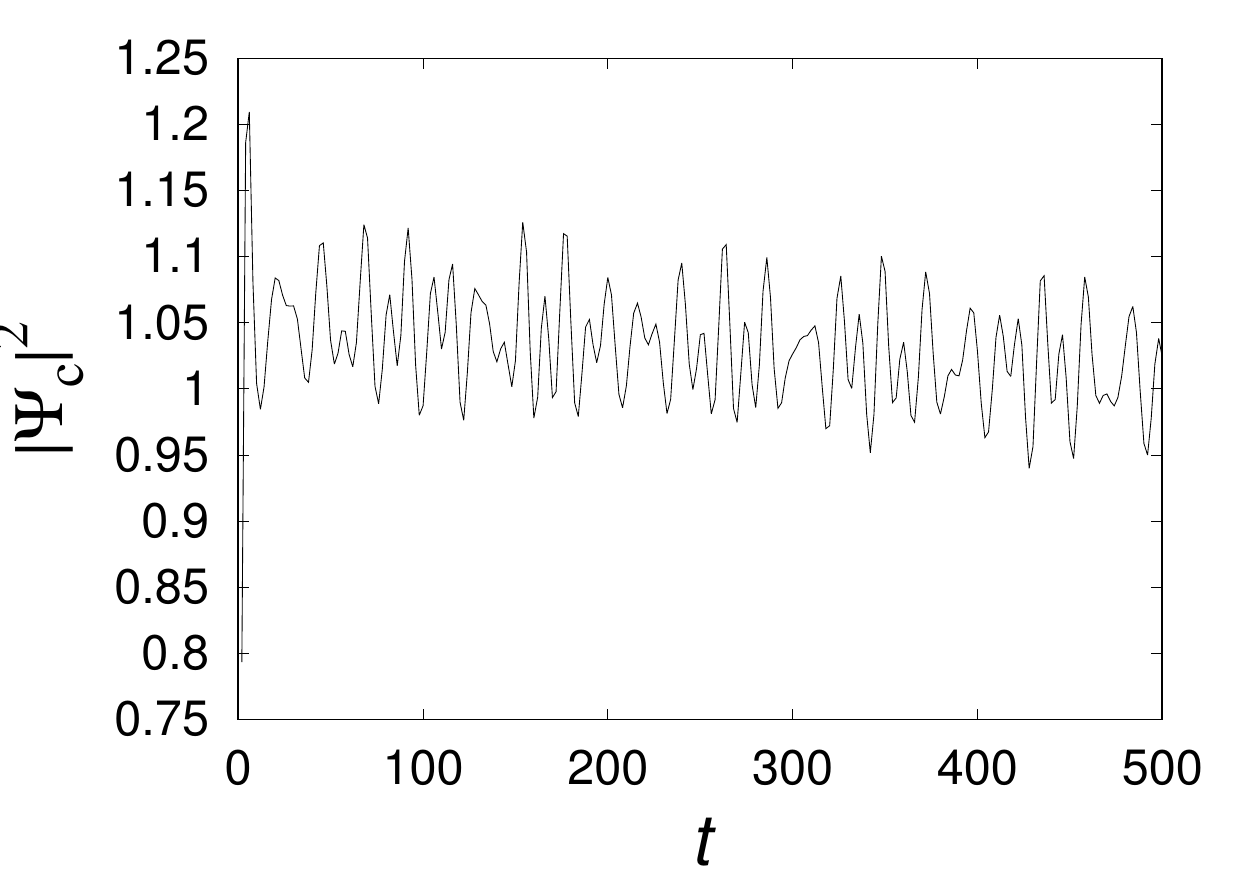}
\includegraphics[width=4.15cm]{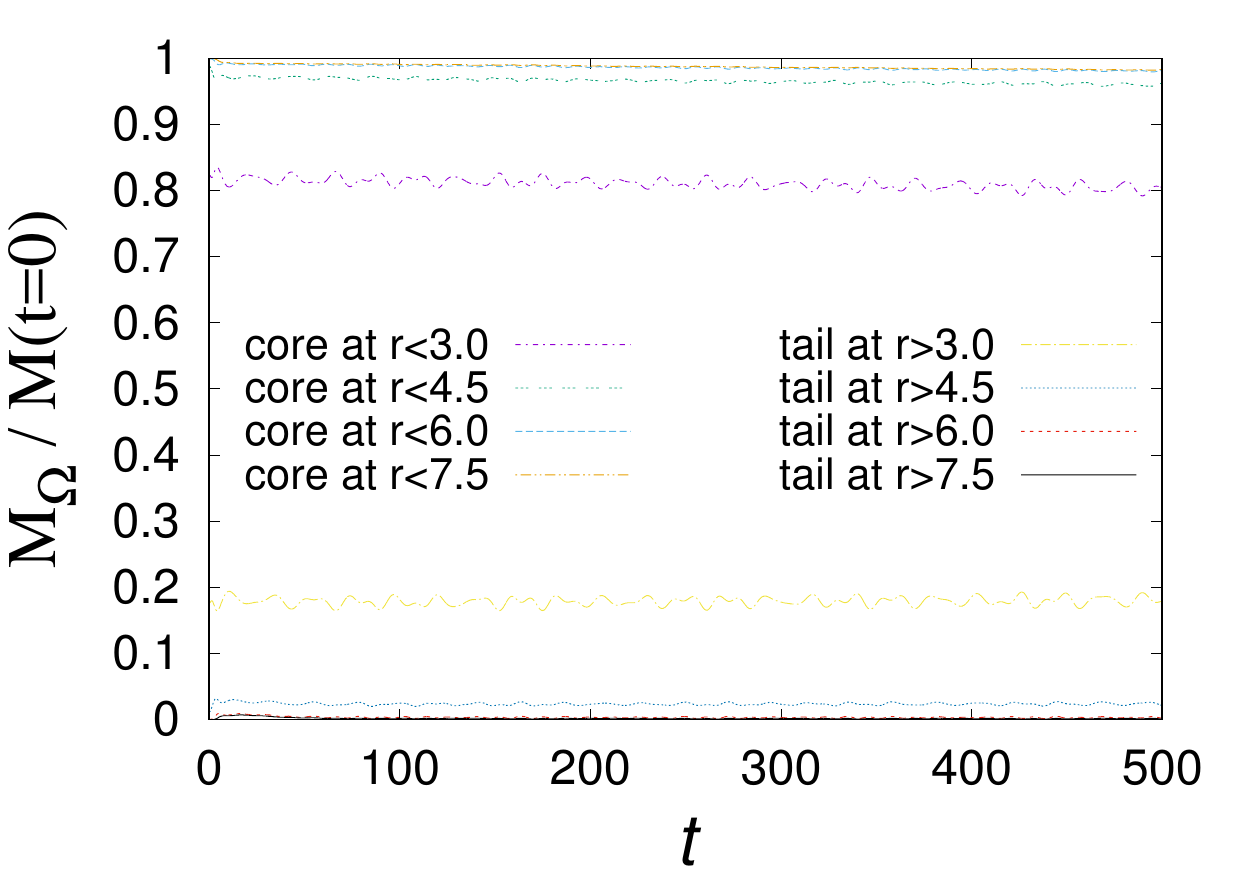}
\caption{\label{fig:isolated gaussian40} 
Evolution of a Gaussian in the domain $[-40,40]^3$ using isolated boundary conditions. 
(Top) Snapshots of the density $|\Psi|^2$ at various times that illustrates the core formation and the restless behavior of the density outside of the core. Also shown is the time average of the density profile from $t$=250 to $t$=500 together with the  solitonic fitting profile.
(Bottom) On the left the central density as function of time that shows the oscillations of the core and on the right the mass of core and tail with respect to their initial value using various values of the core radius $r_c=3, 4.5, 6.0$ and $7.0$.}
\end{figure}

The evolution of the fluctuation using isolation boundary conditions is illustrated with the results in Figures \ref{fig:isolated gaussian20} and \ref{fig:isolated gaussian40}, that respectively correspond to the use of domains $[-20,20]^3$ and $[-40,40]^3$. In the two cases a core is formed that can be fit with the empirical formula (\ref{eq:coreprofile}), and coincides with the equilibrium density profile. The tail density falls rapidly towards the boundary where the sponge absorbs the density. The central density tends to stabilize, whereas the mass in the time window used decreases linearly but slowly, slower in the big domain, which illustrates the effects of the sponge, that is, the dynamical behavior of the tail region permanently pumps matter, even if in small quantities, toward the boundary, where it is absorbed.

\begin{figure}
\includegraphics[width=4.15cm]{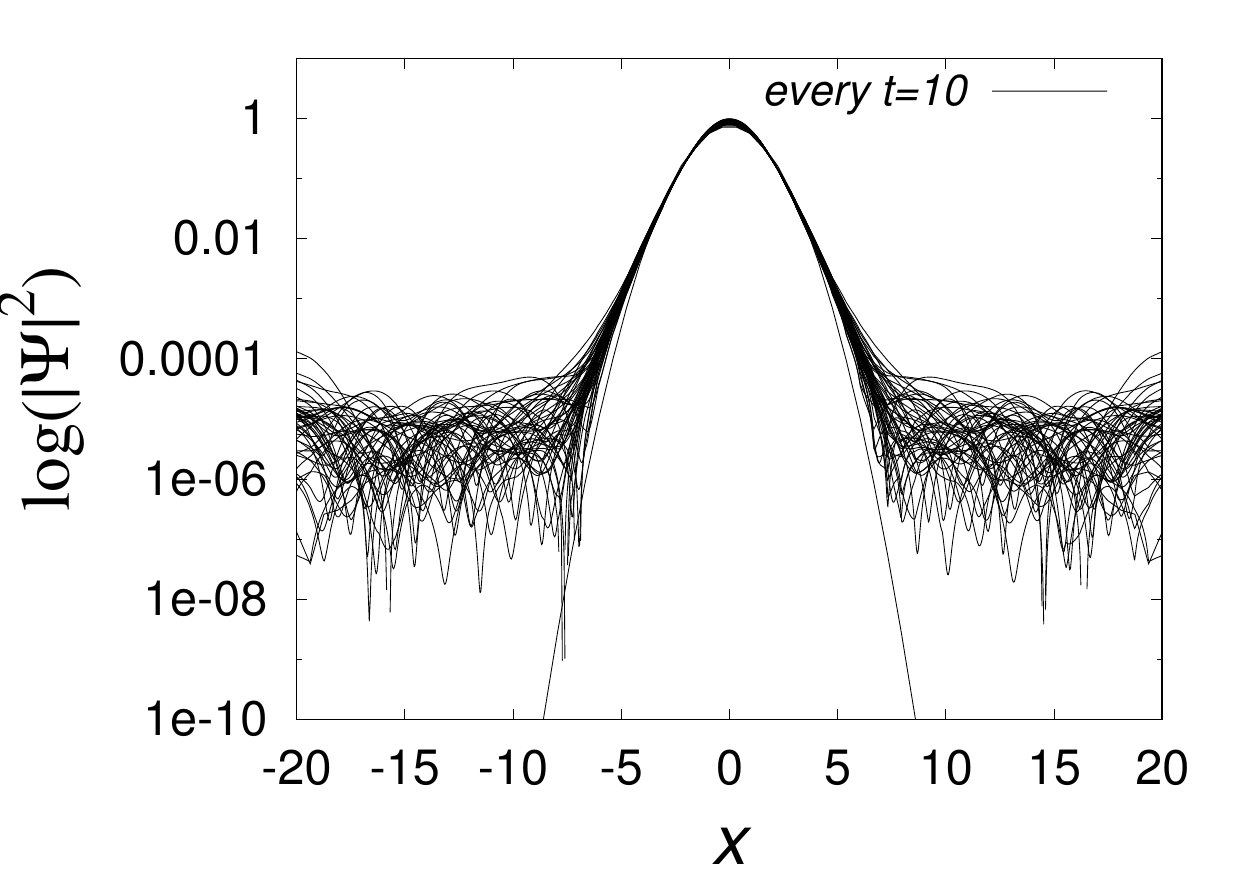}
\includegraphics[width=4.15cm]{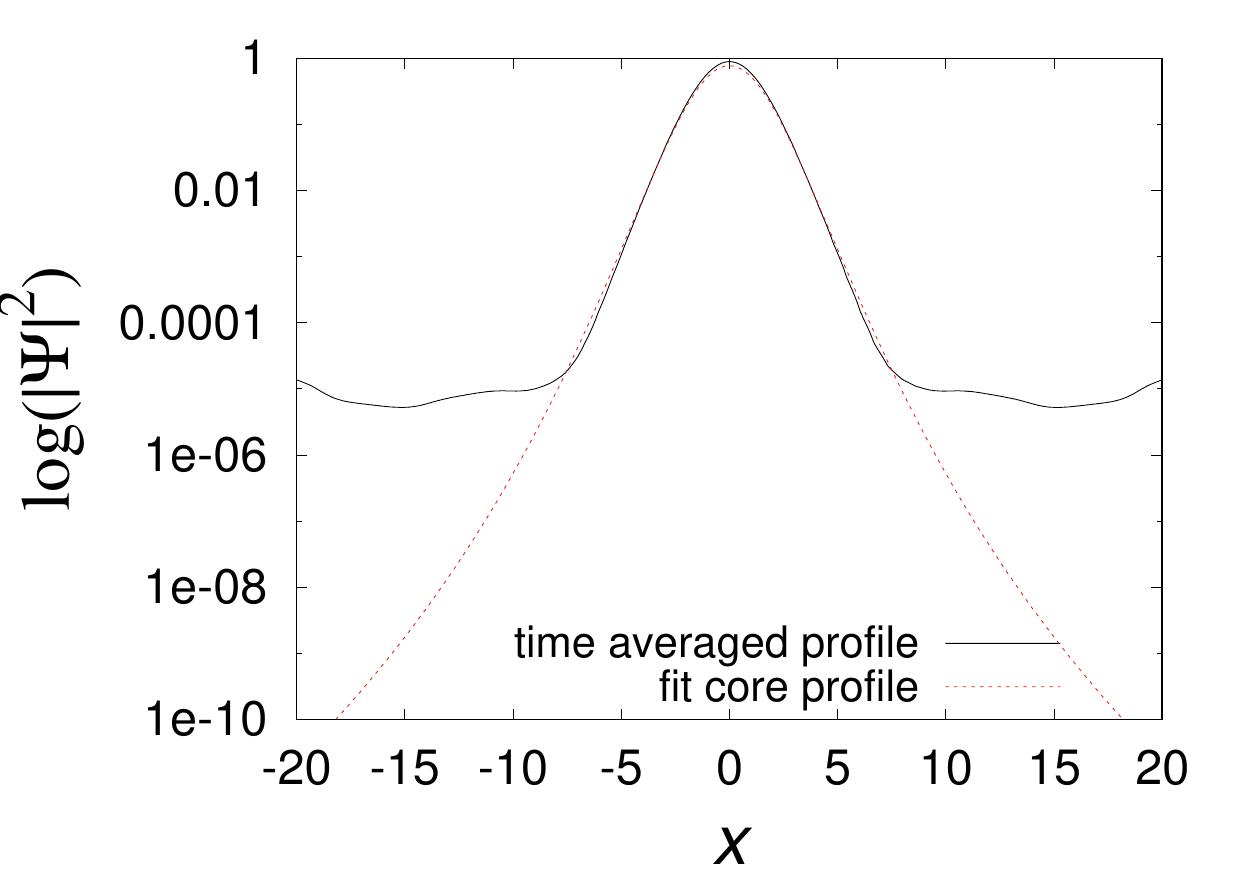}
\includegraphics[width=4.15cm]{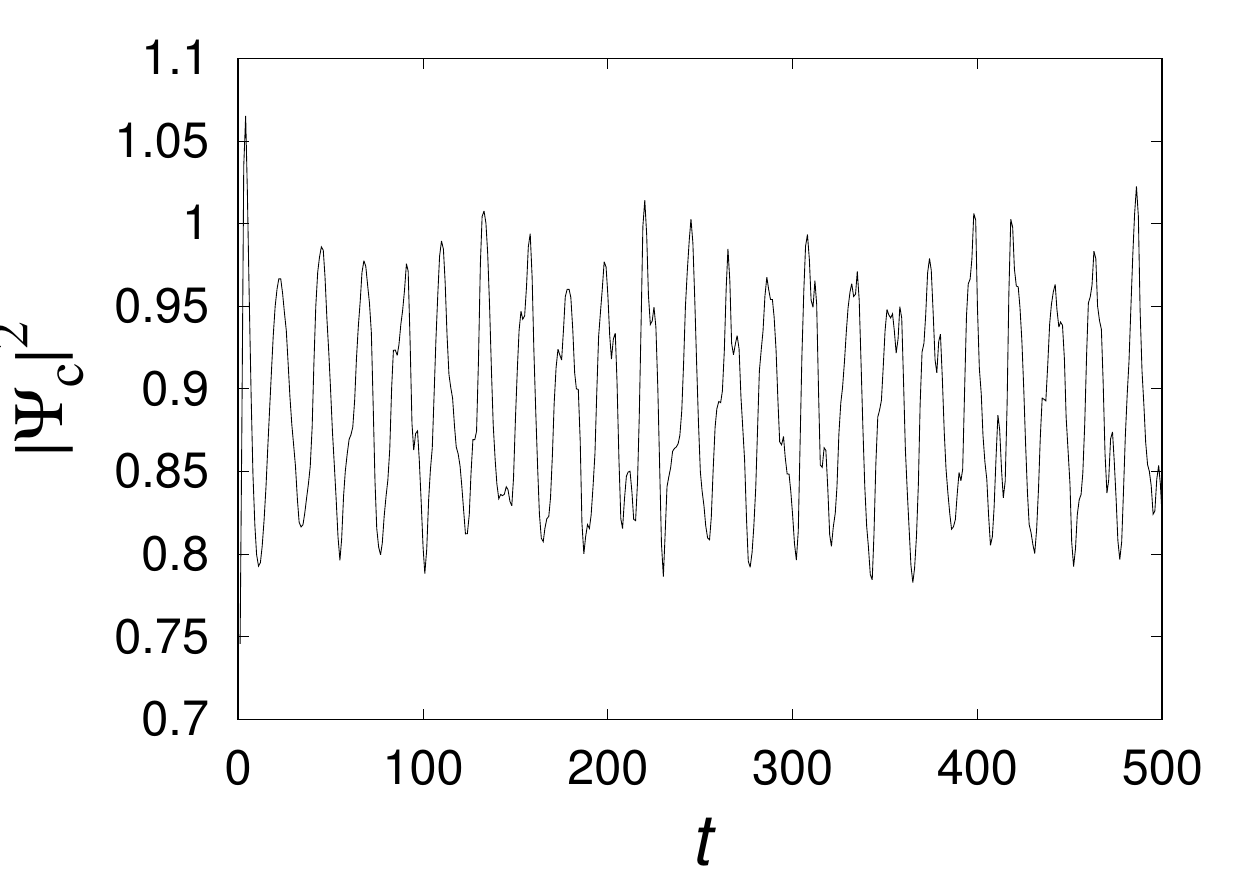}
\includegraphics[width=4.15cm]{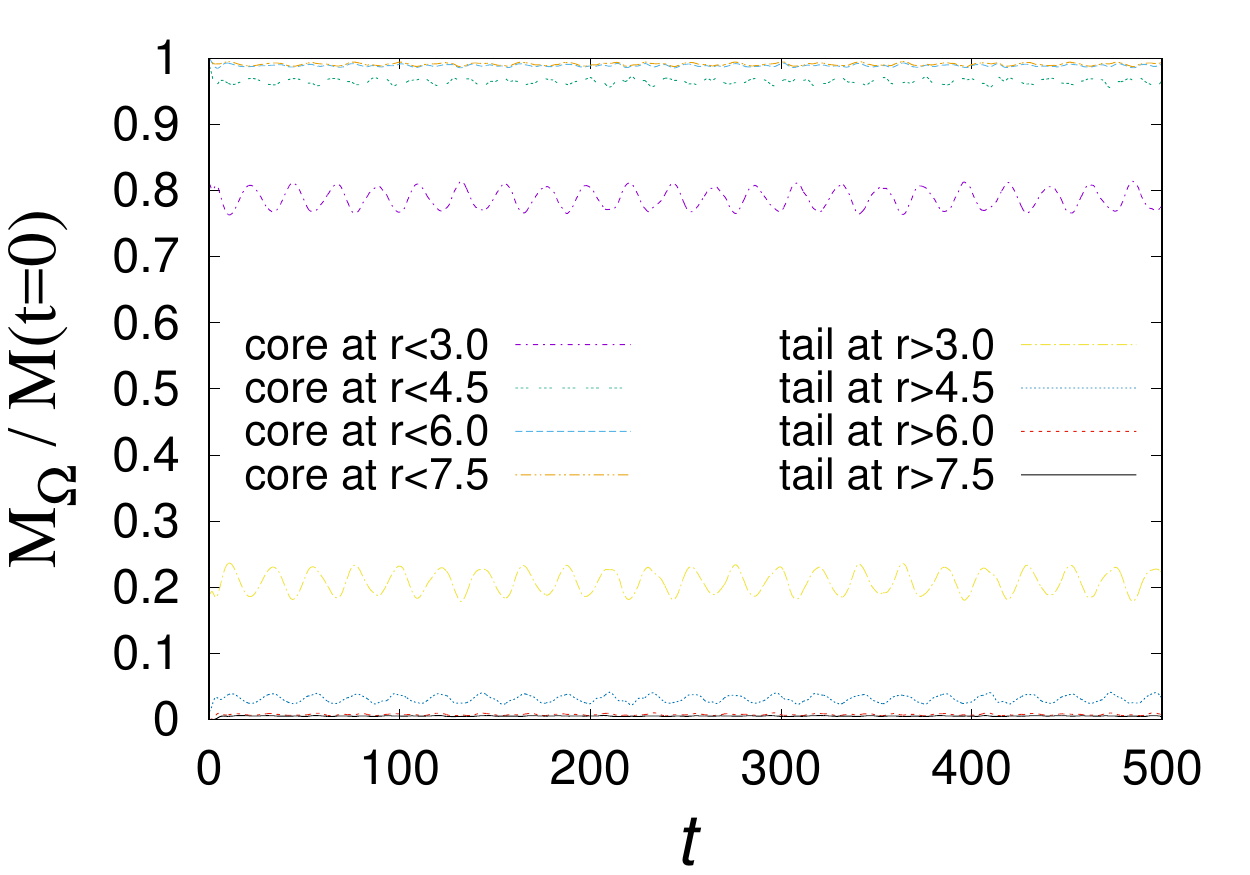}
\caption{\label{fig:periodic gaussian20} 
Evolution of a Gaussian in the domain $[-20,20]^3$ with periodic boundary conditions. 
(Top) Snapshots of the density $|\Psi|^2$ at various times that illustrates core and the restless behavior of the density outside of the core. Also shown is the time average of the density profile from $t$=250 to $t$=500 next to the initial configuration.
(Bottom)  On the left the central density as function of time that shows the oscillations of the core and on the right the of mass of core and tail and with respect to their initial time using various values of the core radius $r_c=3, 4.5, 6.0$ and $7.0$.}
\end{figure}

\begin{figure}
\includegraphics[width=4.15cm]{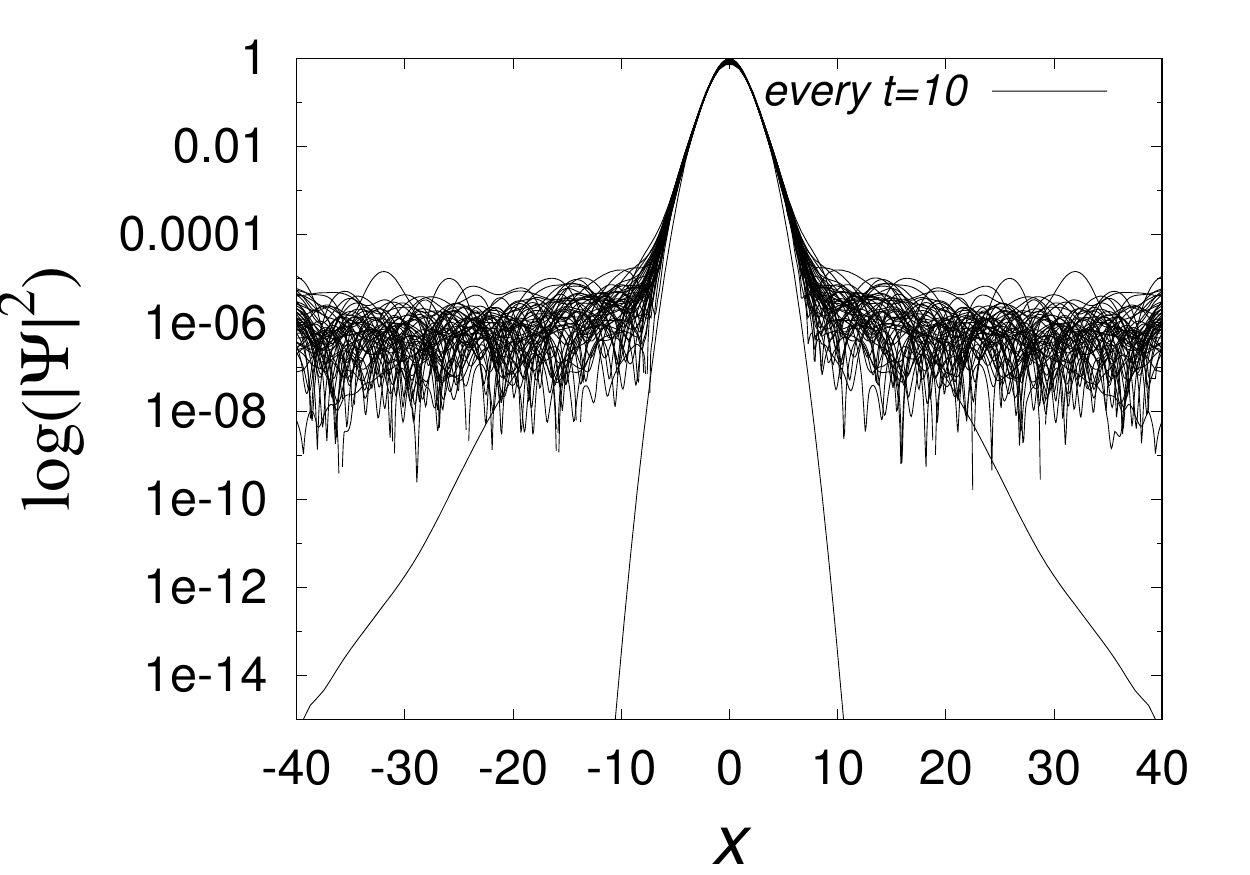}
\includegraphics[width=4.15cm]{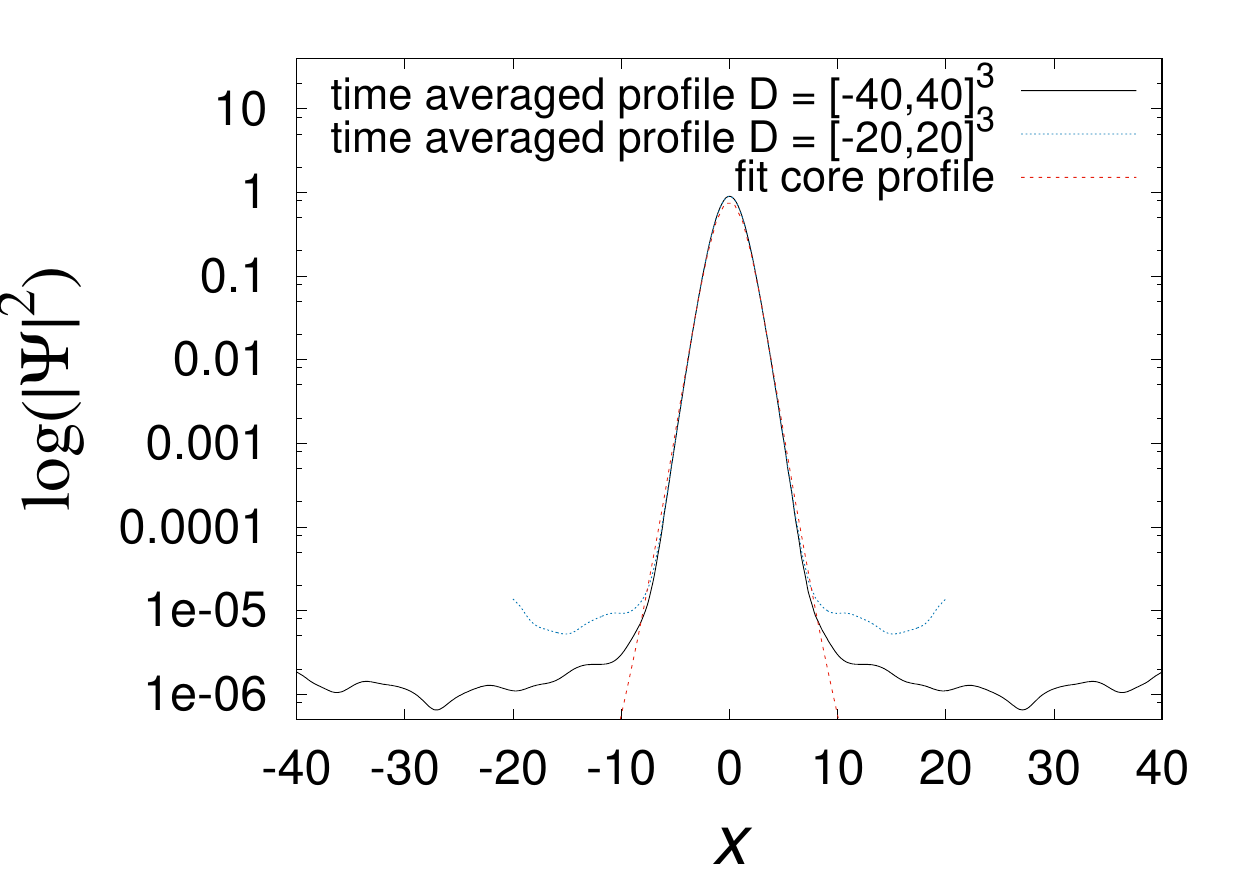}
\includegraphics[width=4.15cm]{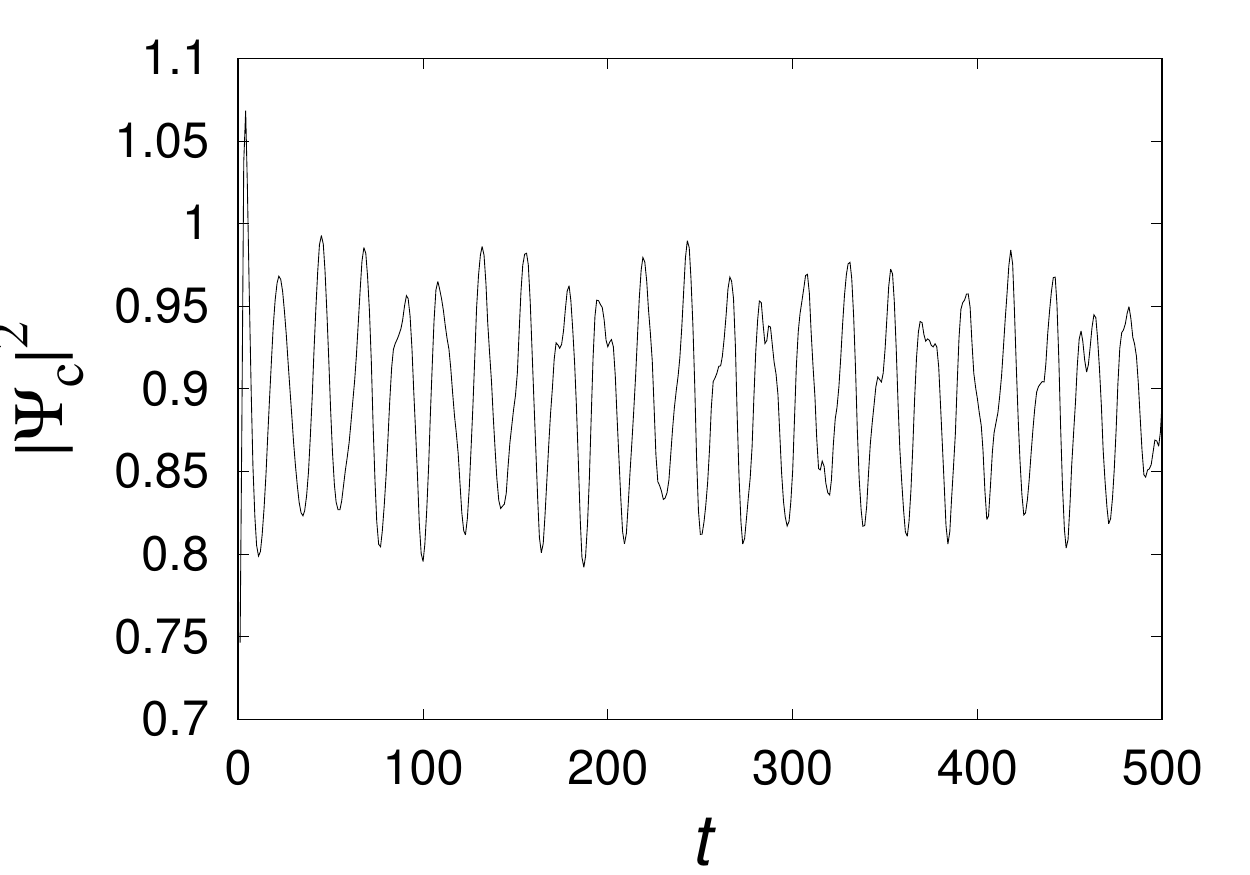}
\includegraphics[width=4.15cm]{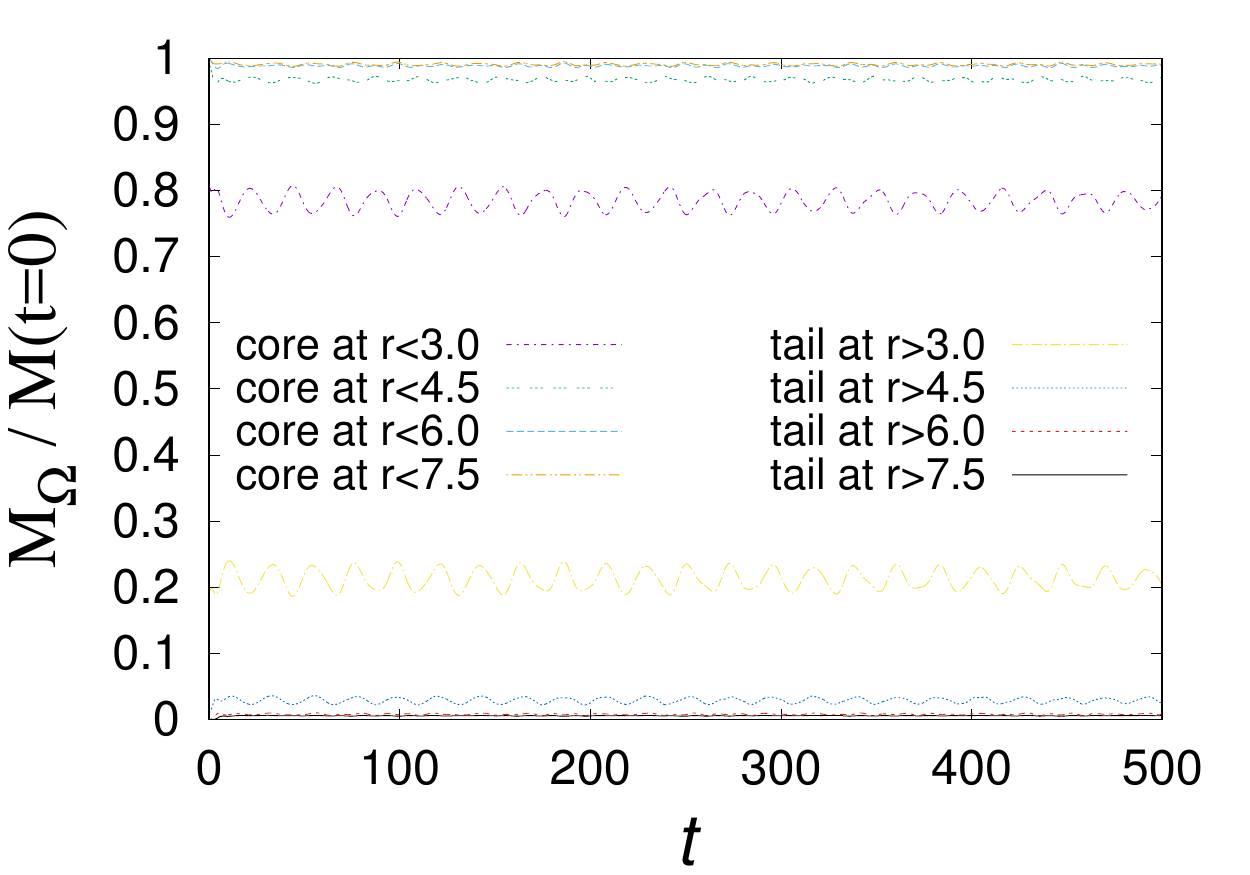}
\caption{\label{fig:periodic gaussian40} 
Evolution of a Gaussian in the domain $[-40,40]^3$ with periodic boundary conditions. 
(Top) Snapshots of the density $|\Psi|^2$ at various times that illustrates core and the restless behavior of the density outside of the core. Also shown is the time average of the density profile from $t$=250 to $t$=500 next to the initial configuration.
(Bottom)  On the left the central density as function of time that shows the oscillations of the core and on the right the of mass of core and tail and with respect to their initial time using various values of the core radius $r_c=3, 4.5, 6.0$ and $7.0$.}
\end{figure}

The results are different when periodic boundary conditions are used. The results obtained for the relaxation of the Gaussian pulse are shown in Figures \ref{fig:periodic gaussian20} and \ref{fig:periodic gaussian40}, corresponding to the use of domains $[-20,20]^3$ and $[-40,40]^3$. This time the initial conditions are not the soliton itself, but the soliton gets formed during the evolution with the profile (\ref{eq:coreprofile}). The density in the tail region is restless with an endless motion that in average distributes in a nearly constant profile, unlike the NFW decay found in structure formation halos \cite{Schive:2014dra,Mocz:2017wlg}. We observe that the Gaussian pulse does not fragment, whereas the solitonic cores obtained from structure formation simulations result from the interference of multiple density fluctuations and are the superposition of many smaller fluctuations as shown later on in this paper. From these two simulations one can infer that the matter outside of the core distributes with a nearly constant profile, with a value that depends on the volume out of the core, which in turn depends on the domain size. A clear difference in comparison with the isolated case is that mass is not being lost during the evolution as expected from the topology of the domain.

%
\subsection{Free-fall head-on merger}
\label{subsec:headon}

Another scenario that may be affected by the change of topology of the domain is the merger of two structures, as the gravitational potential repeats itself in copies outside of the domain along the three Cartesian directions. For this study, we focused on the head-on collision of two ground state equilibrium configurations in free-fall.

To illustrate this scenario, we conducted simulations in the domains $[-20,20]^3$ and $[-40,40]^3$ using the same space and time resolution as in the previous examples. We selected three cases with initial positions of the configurations at A) $(\pm 5,0,0)$, B) $(\pm 10,0,0)$ and C) $(\pm 15,0,0)$. It is expected that the domain outside of the box, which is plagued with a network of similar binary configurations, will have an effect on the collision dynamics. For comparison, we also simulated the merger using isolated boundary conditions.

\begin{figure}
\includegraphics[width=4.15cm]{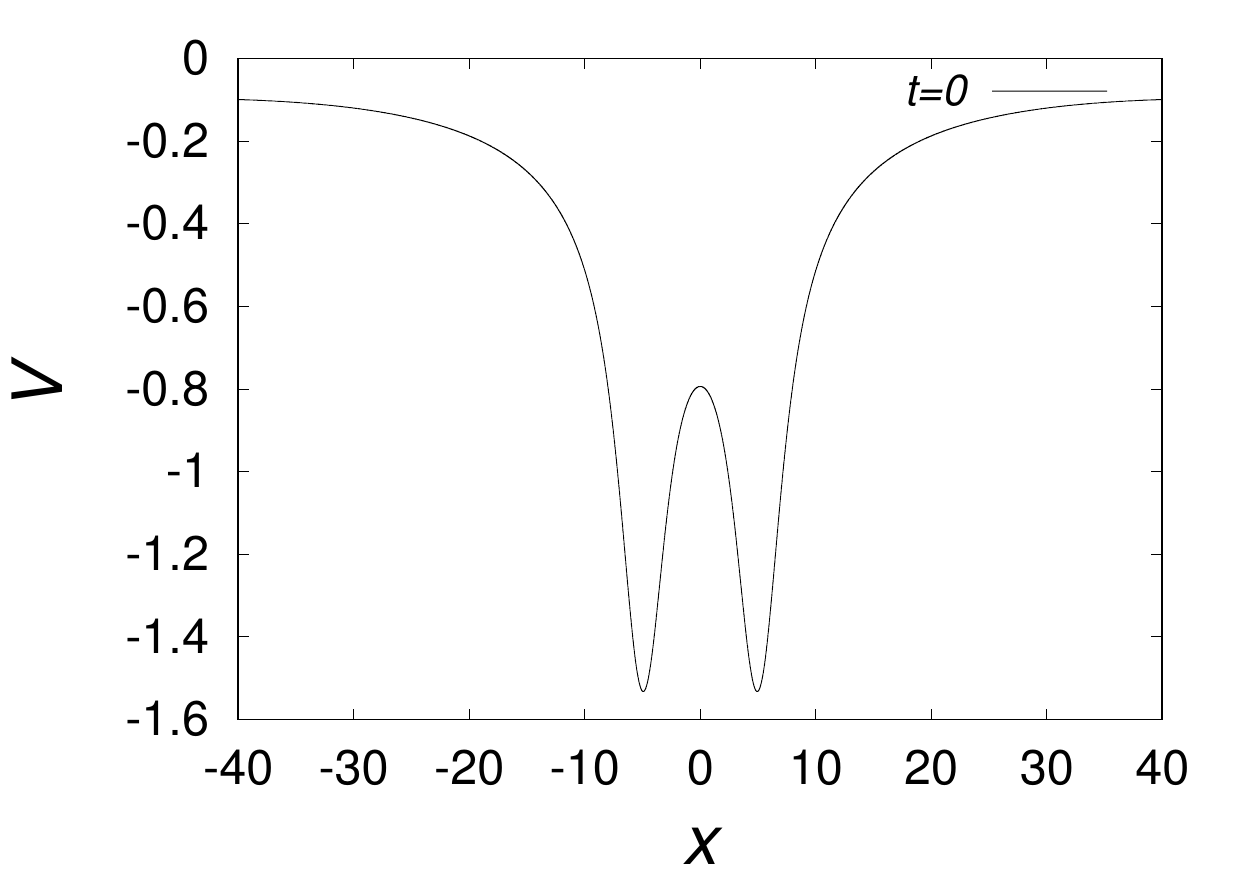}
\includegraphics[width=4.15cm]{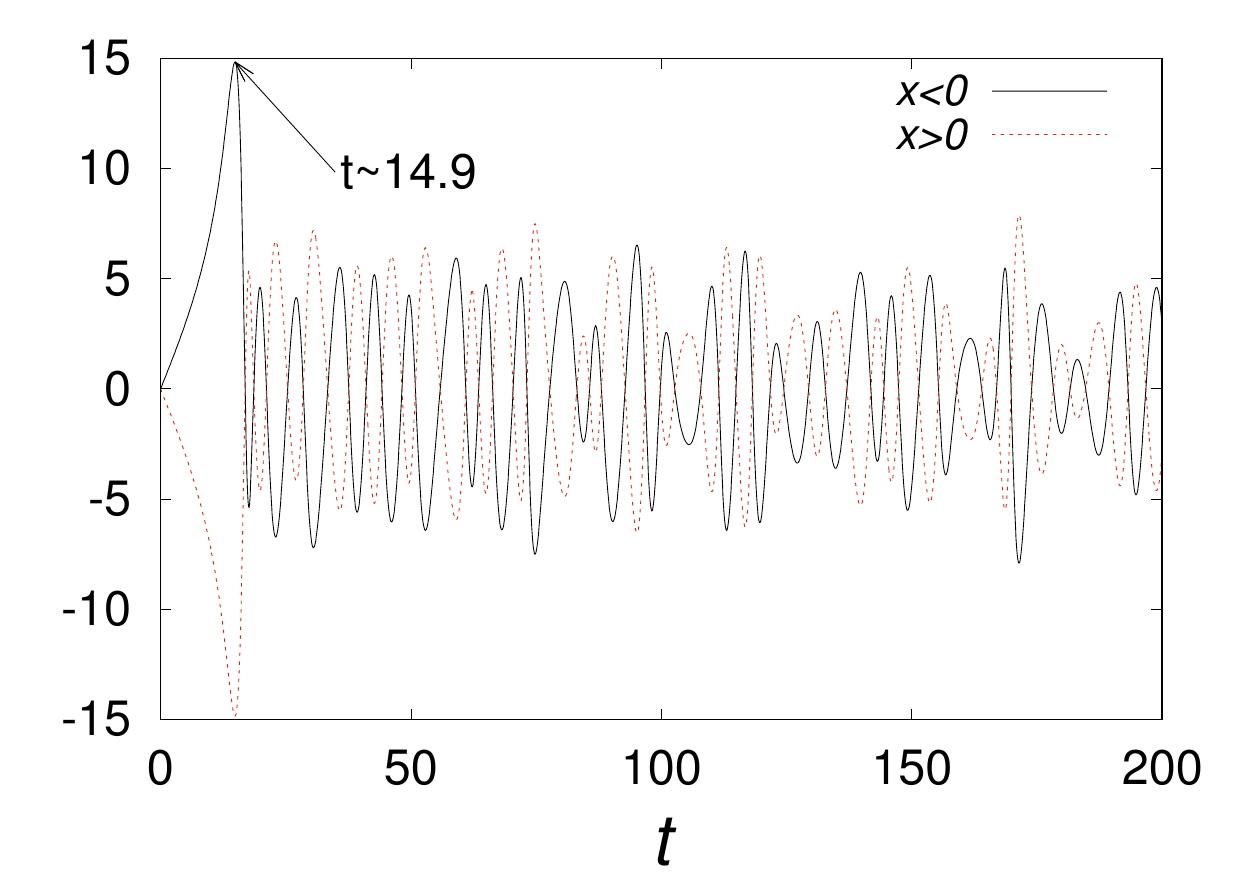}
\includegraphics[width=4.15cm]{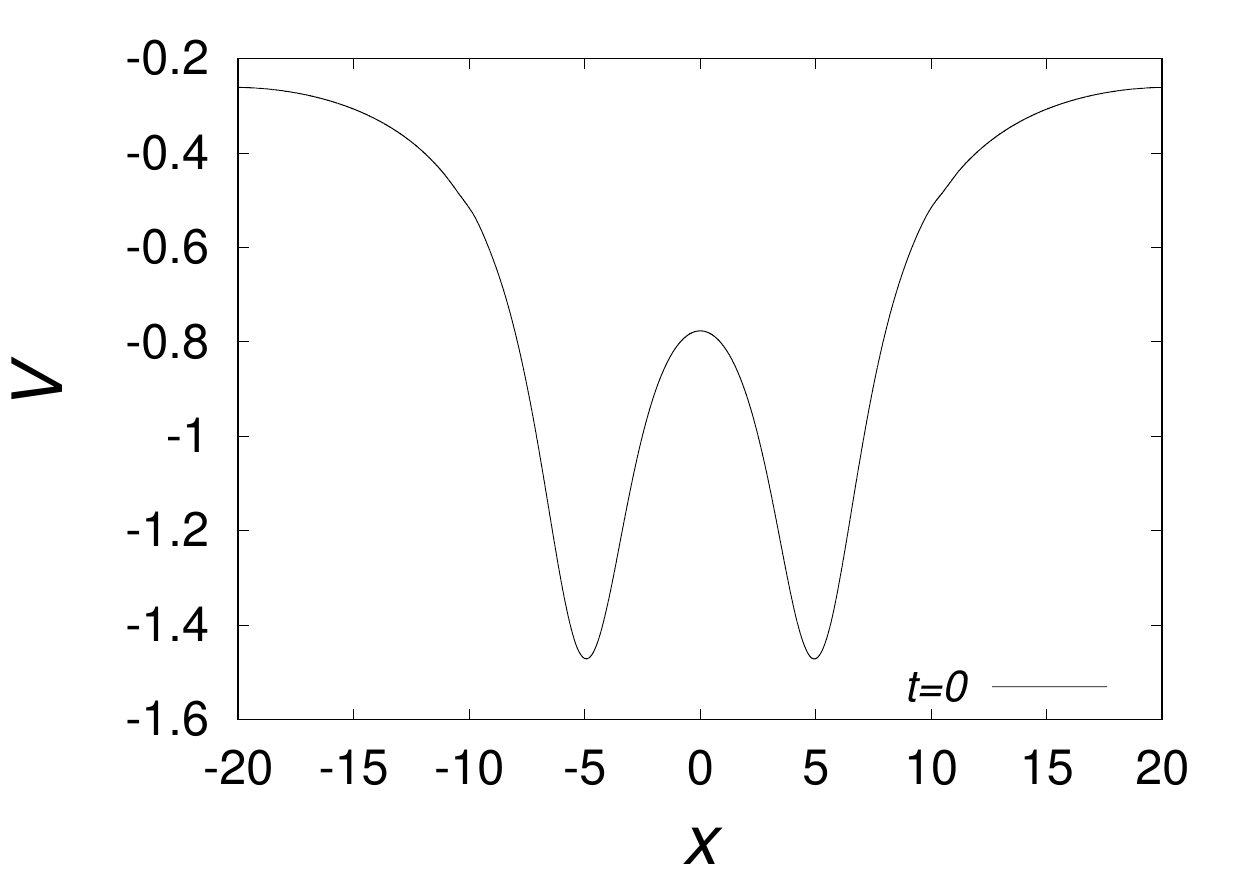}
\includegraphics[width=4.15cm]{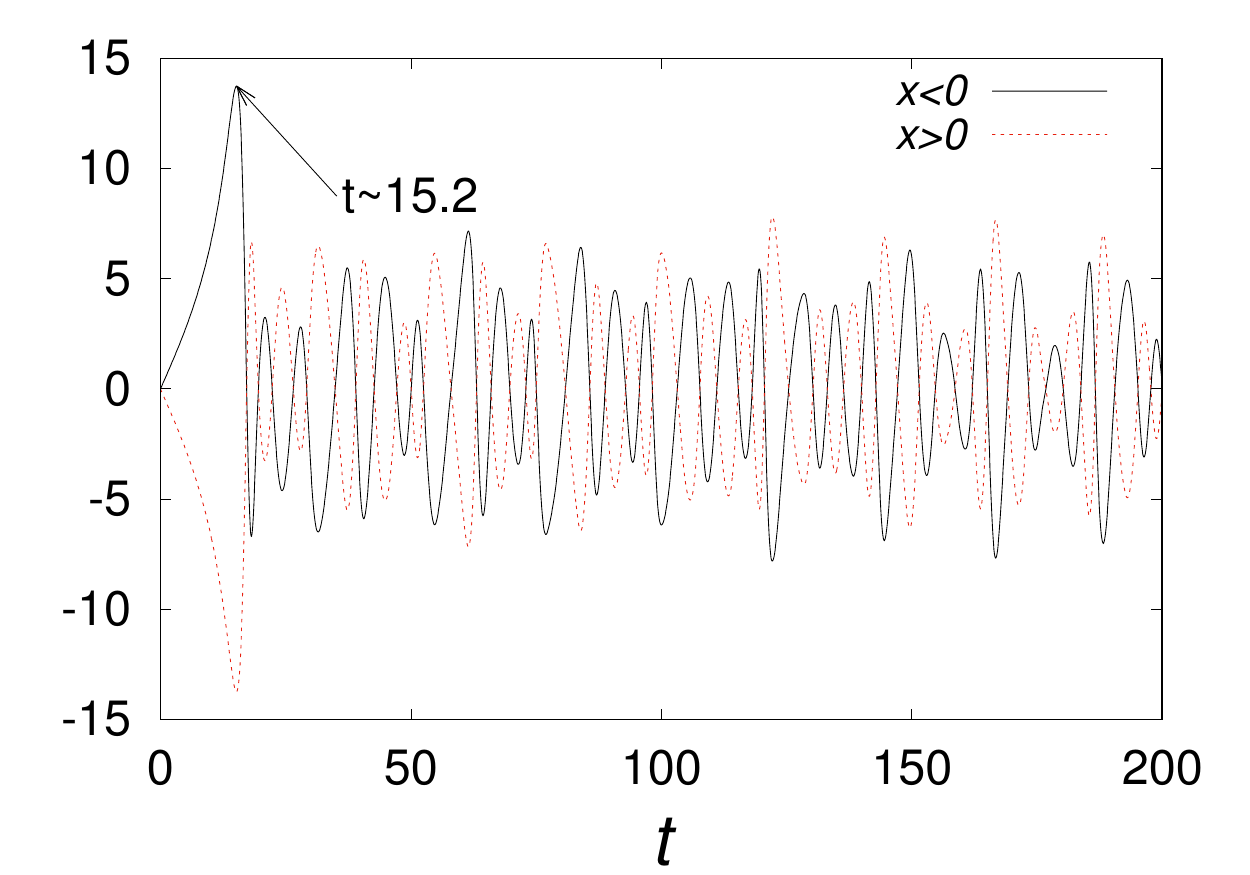}
\includegraphics[width=4.15cm]{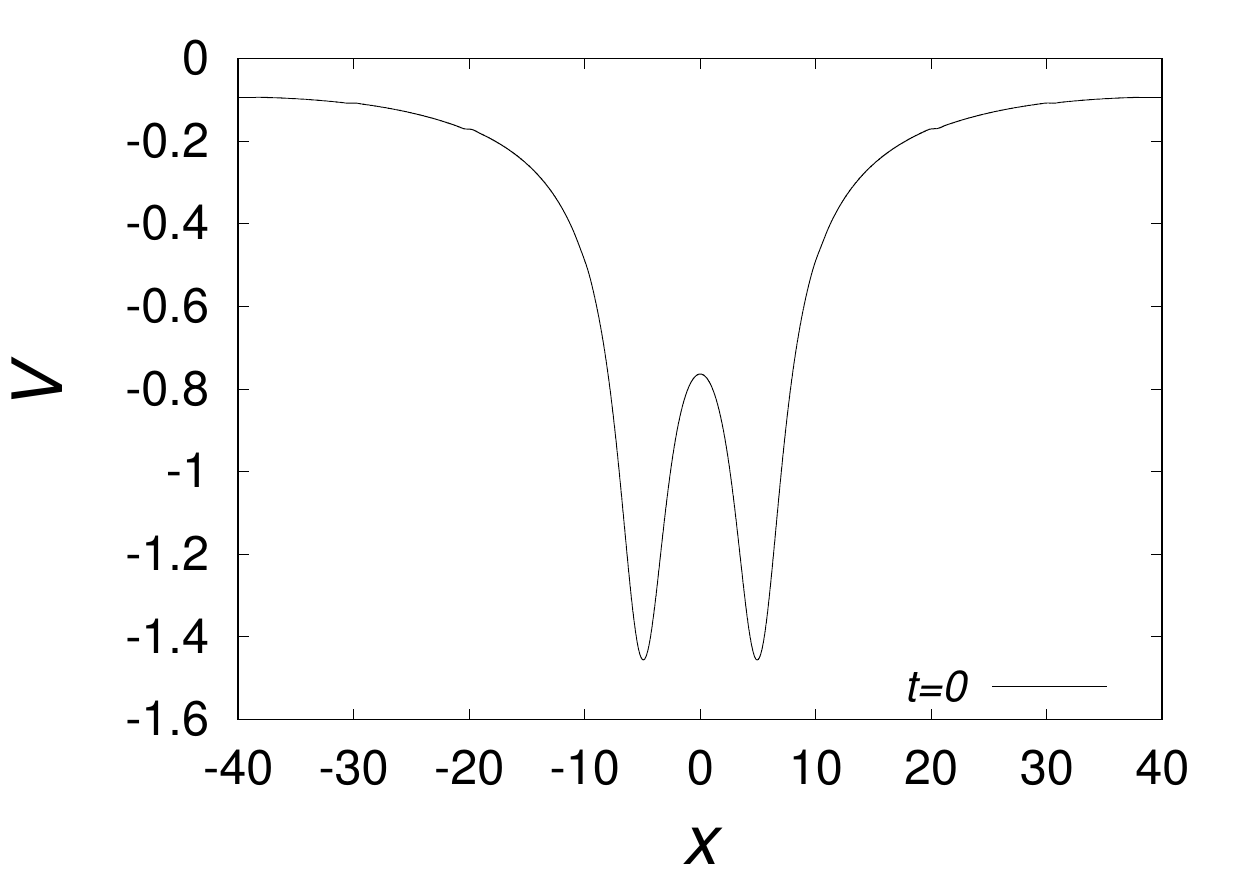}
\includegraphics[width=4.15cm]{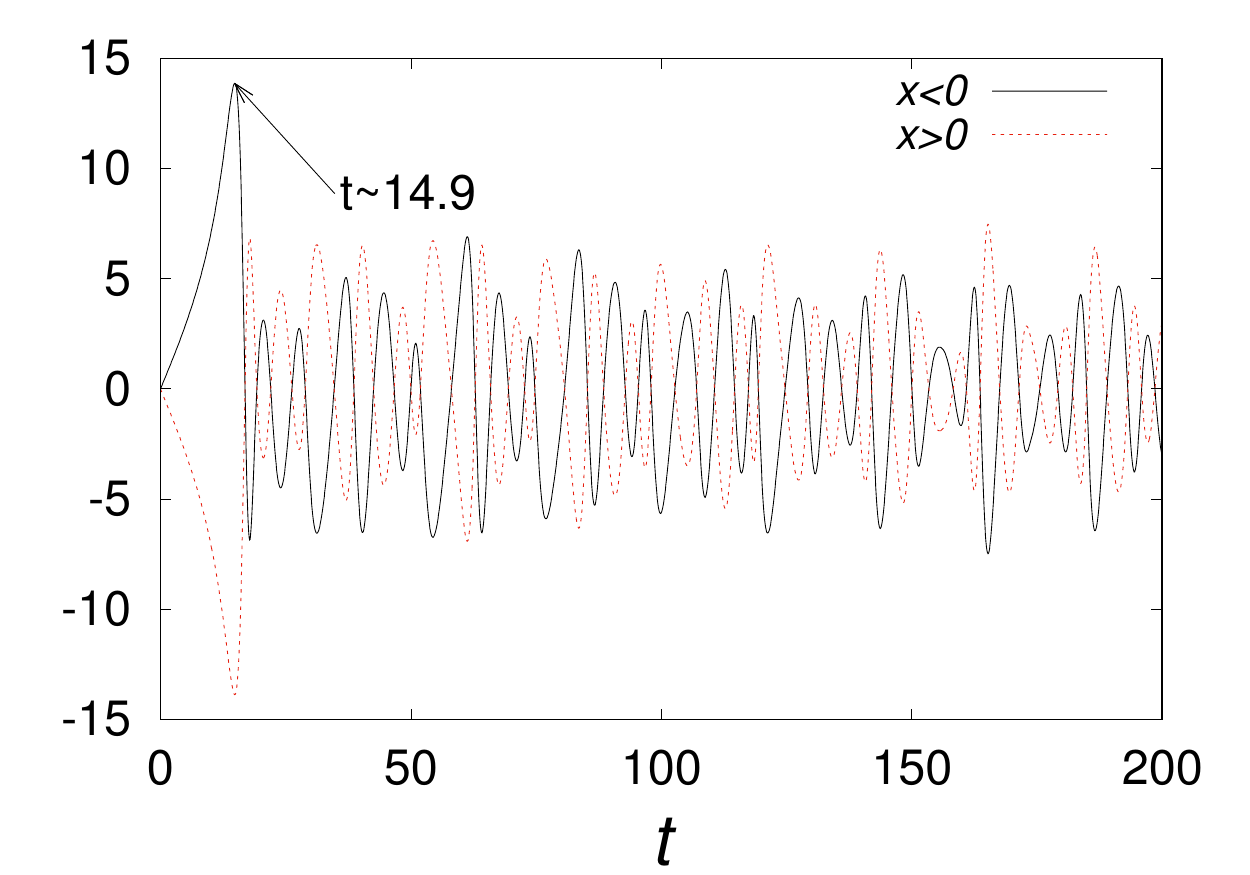}
\caption{\label{fig:head-onA} 
Case A. Head-on momentum $\langle p_x \rangle$ integrated in the semi-domains $x<0$ and $x>0$. At the top we show the results for the free-fall head-on merger using isolation boundary conditions. In second and third rows we show the momentum in the case of a periodic domain $[-20,20]^3$ and $[-40,40]^3$ respectively. The magnitude of the momentum is smaller when using a periodic domain, whereas the time of the maximum momentum is affected by the domain size.}
\end{figure}

The simulations results for Case A are presented in Figure \ref{fig:head-onA}. The left column displays the gravitational potential along the $x-$direction, while the right column shows the head-on momentum $\langle p_x \rangle$ integrated over the half-domains $x<0$ and $x>0$. In the first row, the results using isolation boundary conditions are presented, and the merger time is identified at $t\sim 14.9$, which is defined as the moment when the maximum head-on momentum is acquired. In the second and third rows, the results using periodic boundary conditions with domains $[-20,20]^3$ and $[-40,40]^3$ are shown, respectively. It can be observed that the merger time is affected by the periodic domain size; when using a small domain, the pull of the neighboring gravitational potentials retards the merger time. Furthermore, the magnitude of the momentum is also affected by the periodic domain.

Case B presents a different situation. It is a borderline case because the periodic boundary conditions imply the existence of a similar binary configuration on neighboring domains located at exactly the same distance, of 20 units of length along the $x-$axis. Binary configurations along the $y$ and $z$ axes will be separated by 40 length units from one another. This means that the binary configuration with periodic boundary conditions actually represents an infinite array of cores along the $x-$axis, equally spaced by 20 units of length and by 40 units along $y$ and $z$. The gravitational field is expected to be compensated along the $x-$direction in some way.

\begin{figure}
\includegraphics[width=4.15cm]{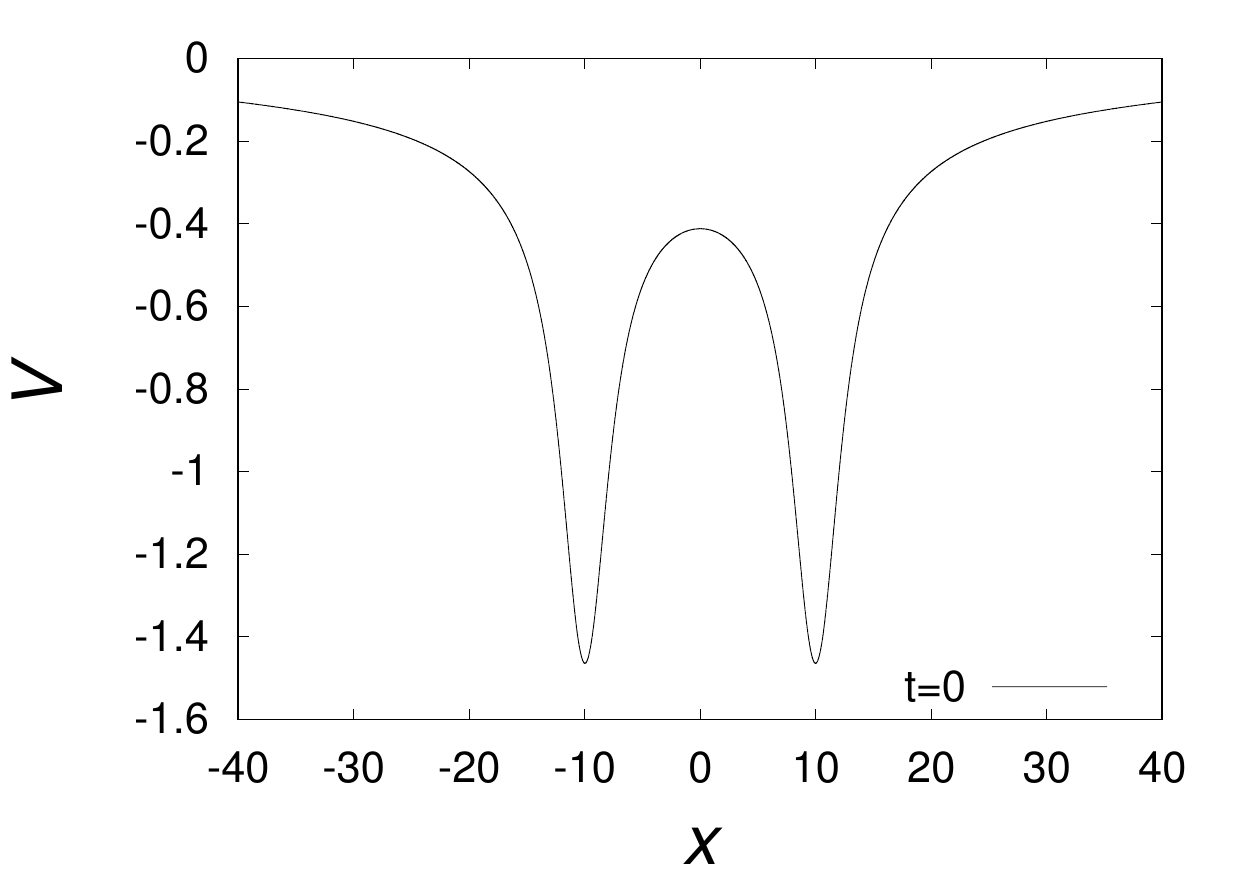}
\includegraphics[width=4.15cm]{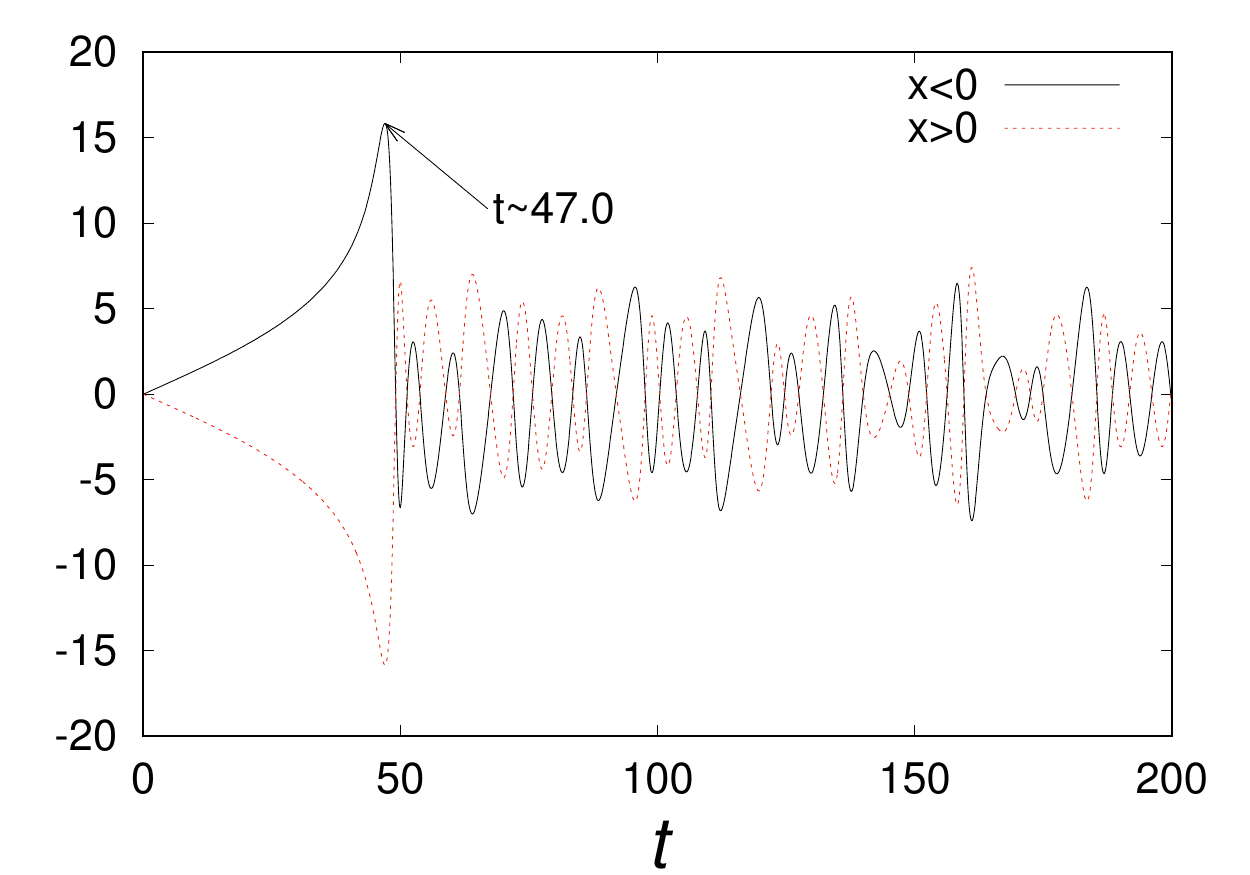}
\includegraphics[width=4.15cm]{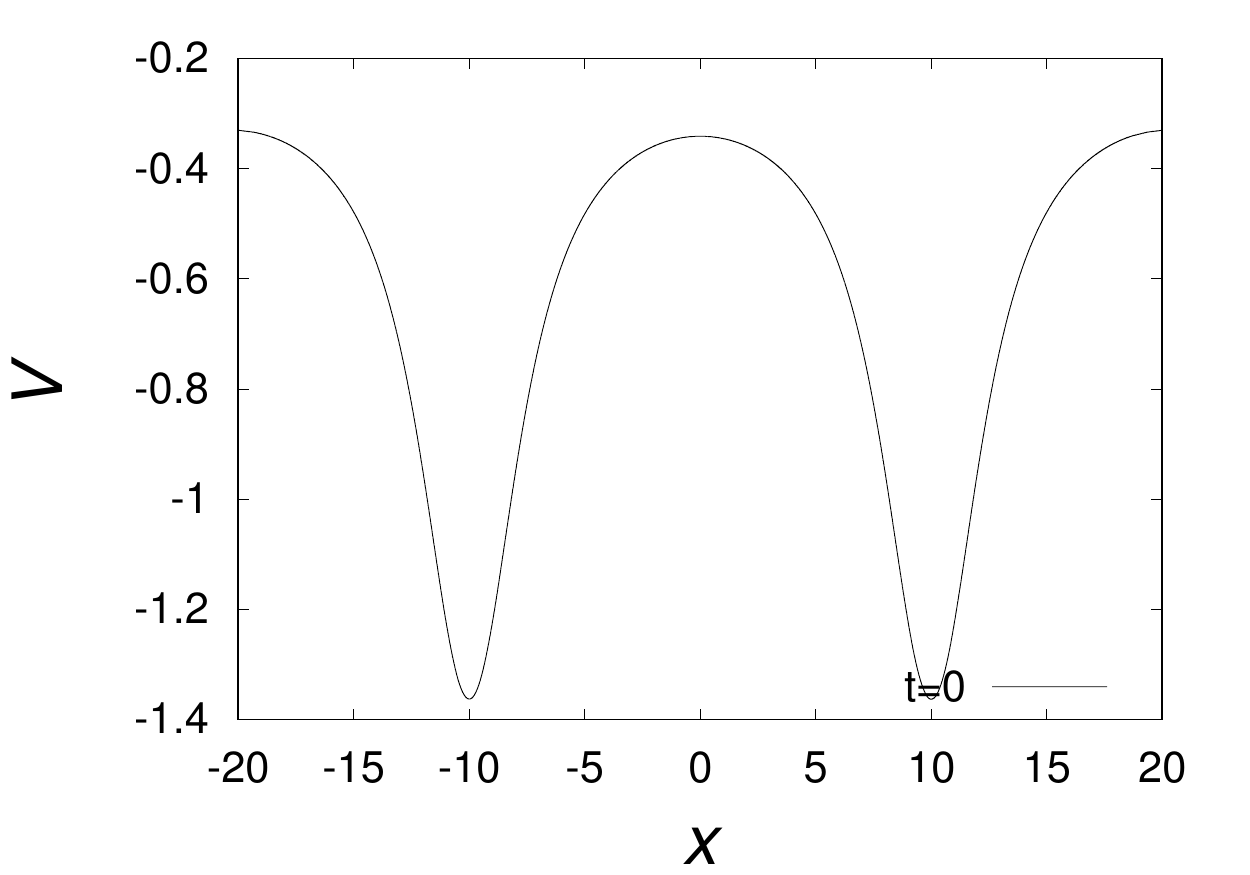}
\includegraphics[width=4.15cm]{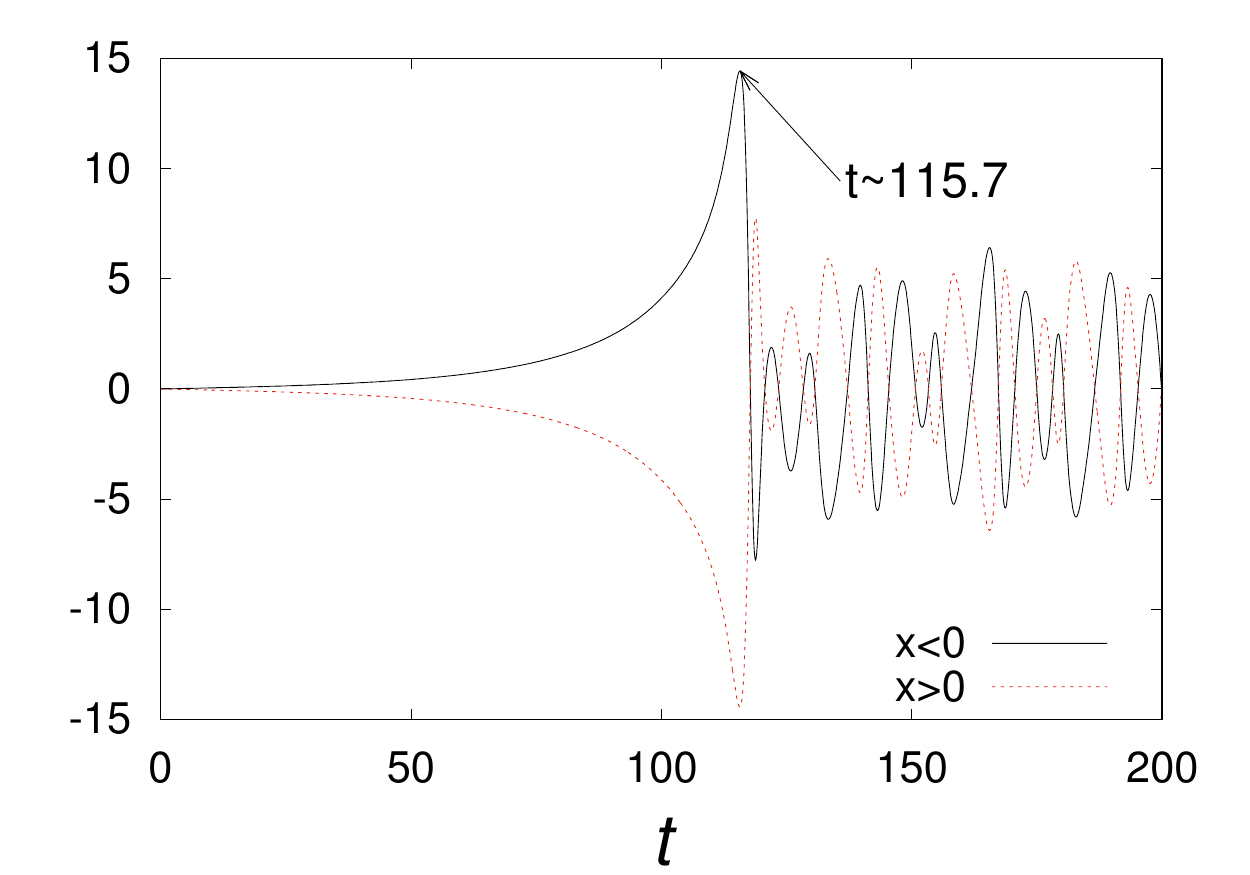}
\includegraphics[width=4.15cm]{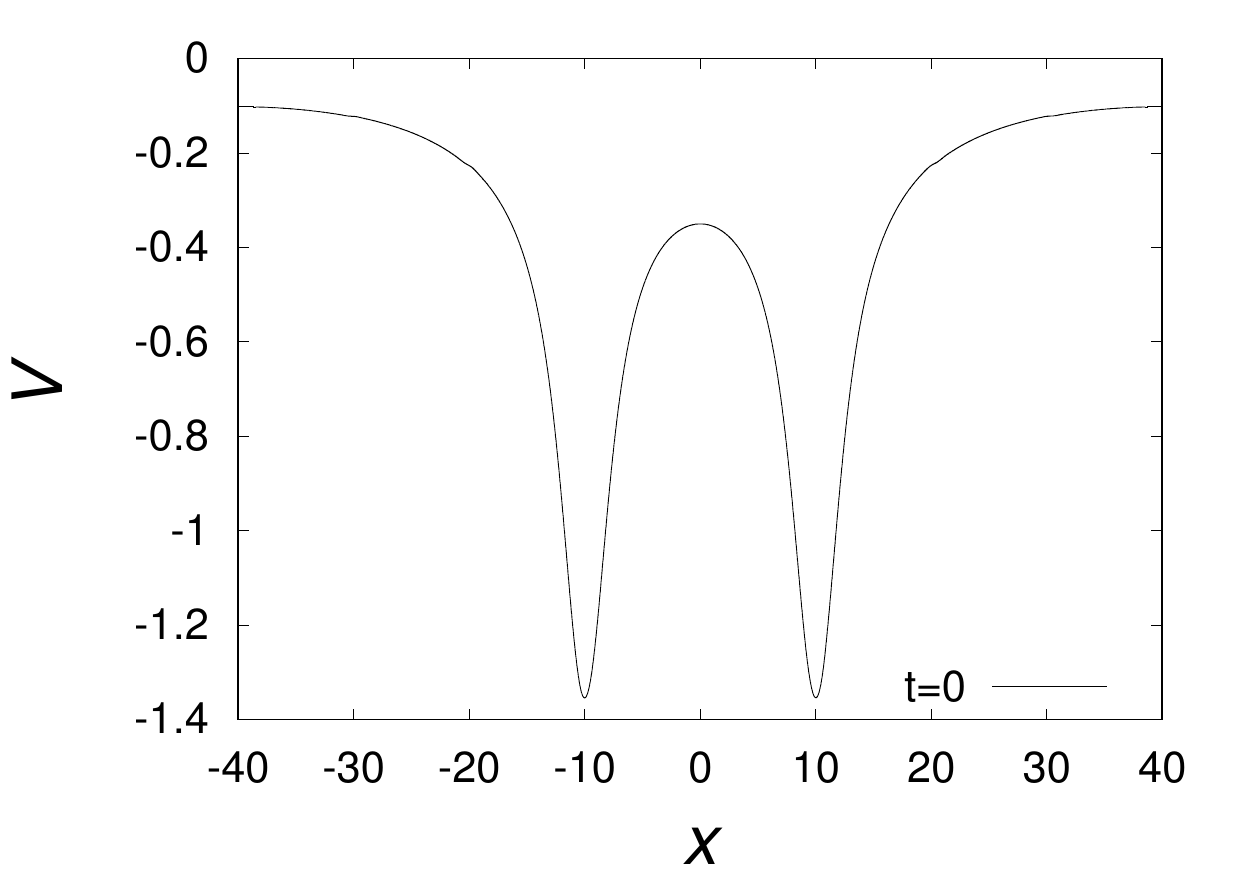}
\includegraphics[width=4.15cm]{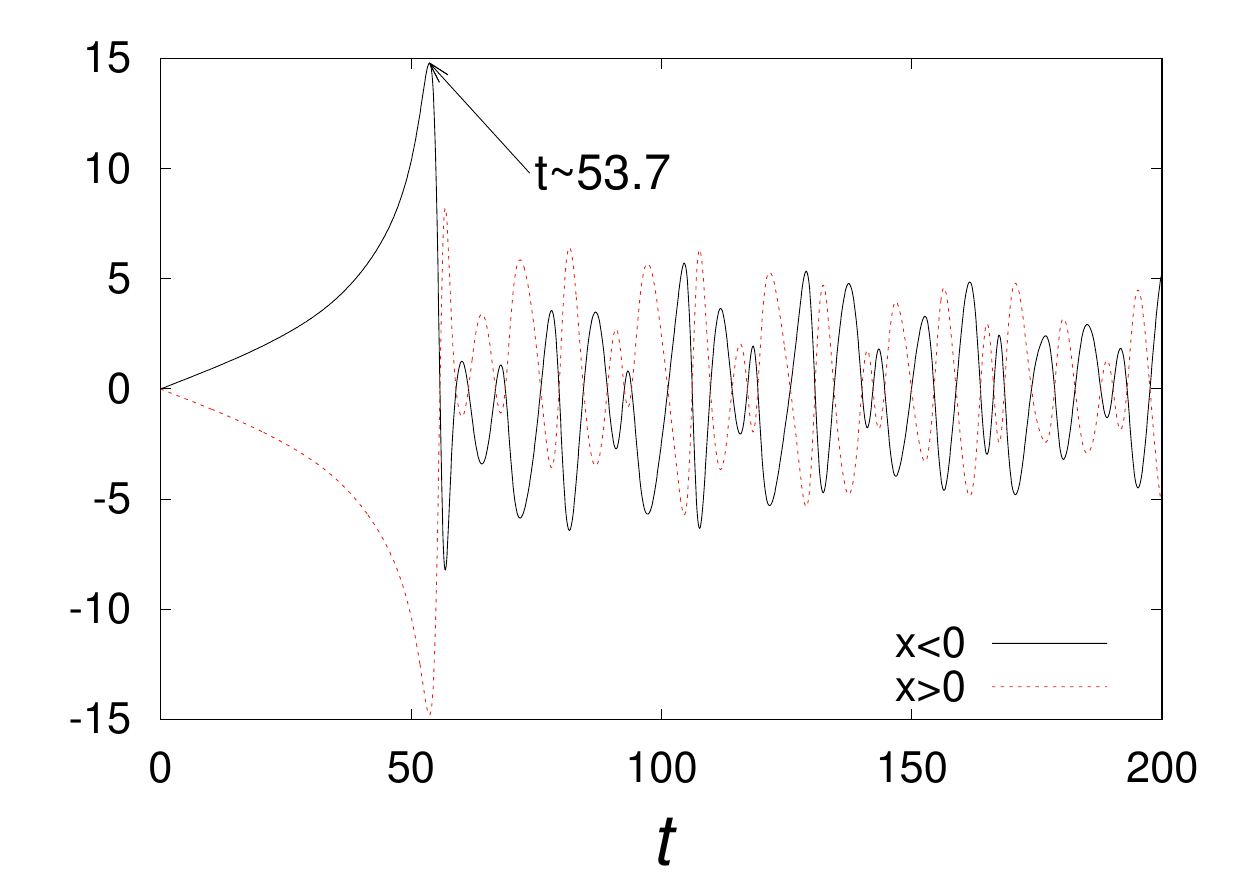}
\caption{\label{fig:head-onB} 
Case B. Form top to bottom results for the isolated domain  $[-20,20]^3$, periodic domain  $[-20,20]^3$ and periodic domain  $[-40,40]^3$. At the left we show the gravitational potential at initial time along the $x-$axis, in order to illustrate its local minimums and behavior at the boundary. Particularly interesting is the second row, where the gravitational potentials of the cores are possibly expected to compensate each other along the $x-$direction. At the right we show $\langle p_x \rangle$ integrated in the semi-domains $x<0$ and $x>0$, as function of time.}
\end{figure}

The results for Case B are shown in Figure \ref{fig:head-onB}. On the left, we present the gravitational potential along the $x-$direction, while on the right, we display $\langle p_x \rangle$ for the simulations with the isolated domain, periodic domain $[-20,20]^3$, and periodic domain $[-40,40]^3$, respectively. The first row corresponds to the head-on merger used as a control case that is not affected by the periodic domain.

The second row is the most noteworthy, as it takes a longer time for the cores to merge. This scenario requires further explanation since one might expect that the cores would never merge, given that they have infinite copies of themselves along the $x-$direction, and the gravitational effects along this direction would compensate and prevent the merger. However, the use of periodic boundary conditions introduces a subtlety, as seen in the left plots, where the gravitational potential is not zero at the center of the domain as it is at the boundary face. This effect has not been discussed in the literature on FDM simulations using periodic conditions, and it could have a significant influence on the results. It would be interesting to discuss and compare among codes and numerical implementations. Despite this asymmetry, the cores ultimately collide at the center of the domain after 115 time units.

The third row of results shows that with a bigger periodic domain, the collision time is more similar to that of the isolated case. This is because the potential wells of the cores of neighboring domains along the $x-$direction are further away from the binary system.

\begin{figure}
\includegraphics[width=4.15cm]{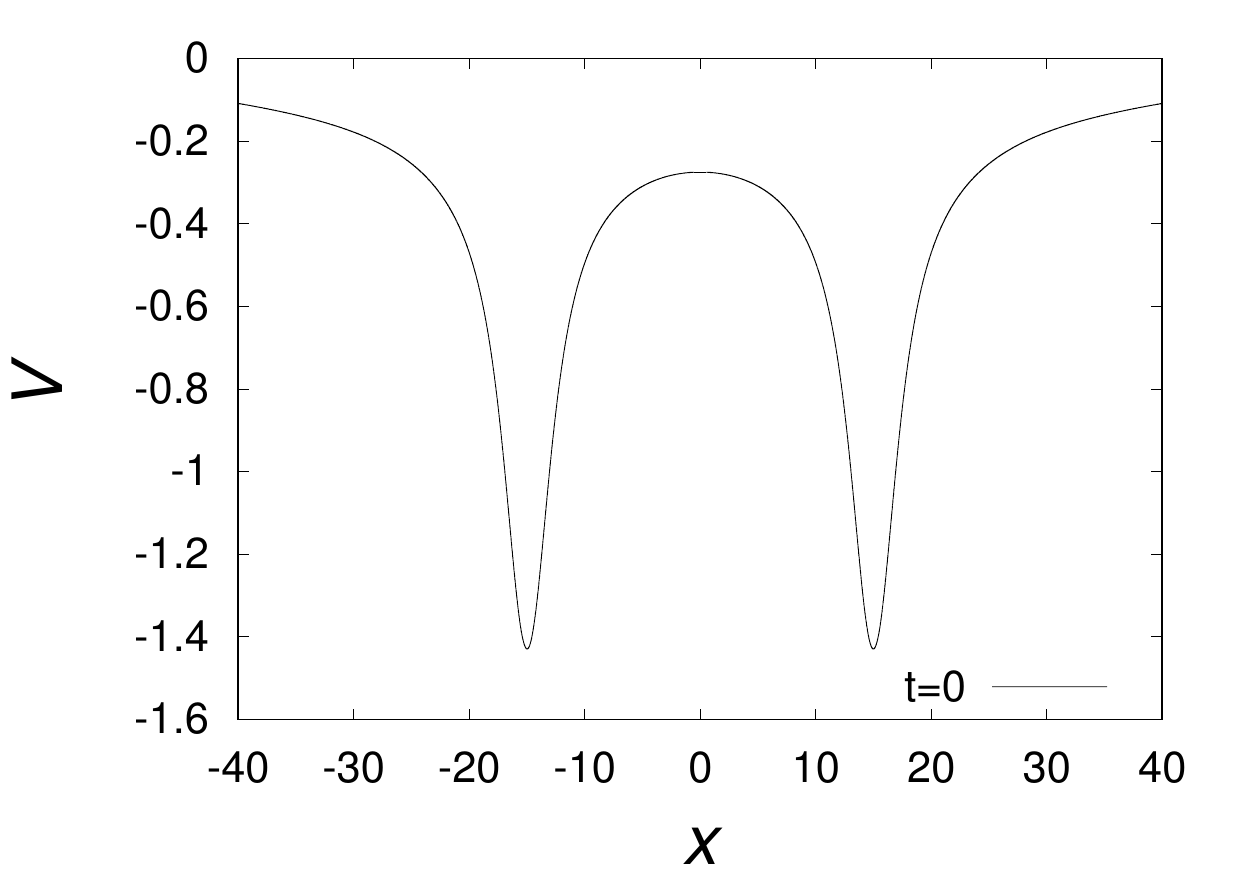}
\includegraphics[width=4.15cm]{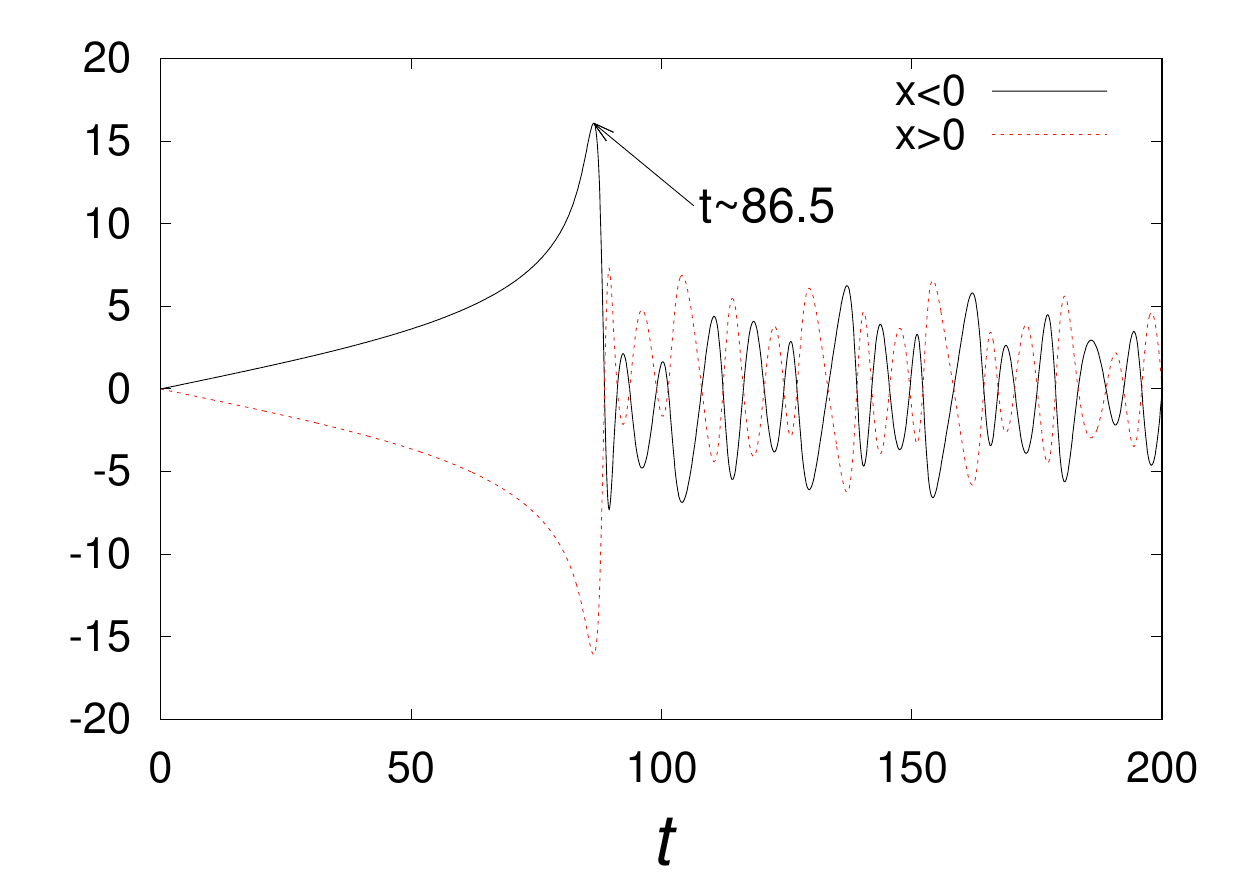}
\includegraphics[width=4.15cm]{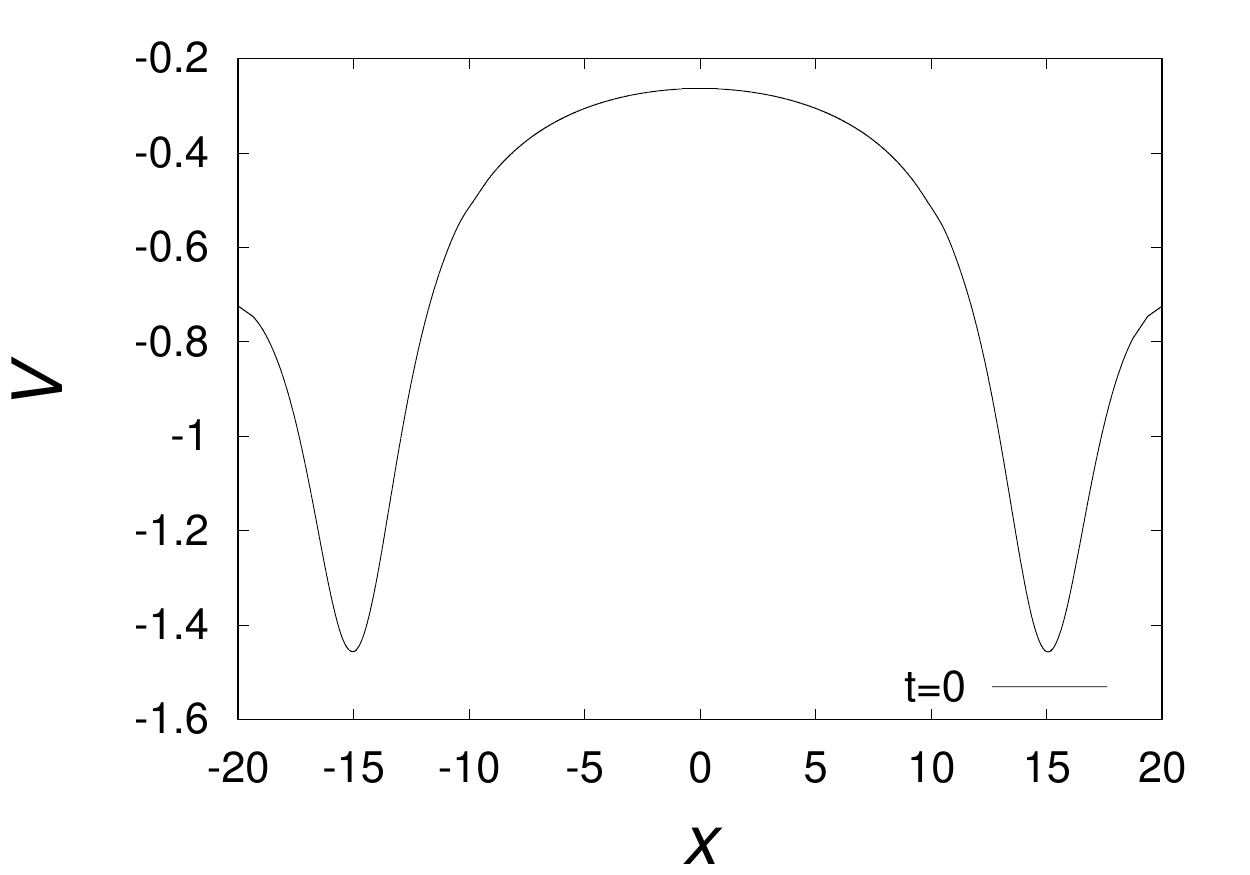}
\includegraphics[width=4.15cm]{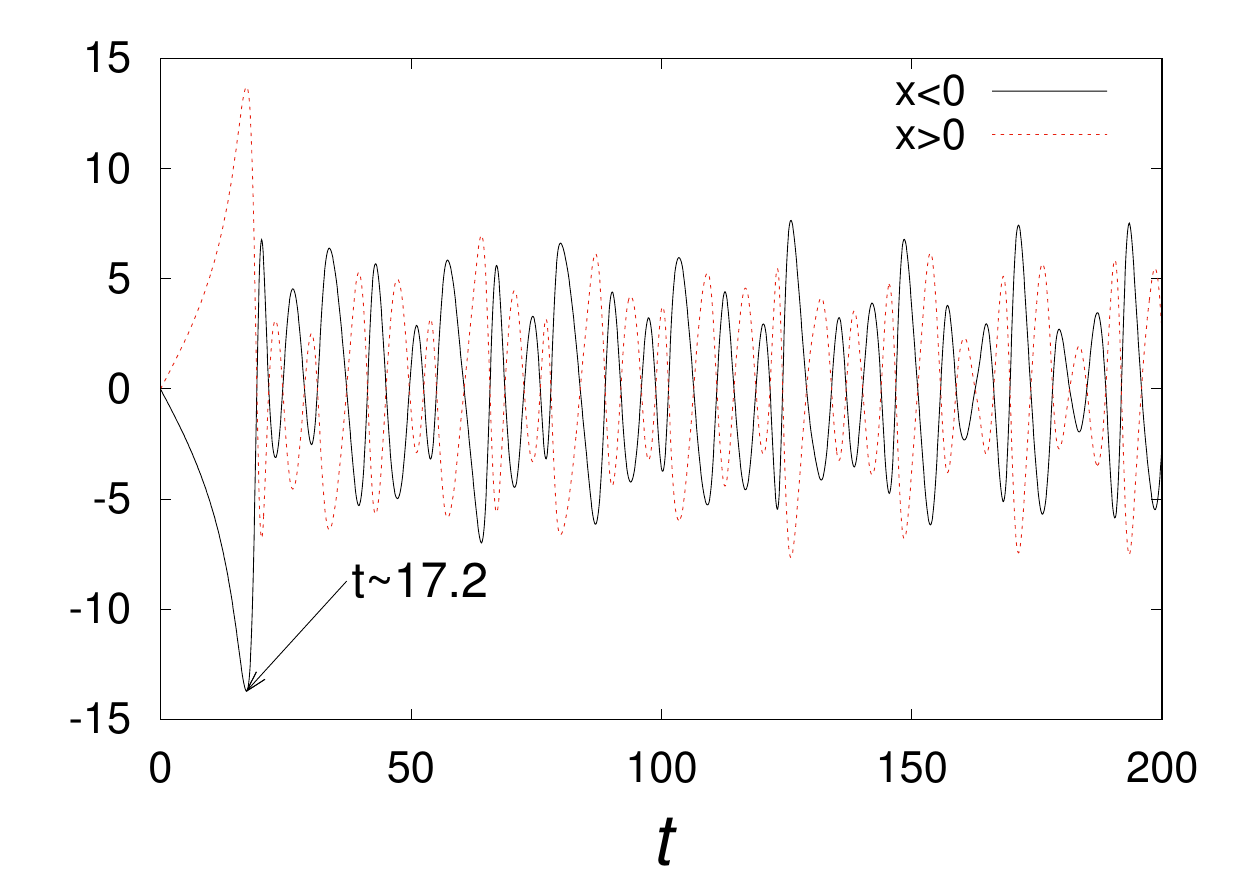}
\includegraphics[width=4.15cm]{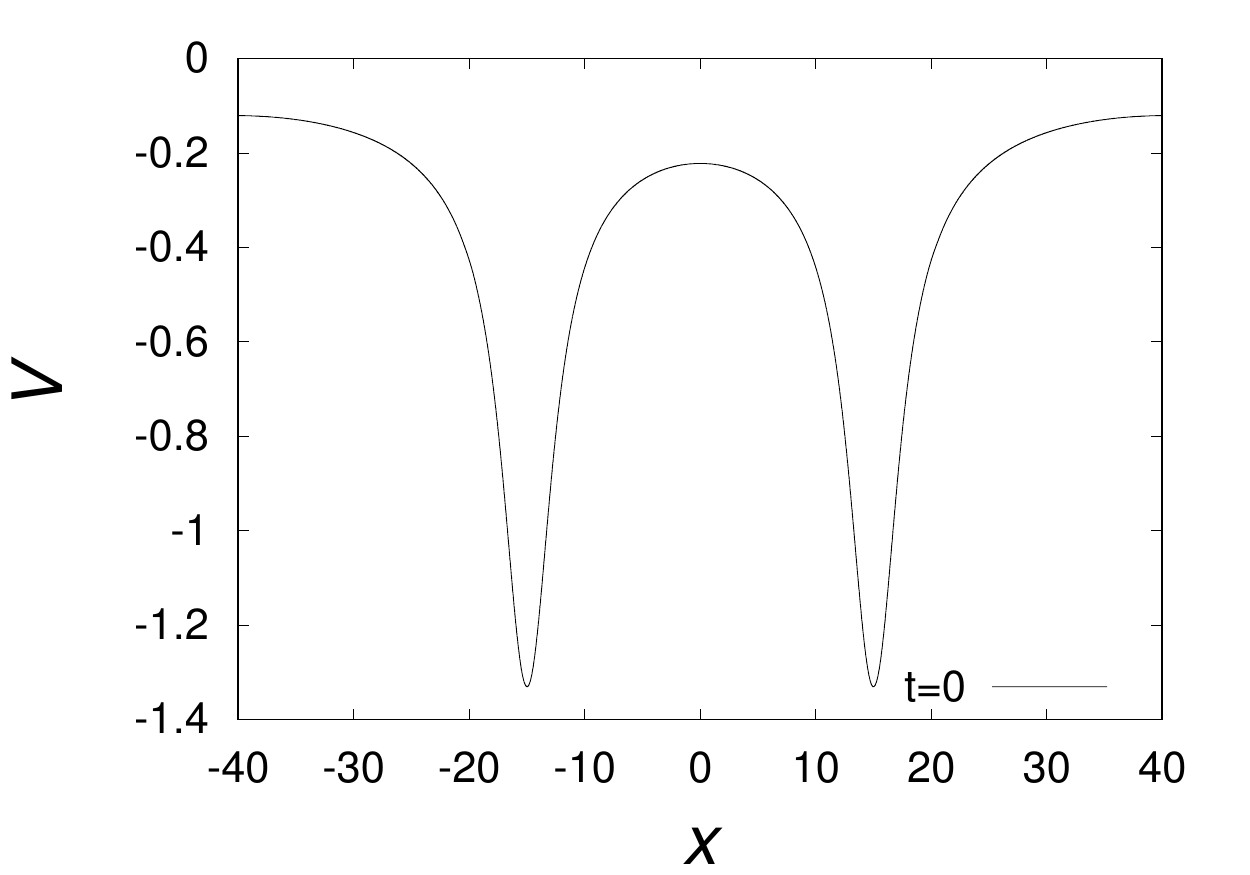}
\includegraphics[width=4.15cm]{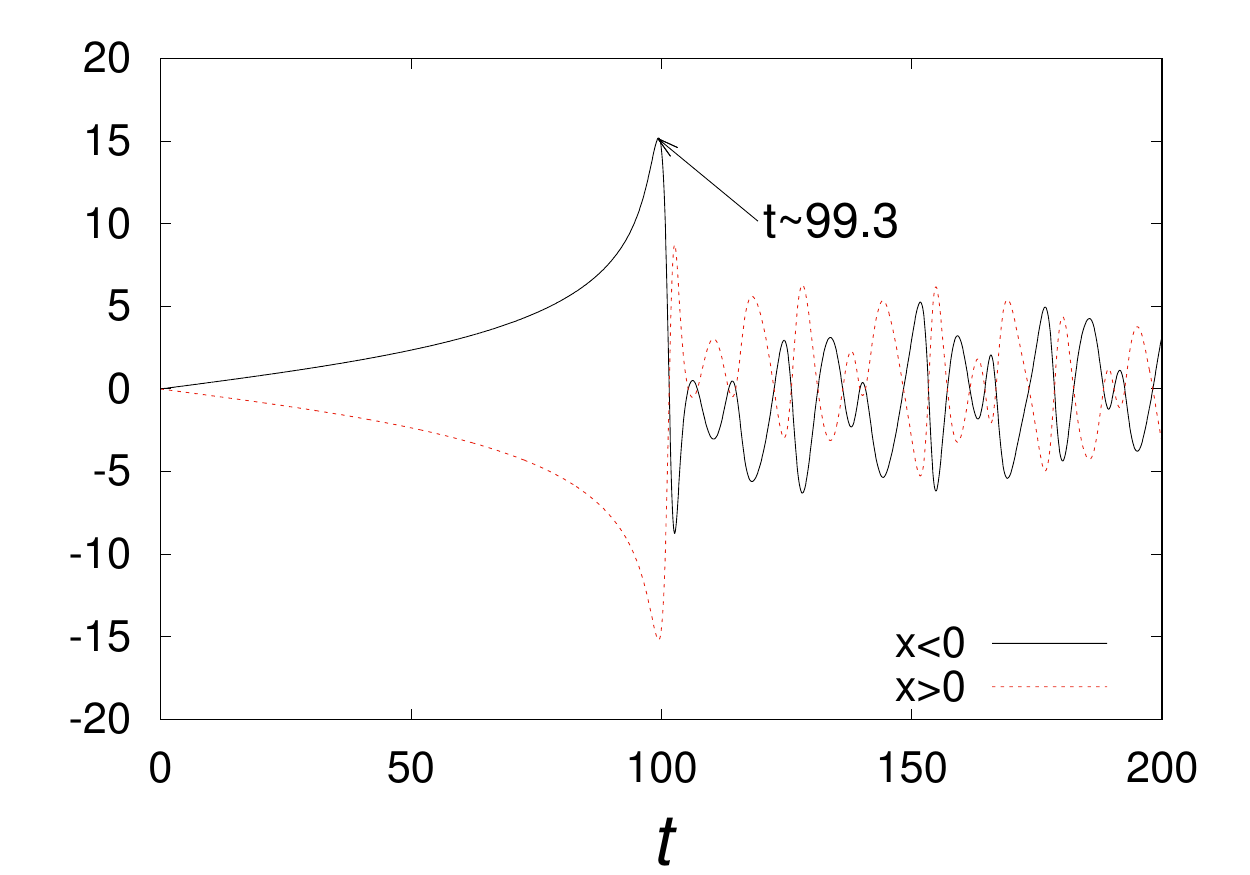}
\caption{\label{fig:head-onC} 
Case C. Form top to bottom results for the isolated domain  $[-20,20]^3$, periodic domain  $[-20,20]^3$ and periodic domain  $[-40,40]^3$. At the left we show the gravitational potential at initial time along the $x-$axis. At the right we show $\langle p_x \rangle$ integrated in the semi-domains $x<0$ and $x>0$, as function of time.}
\end{figure}

Case C shows a more significant contribution of the boundary conditions. Instead of being pulled towards each other, the two solitons are pulled by their equivalent counterparts from the neighboring domains along the $x$-direction. The results are presented in Figure \ref{fig:head-onC}. It is worth noting that in the isolated scenario, the solitons collide frontally, whereas in the periodic domain, they collide from the ``behind'', as evidenced by the linear momentum, which is positive for $x>0$ and negative for $x<0$. This indicates that initially they are moving apart from each other.

\subsection{Merger in orbit}

This is a case where the collision between two solitons has orbital angular momentum. We present an illustrative example for the initial conditions with the solitons centered at $(-10,10,0)$ and $(10,-10,0)$ and initial velocities $v_{x0}=0.1$ and $ -0.1$ respectively. We evolve the system in the domain $D=[-40,40]^3$ using isolated boundary conditions, and domains $D=[-20,20]^3$ and $[-40,40]^3$ using periodic boundary conditions. 

\begin{figure}
\includegraphics[width=8cm]{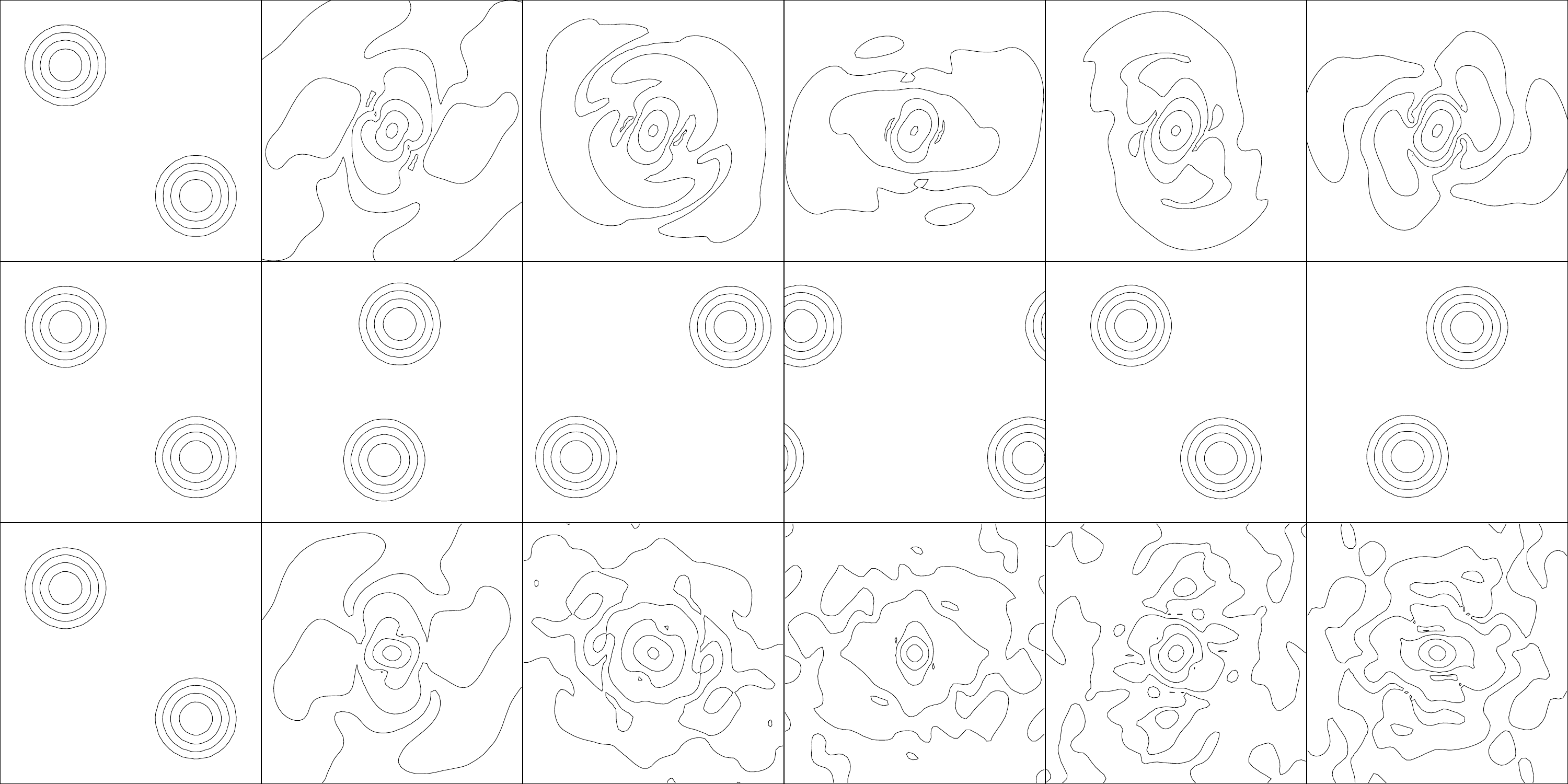}
\caption{Snapshots of the density level curves $|\Psi|^2$ at times $t=0,100,200,300,400$ and 500 for isolated boundary conditions in the domain $D=[-40,40]^3$ (top) , periodic boundary conditions in the domains $D=[-20,20]^3$ (middle)  and  $D=[-40,40]^3$ (bottom).}
\label{fig: merger in orbit}
\end{figure}

As observed in previous studies such as \cite{Schwabe:2016,GuzmanAlvarezGonzalez2021}, when the system is isolated, a significant portion of matter and angular momentum is radiated away. However, using a periodic domain allows for the re-entry of matter and angular momentum, which can then combine with the binary system. Figure \ref{fig: merger in orbit} displays snapshots of isocontours of the density $|\Psi|^2$ at times $t=0,100,200,300,400,$ and $500$, from left to right, respectively.

\begin{figure}
\includegraphics[width=4.15cm]{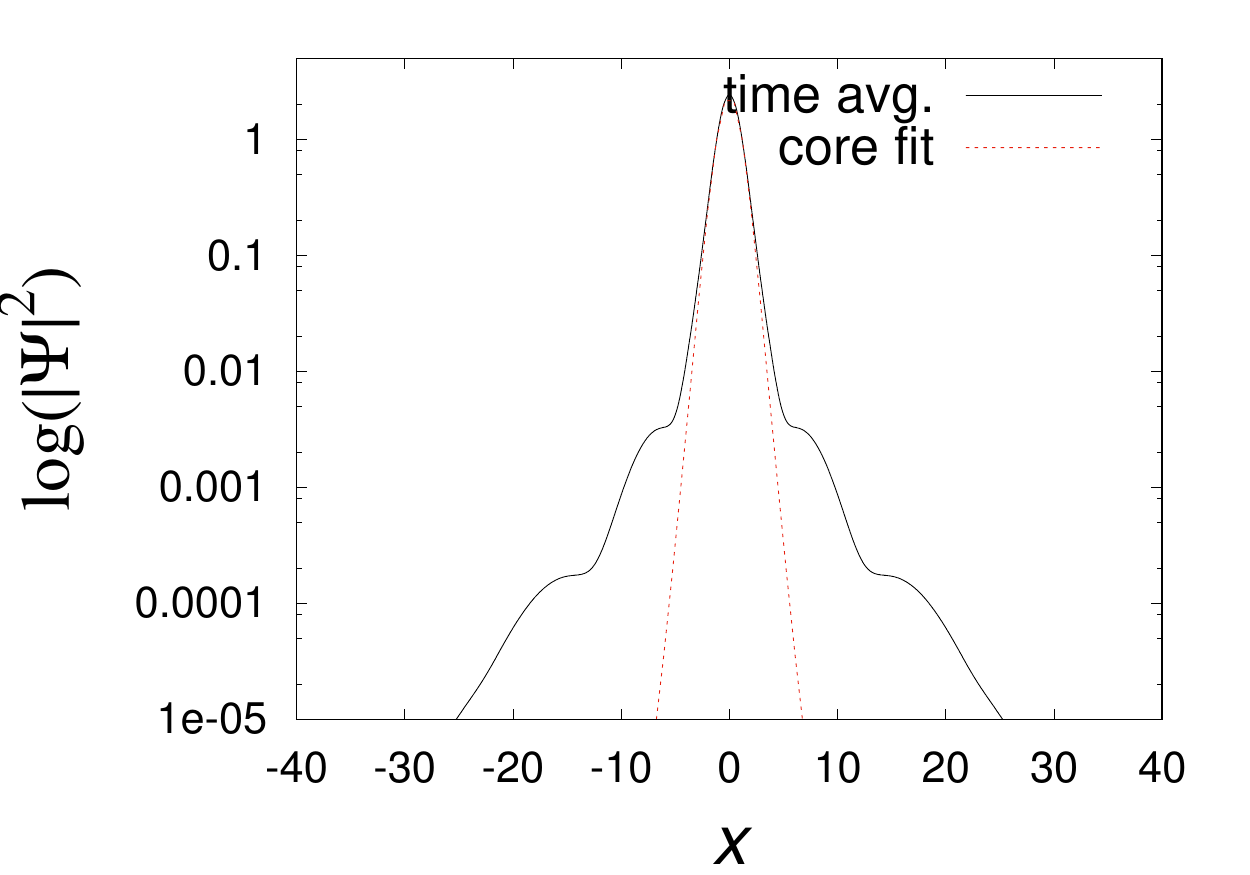}
\includegraphics[width=4.15cm]{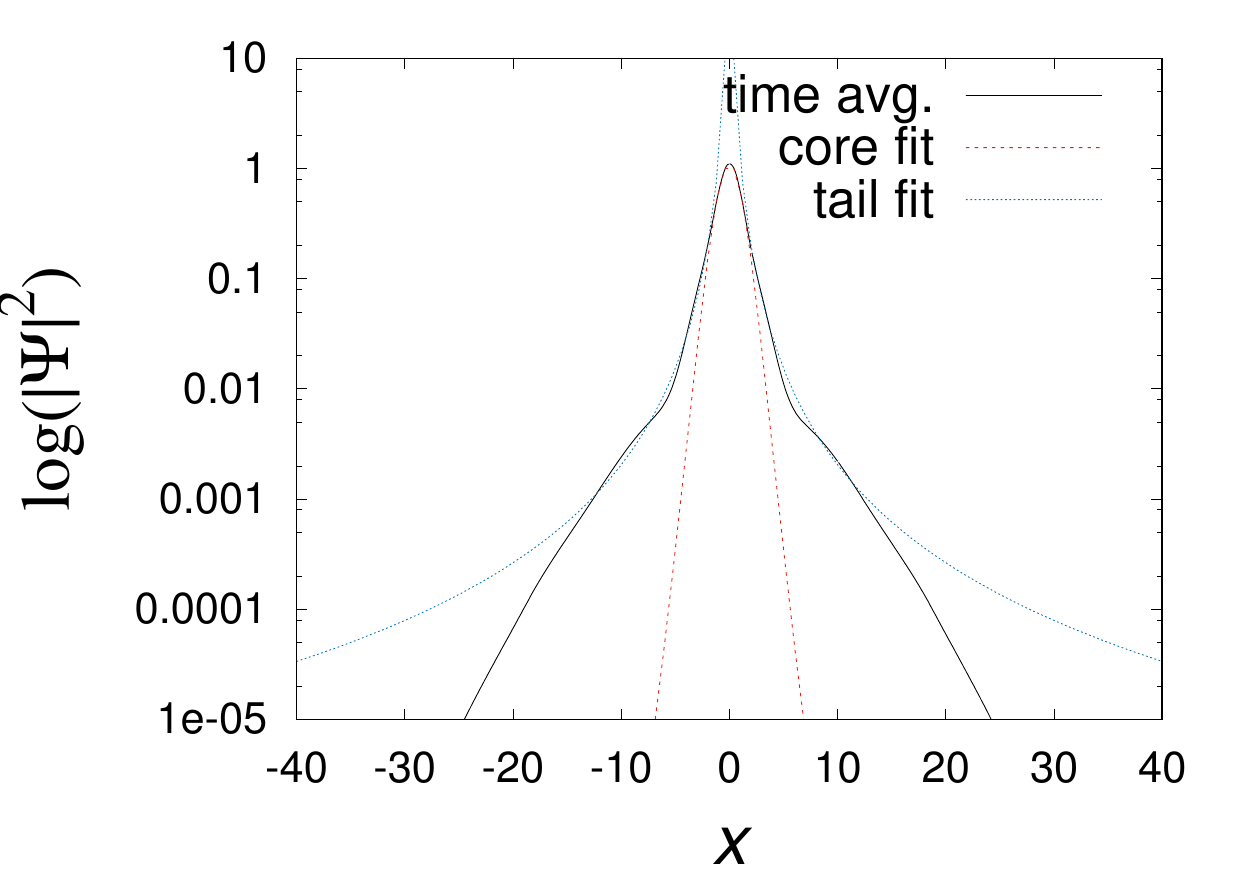}
\caption{Time average of the density calculated over the interval $t\in[250,500]$ once the two equilibrium configurations have merged, simulated in the domain $[-40,40]^3$. On the left the case of isolated boundary conditions and on the right the case of periodic boundary conditions. Notice that the periodic domain allows the formation of a tail with polynomial profile.}
\label{fig: binary densities avg}
\end{figure}

At the top of the Figure, we show the results using the isolated domain, indicating how the final configuration rotates and remains stable. In the middle row, we present snapshots of the evolution with periodic boundary conditions in the small domain. This can be ideally seen as an infinite network of equilibrium configurations, where the symmetry of the system and the domain size make the sum of the external forces near zero, leaving only the equilibrium configurations in uniform rectilinear motion, which is ideally eternal. This can be seen similarly to the boosted configuration presented in the Appendix. In the bottom row, we present snapshots of the density in the large periodic domain, which illustrates how the periodicity of the system enables the high kinetic energy matter expelled during the merger to re-enter and spread throughout the entire domain, ultimately leading to the formation of tail profiles. Notably, in both the isolated and periodic boundary condition cases for the domain $D=[-40,40]^3$, a single density distribution forms at the origin of coordinates, with its averaged profile along the $x$-axis shown in Figure \ref{fig: binary densities avg}. This profile is well-fitted by the core profile (\ref{eq:coreprofile}).

\begin{figure}
\includegraphics[width=4.15cm]{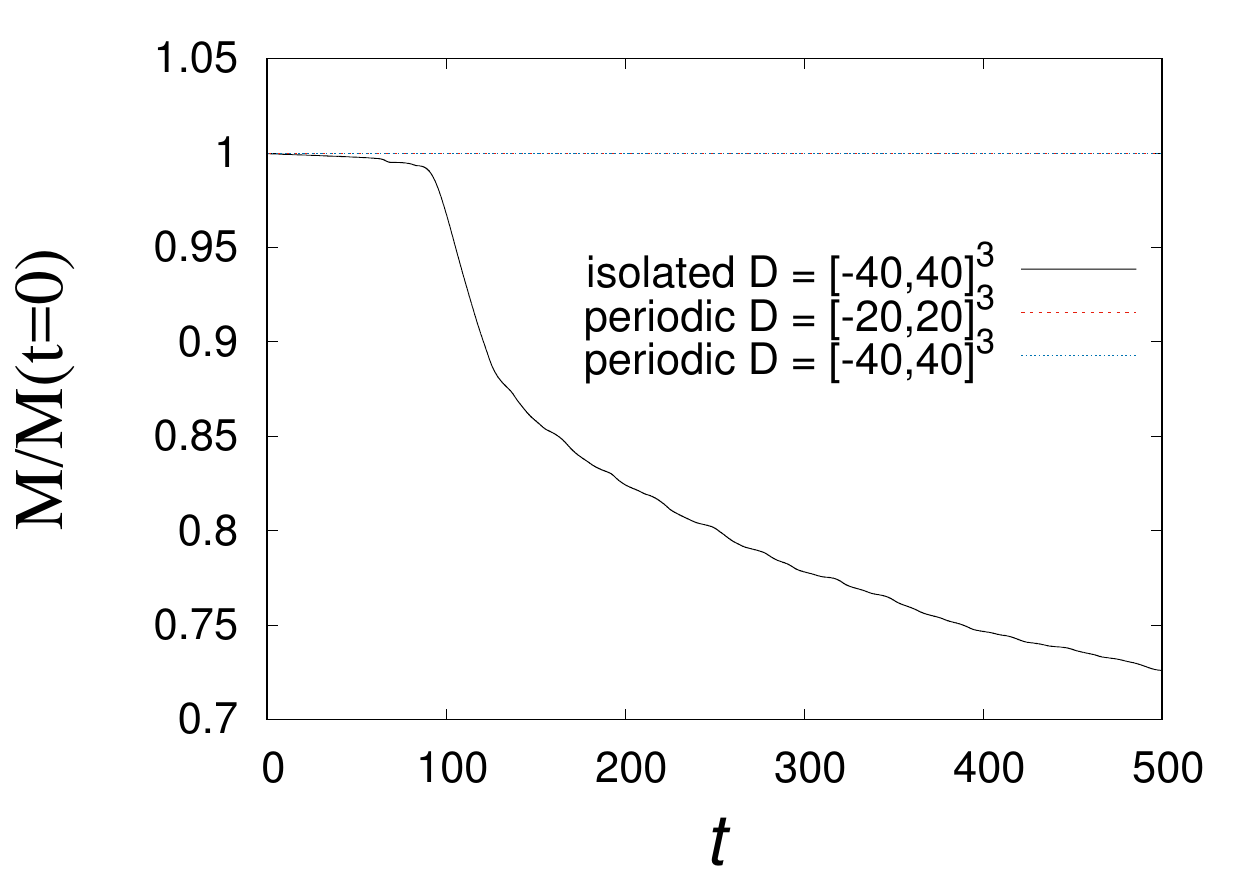}
\includegraphics[width=4.15cm]{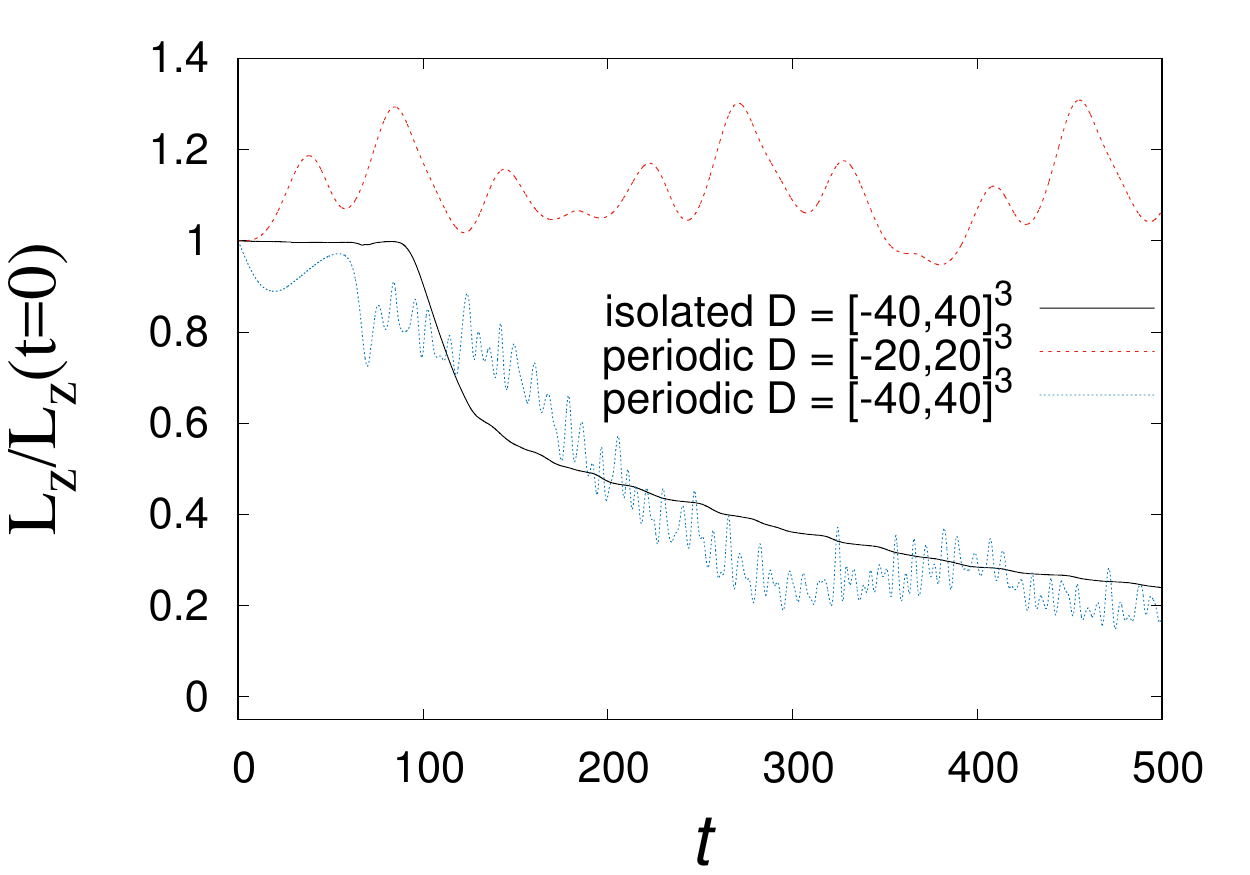}
\caption{Evolution of a binary system with non-zero angular momentum, isolated boundary conditions in the domain $D=[-40,40]^3$ (solid black lines) and with periodic boundary conditions in the domains $D=[- 20,20]^3$ (dotted red lines) and $D=[-40,40]^3$ (dashed blue lines). Evolution of the total mass $M$ and angular momentum $L_z$ as a functions of time.}
\label{fig: binary diagnostics}
\end{figure}

Figure \ref{fig: binary diagnostics} shows the diagnostics for this system. The mass as a function of time indicates that in the periodic domain, the mass is conserved, while in the isolated case, there is a mass loss of approximately $28\%$ radiated out of the domain. Additionally, in the isolated domain, the angular momentum is carried out of the domain with the matter, while in the periodic domain, the reentrance is noticeable in the small domain. However, we do not observe any trend of $L_z$ in terms of the domain size.

\subsection{Merger of multiple cores}
\label{subsec:multiple}

We investigate a final, more complex problem related to core formation. Drawing inspiration from previous studies \cite{Schwabe:2016,Mocz:2017wlg,Li_2021,KangLiu2023}, we simulate the merger of $30$ equilibrium configurations with an ultralight bosonic mass of $m_B = 10^{-22}eV/c^2$. These equilibrium configurations have random masses ranging from $2.6\times 10^8 M_{\odot}$ to $26\times 10^8 M_{\odot}$ and are initially positioned randomly within a cube of side length $30$ kpc, enabling comparison between the periodic and isolated boundary conditions. In the periodic domain case, we evolve the system in both a small cubic domain and a large cubic domain, with sides of 80 kpc and 100 kpc, respectively. We use the same random initial positions, configurations, and resolution in both cases, allowing us to isolate the effects of domain size on the dynamics of the system.

\begin{figure}
\centering
\includegraphics[width=4.15cm]{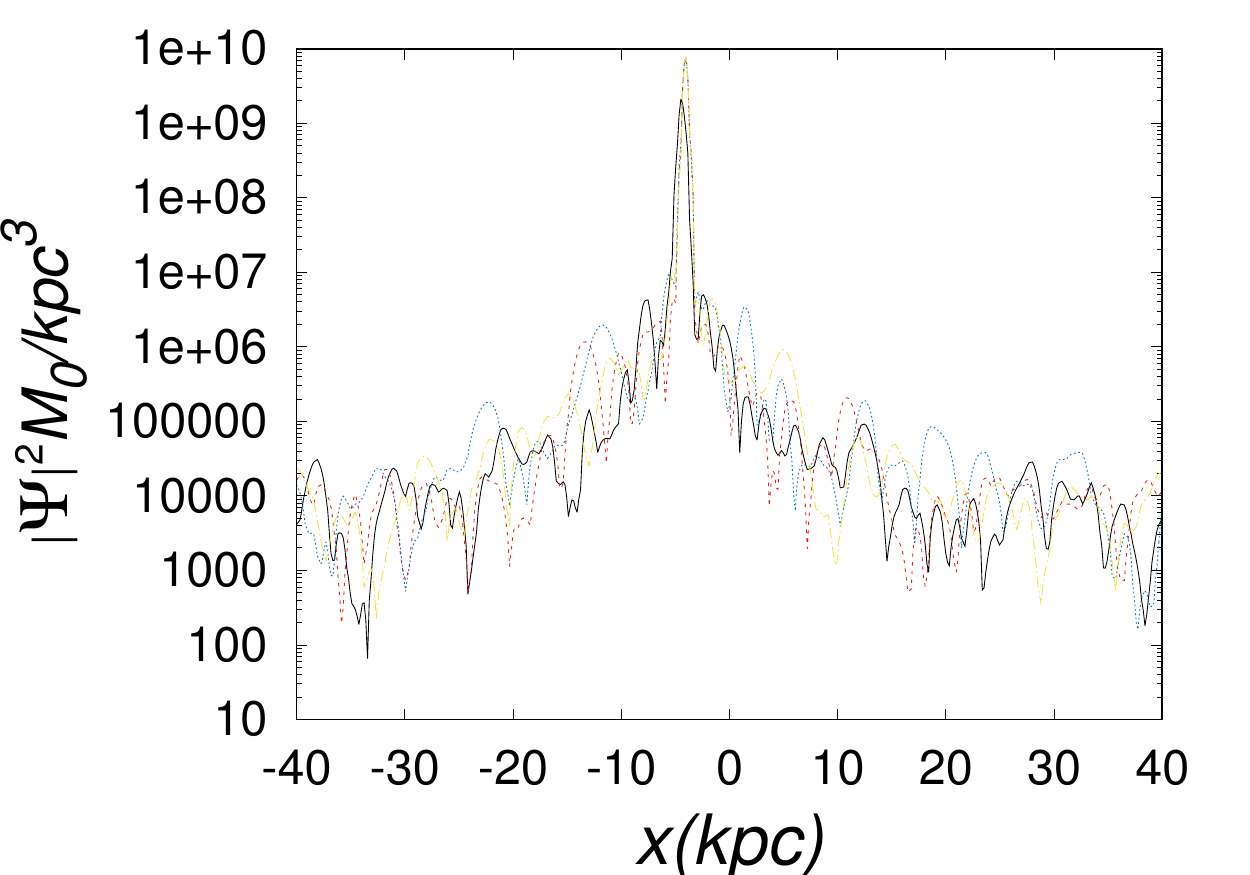}
\includegraphics[width=4.15cm]{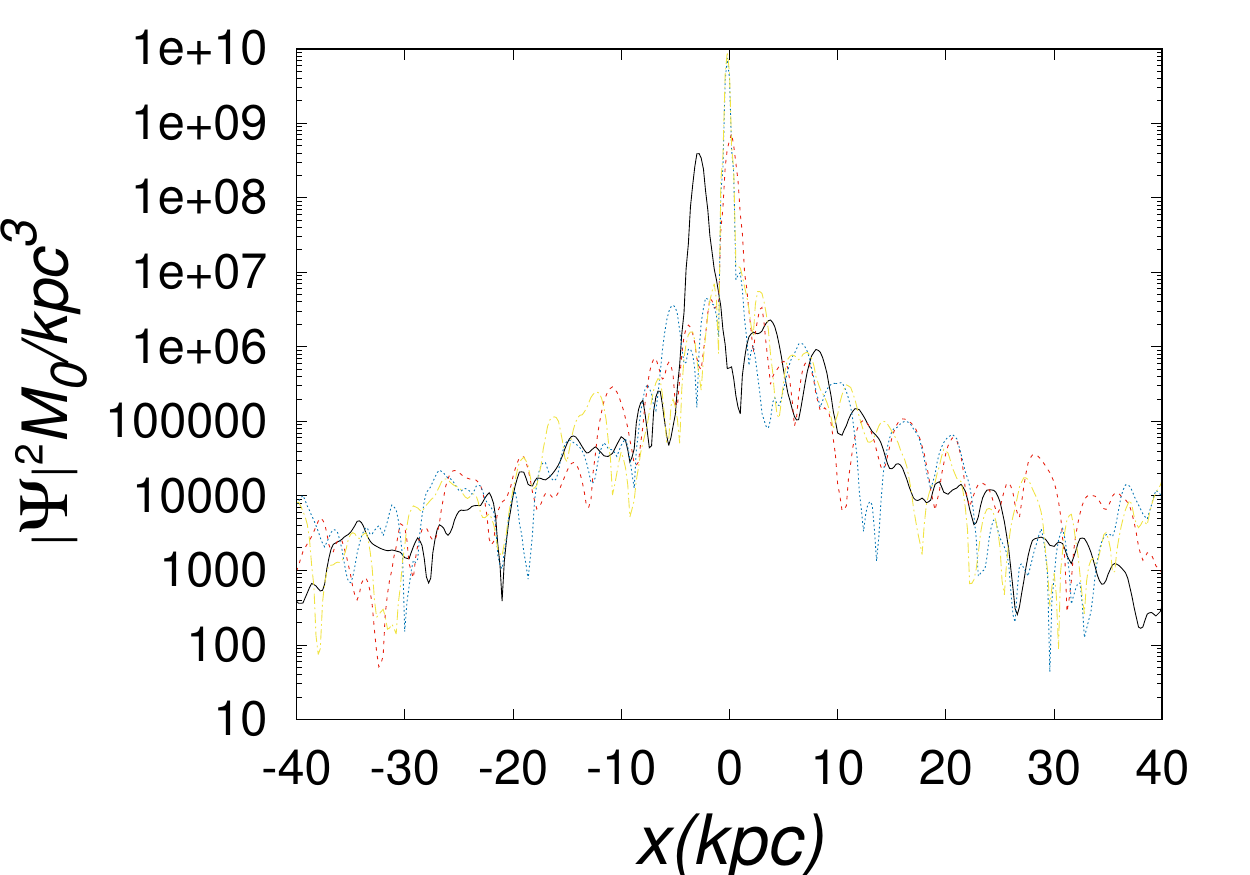}
\includegraphics[width=4.15cm]{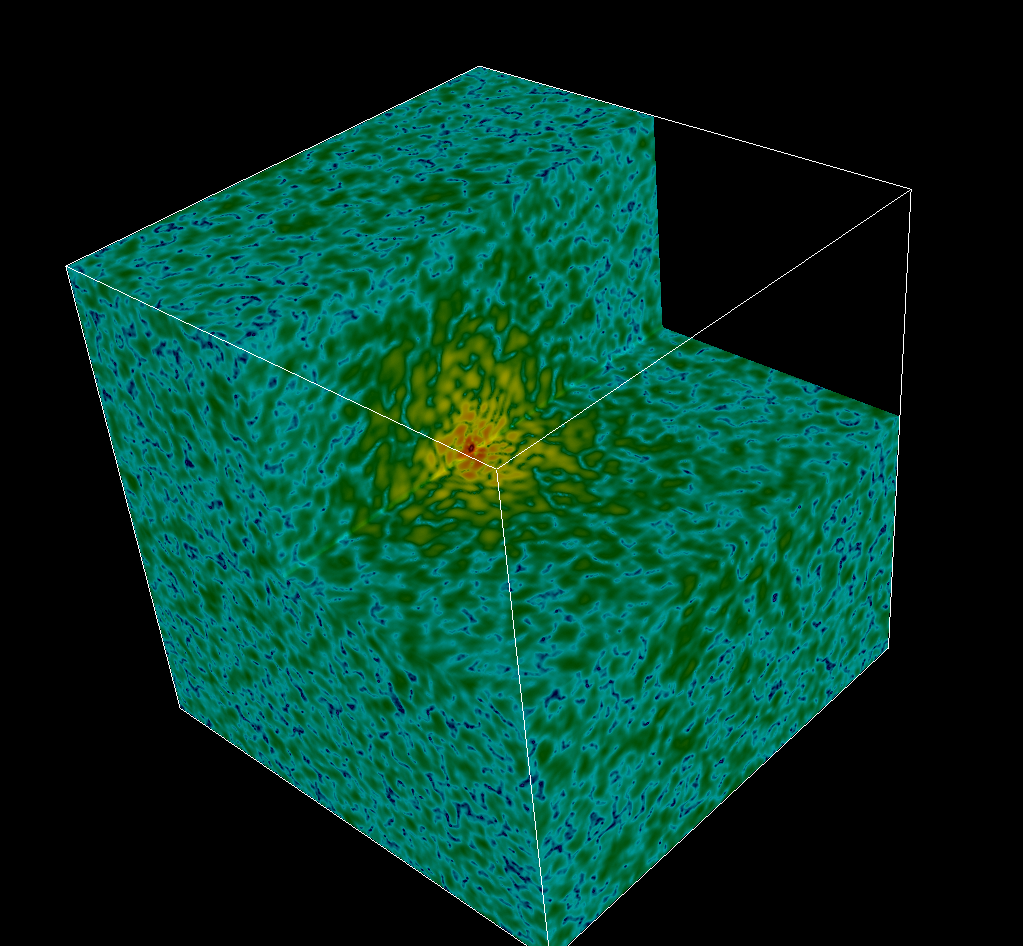}
\includegraphics[width=4.15cm]{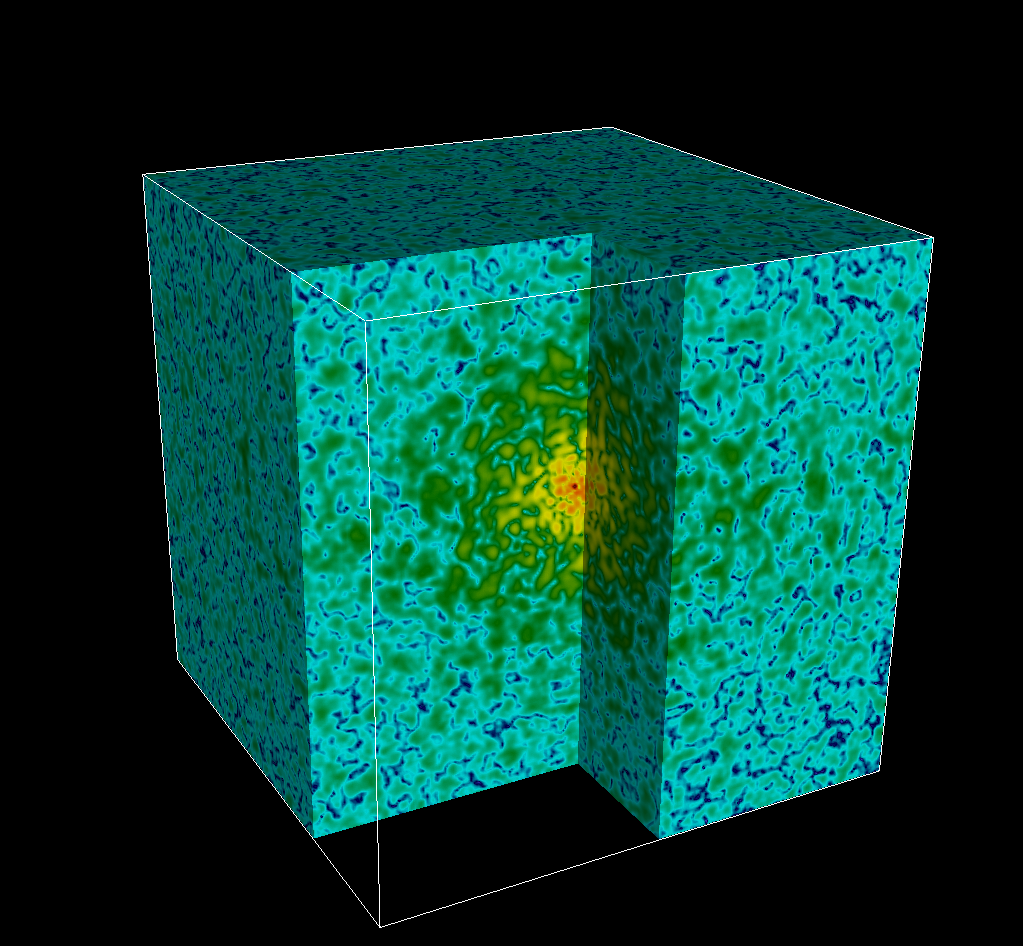}
\includegraphics[width=4.15cm]{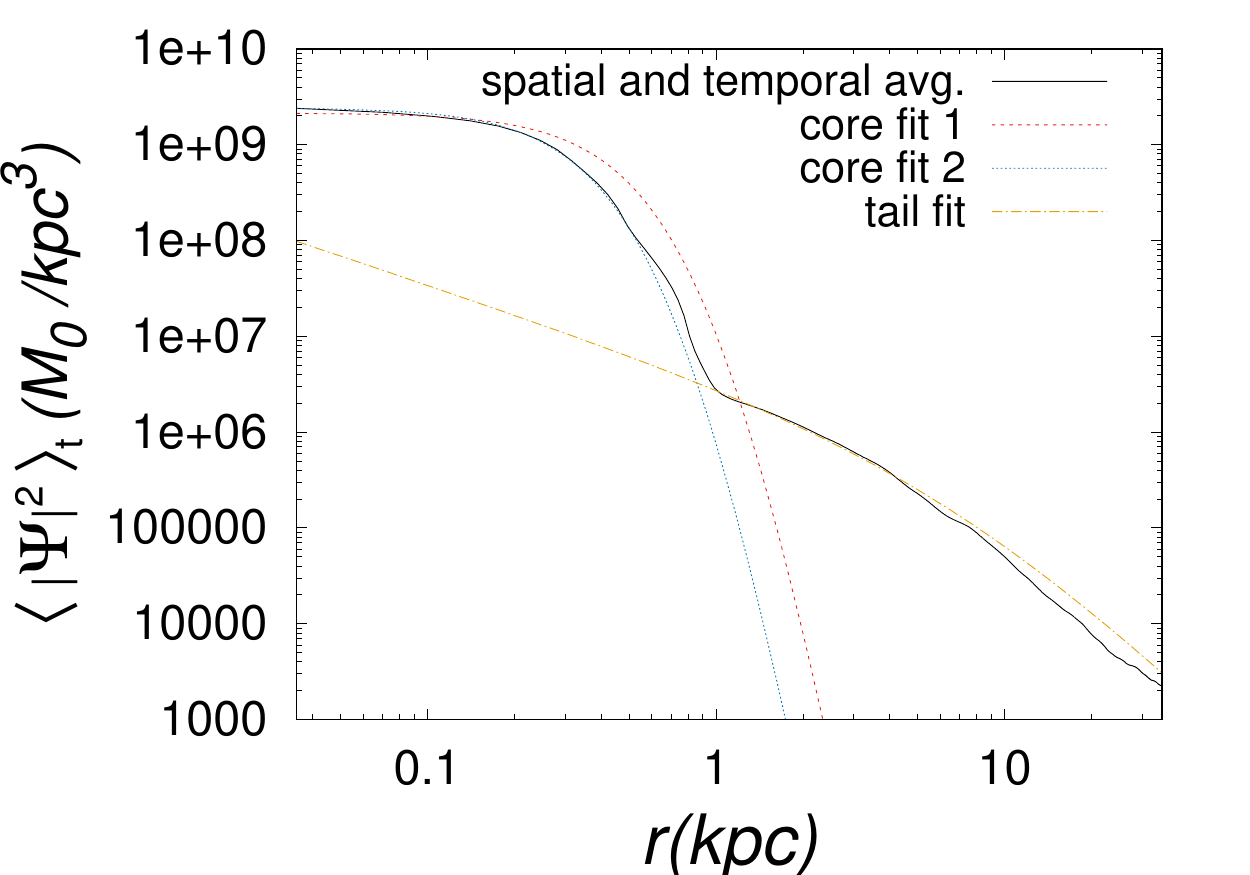}
\includegraphics[width=4.15cm]{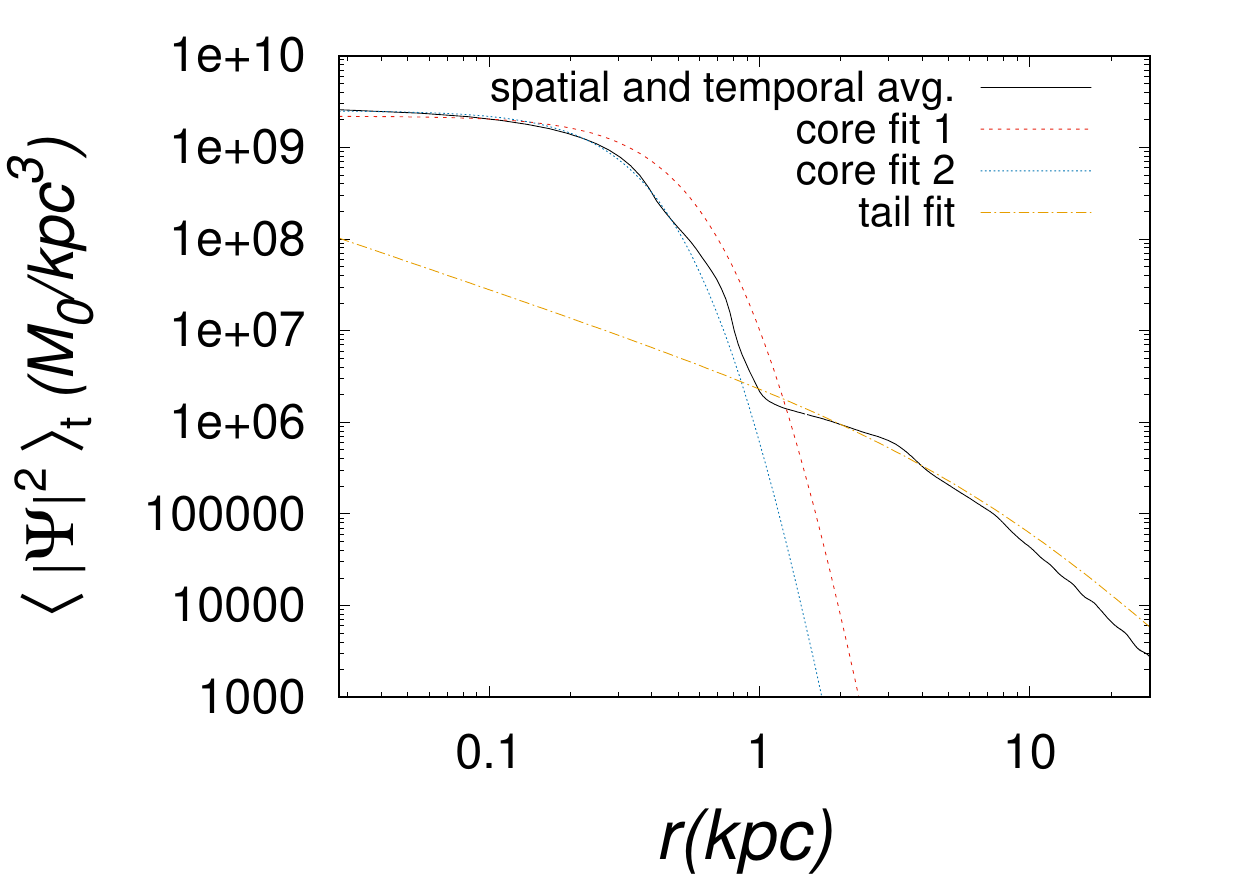}
\caption{On the left/right we show results for the {\it small/big} domain. In the first row we show a few snapshots of the density along a line parallel to the $x-$axis that passes through the position of maximum density, which show how dynamic the system is. In the second row we show a volume view of the density illustrating the distribution of the bosonic cloud. In the third row we show the fit of density using formula (\ref{eq:coreprofile}) for the core and (\ref{eq:tailprofile}) for the NFW tail.}
\label{fig:multiplefit}
\end{figure}

The results are summarized in Figure \ref{fig:multiplefit}. On the left/right we show results obtained from simulations on  the small/big domain. At the top we show some snapshots of the density projected on the $x-$axis, that show the dynamic behavior and interference patterns that change with time. In the mid row we show a snapshot of density in three-dimensions, at a time when the core is already formed. Finally, we calculate an average of the density in time and along various directions from the center of the core, in order to fit the core-tail structure that we show in the third row. The core is fitted using the function (\ref{eq:coreprofile}), using two methods. In the {\it first method} we fit the core with $r_c$ and $\rho_{0,core}$ as free parameters, the best fittings obtained with  
($r_c \sim 0.228$kpc,
$\rho_{0,core} \sim 2.42 \times10^{9} M_{\odot} /\textup{kpc}^{3}$) 
and
($r_c \sim 0.223$kpc,
$\rho_{0,core} \sim 2.51\times10^{9} M_{\odot} /\textup{kpc}^{3}$)
in the small and big domains respectively, drawn with the blue line. 
The {\it second method} enforces the scaling relation 
$M r_c\sim$constant \st{to hold}, which implies a constraint on the two free parameters; in this case the fitting parameters are 
($r_c \sim 0.311$kpc,
$\rho_{0,core} \sim 2.13 \times10^{9} M_{\odot} /\textup{kpc}^{3}$) 
and
($r_c \sim 0.309$kpc,
$\rho_{0,core} \sim 2.19\times10^{9} M_{\odot} /\textup{kpc}^{3}$)
in the small and big domains respectively, whose profiles are represented with red lines. Finally, the tail is fitted with the NFW profile \cite{NFW}

\begin{equation}
\rho_{tail} = \dfrac{\rho_{0,tail}}{\frac{r}{R_s}\left(1+\frac{r}{R_s}\right)^2}\label{eq:tailprofile}
\end{equation}

\noindent with fitting parameters $\rho_{0,tail} \sim 4.6525\times 10^{5} M_\odot  / \textup{kpc}^3 $, $R_s \sim 7.473$ kpc for the small domain and $\rho_{0,tail} \sim 3.3\times 10^{5} M_\odot/\textup{kpc}^3$, $R_s \sim 8.67$ kpc for the big domain. The simulation lasts $\sim 12.7$Gyr, a time window within which none of the configurations has yet virialized, which explains why in the periodic domain case, the $M r_c \sim constant$ constraint is not yet satisfied, e.g. as expected according to \cite{KangLiu2023}.

\begin{figure}
\centering
\includegraphics[width=8.15cm]{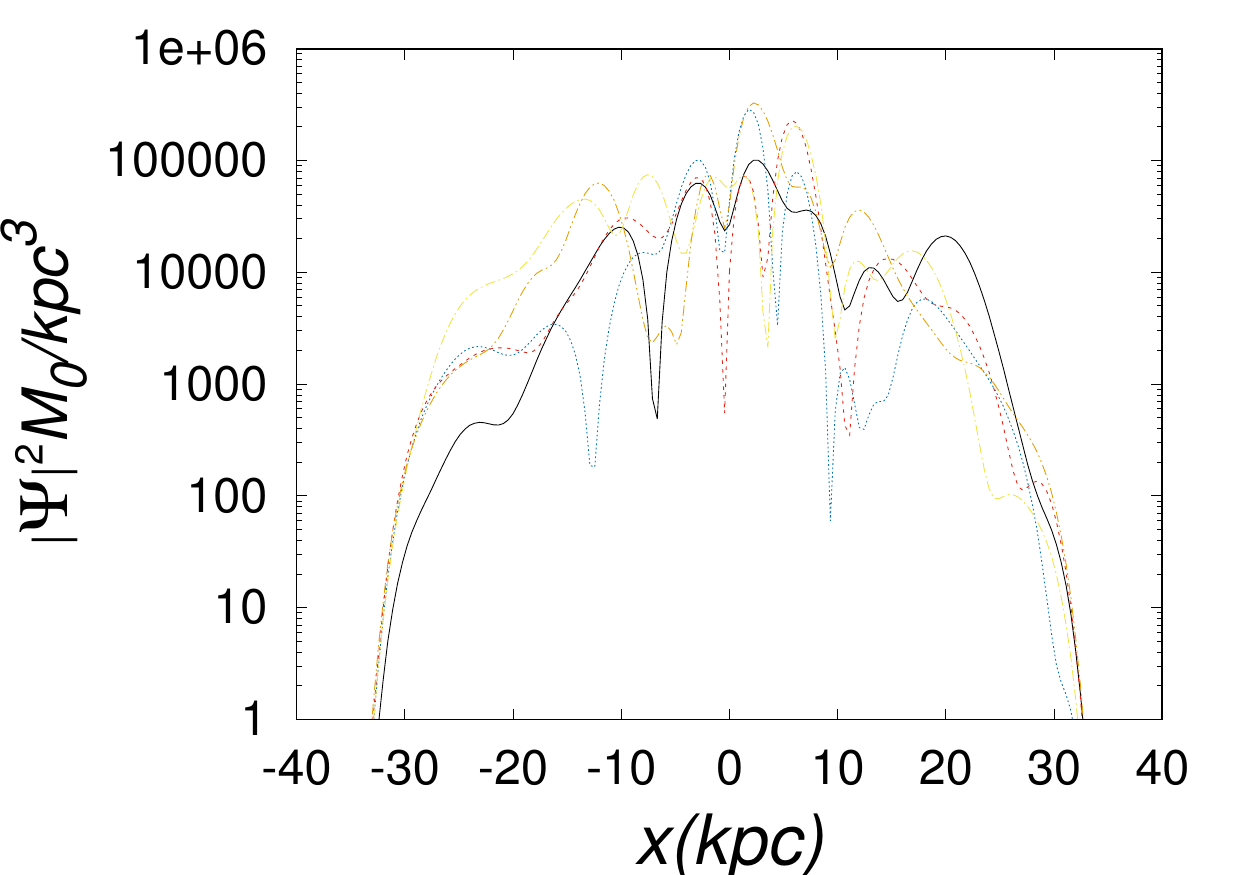}
\includegraphics[width=8.15cm]{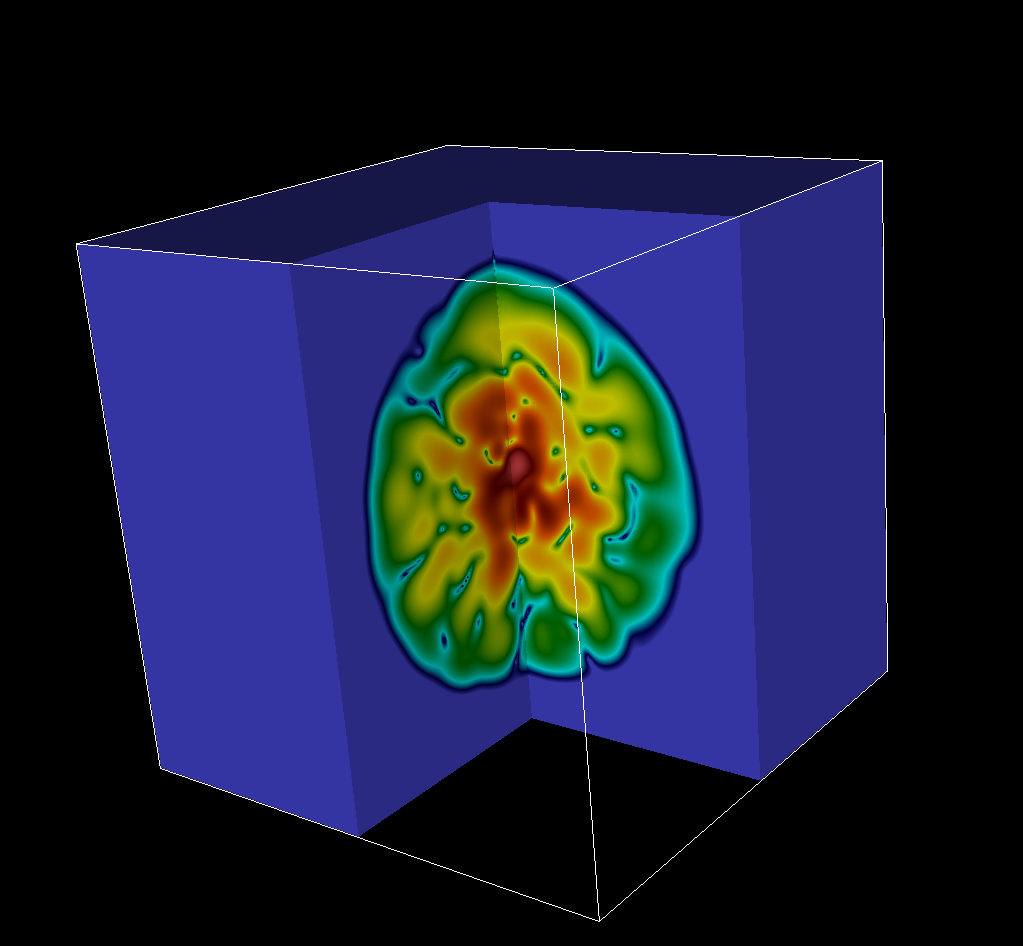}
\includegraphics[width=8.15cm]{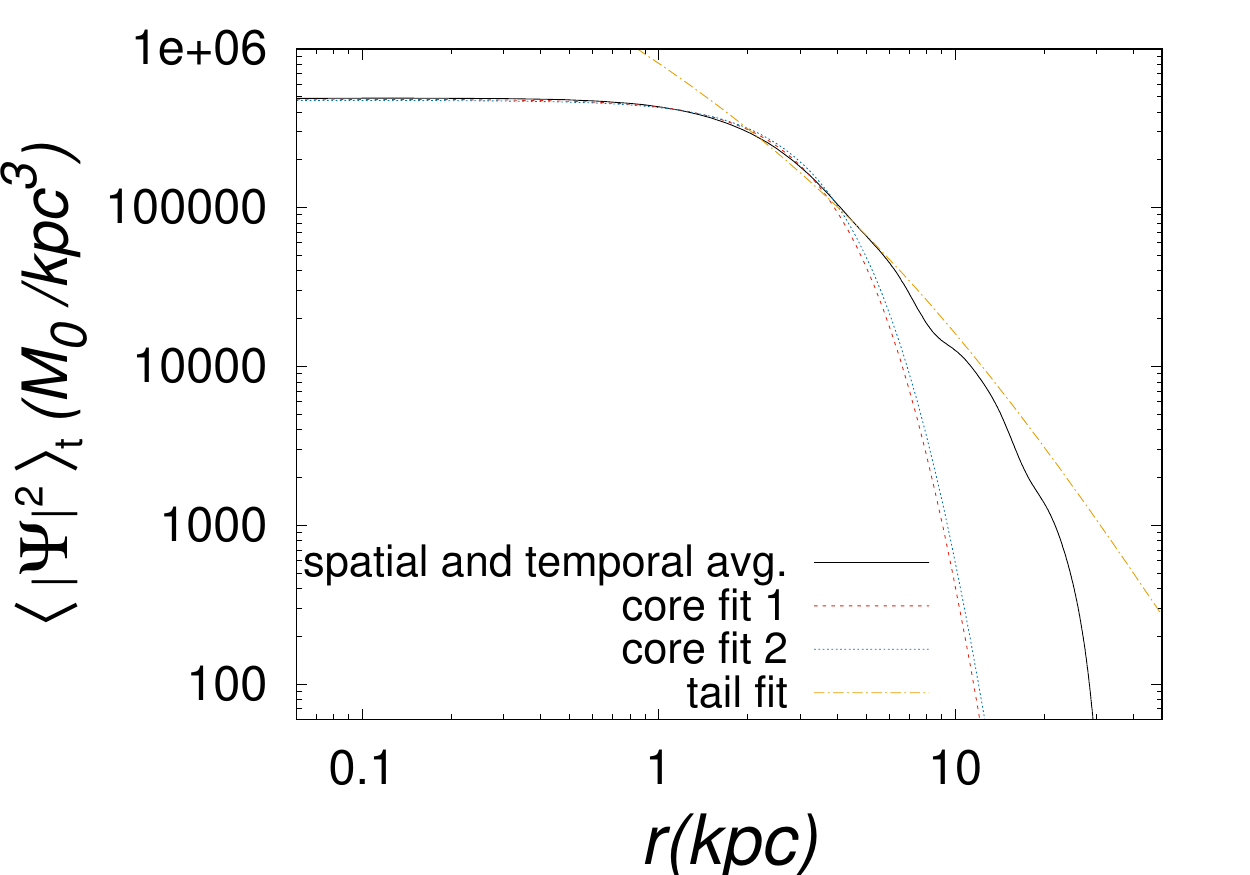}
\caption{(Top) Snapshots of the density along the $x-$axis that illustrate how localized the density is. (Middle) Three dimensional view of the density that shows the concentration of matter inside the sponge region. (Bottom) Density profile and its comparison with fittings at core and tail regions; notice that the core fittings agree.}
\label{fig:multiplefitIsolated}
\end{figure}

For comparison, we simulate this scenario using isolation conditions, where gravitational cooling is expected to drive the configuration toward an equilibrium solitonic profile in asymptotic time. We use the same numerical parameters as the periodic boundary simulation, with a domain of side 80 kpc and 30 solitons initially distributed in a box of side 30 kpc around the center of the domain. The results are summarized in Figure \ref{fig:multiplefitIsolated}, which includes a few snapshots of the density along the $x$-axis, illustrating the concentration of density restricted by the presence of the sponge. Additionally, a volume view of the snapshots highlights the density concentration and the solitonic density profile.

The fitting parameters for the averaged core density profile in Eq. (\ref{eq:coreprofile}) using the first method with the two fitting parameters free are ($r_c = 2.64$ kpc,$\rho_{0,core} = 4.74\times 10^{5} {\rm M_0/kpc^3}$), whereas using the second method gives 
($r_c = 2.54$ kpc,$\rho_{0,core} = 4.81\times 10^{5} {\rm M_0/kpc^3}$)
when enforcing the condition $M r_c\sim$constant. Notice that unlike the periodic domain, in this case the fitting profiles are very similar as illustrated with curves blue and red in the bottom of Figure \ref{fig:multiplefitIsolated}. Finally, for completeness the parameters of a tail with the NFW profile are
$\rho_{0,tail} \sim 1.74\times 10^{5} M_\odot  / \textup{kpc}^3 $, 
$R_s \sim 6.26$ kpc.

This simulation illustrates the dynamics of gravitational cooling in 3-D from initial conditions that are far from spherically symmetric. Previous work
(see e.g. \cite{GuzmanUrena2006}) demonstrated that when initial conditions are spherically symmetric, gravitational cooling drives the configuration asymptotically towards the equilibrium ground-state solution of the solitonic core, with a density profile that drops steeply outside the core.  Some nonsphericities were found to be expelled, as well, in simple axi-symmetric scenarios \cite{BernalGuzman2006b}. 
As a result, it has been hypothesized that the solitonic core corresponding to the ground state is the attractor solution for a wider range of initial conditions, as long as gravitational cooling is permitted to occur and the system is allowed to evolve for a sufficient period to reach a state of relaxation.

Our simulation here represents an attempt to demonstrate this explicitly, in 3-D, from initial conditions that depart strongly from spherical symmetry, with control of gravitational cooling implemented via the isolating effects of the sponge boundary conditions.  While the simulated time of order 12.7 Gyr is not long enough to allow the system to complete its relaxation to the asymptotic state of the ground state solution, it is clearly moving in that direction over time.  By contrast, when the same initial configuration was simulated with periodic BC's, with no sponge to absorb the mass and energy expelled by gravitational cooling, the density profile is fit by a solitonic core surrounded by a power-law envelope, or tail.  Even though it is not fully relaxed by the final time-slice, the isolated case (i.e. with sponge) already shows a solitonic core surrounded by a profile that drops sharply toward large radius, much more steeply than in the case with periodic BC's.

\section{Conclusions}
\label{sec:conclusions}

This paper presents a comparison of the dynamics of FDM cores under different scenarios, utilizing both isolation and periodic boundary conditions.

Our analysis has yielded several observations for each scenario. In the simplest scenario, which involves the evolution of a ground-state equilibrium configuration of the SP system, we observed some distinct differences. Under periodic domain conditions, the density outside of the core, which decays exponentially, is redistributed into a nearly constant profile. Conversely, when using a sponge in the isolated domain, the density is forced to vanish near the boundary faces.

Our study demonstrates that the 3-D dynamical relaxation of a Gaussian density fluctuation near equilibrium results in the formation of a solitonic core, similar to the previously derived 1-D spherical symmetry ground-state equilibrium solution under isolated boundary conditions. However, notable differences emerge beyond the core, particularly in the tail zone. Under isolated domain conditions, the density in the tail decays exponentially with radius, while a nearly constant profile is observed when using a periodic domain.

Significant differences can be observed in the head-on merger of two configurations, particularly when solving the equations in a network of infinitely replicated configurations. In the most extreme case, a collision may occur through the backdoor of the domain.

During a merger of two orbiting configurations under periodic boundary conditions, it is possible that the merger may never occur. Additionally, the resulting configuration from a merger with angular momentum exhibits a tail with a density profile that is neither constant nor exponential, but rather polynomial. Notably, this configuration is connected and not fragmented, in contrast to those obtained from multi-mergers of cores and structure formation scenarios.

Lastly, we simulated the free-fall evolution of several cores to reproduce the formation of a structure with a core-tail density profile. Our simulations were conducted under both periodic boundary and isolation conditions, and we examined the effects of domain size on the density distribution. In the periodic domain, we observed that cores exhibited similar fitting parameters for two different box sizes, while the tail distribution varied. In contrast, under isolated conditions, we confirmed that there was virtually no tail.

The question of whether one BC choice is better or more realistic than the other is problem-specific. The isolating BC's are perhaps most realistic when the object that forms is, itself, physically isolated from other objects and from infall of additional mass, so that its mass is no longer increasing by merger or infall -- leaving it free to expel mass and energy to infinity, therefore, in the process of gravitational cooling, with plenty of time for it to reach the asymptotic equilibrium.  The periodic BC's, on the other hand, are most realistic when the opposite is occurring, that of objects that form by continuous mergers and mass assembly that is ongoing, so it prevents the free escape of mass and energy to infinity. 

Our findings illustrate the quantitative impact of domain size on simulation results using periodic boundary conditions. These effects are worth evaluating in future studies of various astrophysical scenarios, as they can introduce uncertainties to numerical results.


\section*{Acknowledgments}
Iv\'an Alvarez receives support within the CONACyT graduate scholarship program under the CVU 967478. This research is supported by grants CIC-UMSNH-4.9 and CONACyT Ciencias de Frontera Grant No. Sinergias/304001. The runs were carried out in the Big Mamma cluster of the Laboratorio de Inteligencia Artificial y Superc\'omputo, IFM--UMSNH. 
PRS acknowledges support from NASA under Grant No. 80NSSC22K175, and thanks Taha Dawoodbhoy and Luis Padilla for discussion. 

\appendix

\section{Tests of periodic boundary conditions}

As a test of the correct implementation of periodic boundary conditions we evolve a boosted equilibrium configuration with initial speed $v_{x0}=1$, located initially at the center of the domain. When using a periodic domain the configuration should eternally travel crossing the domain periodically. There is no point of comparison between isolated and periodic domains for this scenario, nevertheless it turns out to be a good test for the implementation of periodic conditions.

\begin{figure}
\includegraphics[width=4.15cm]{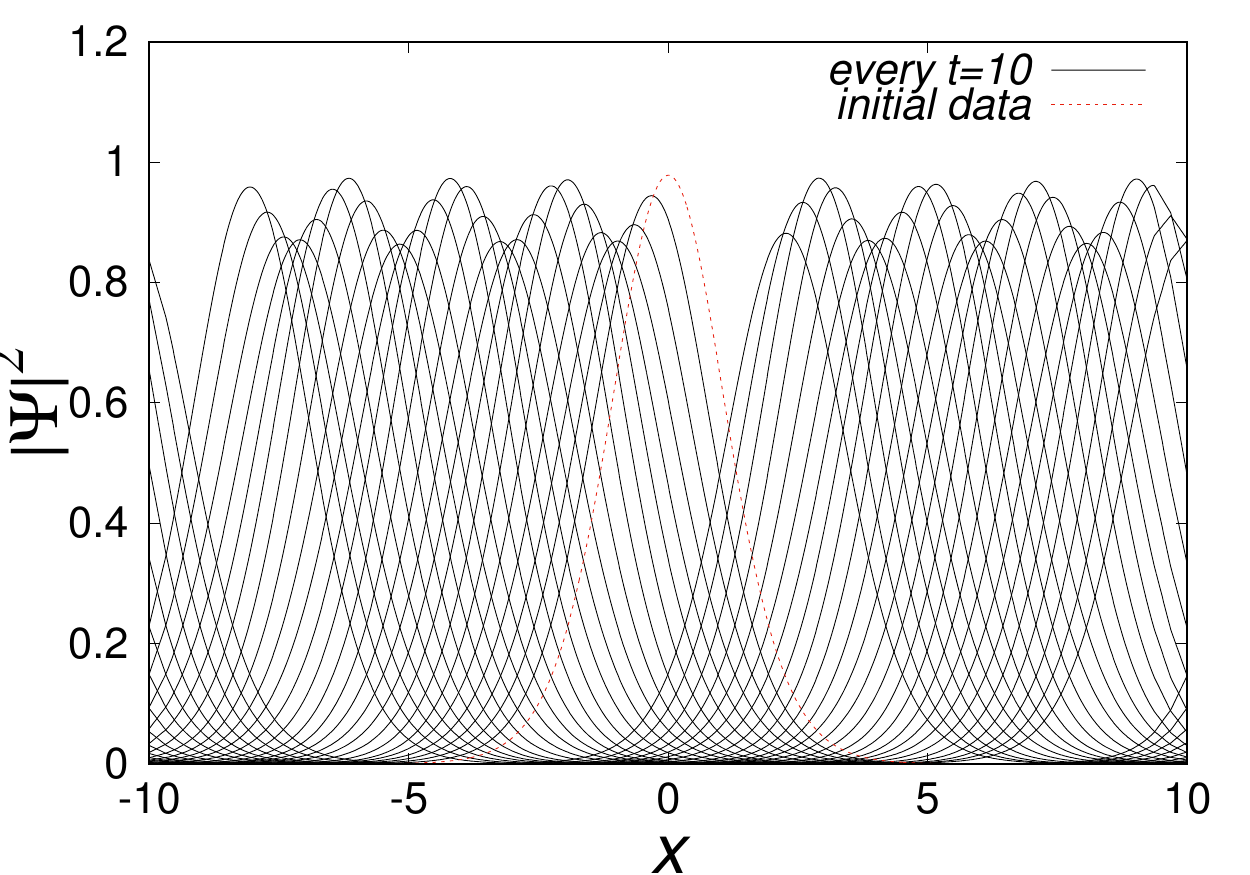}
\includegraphics[width=4.15cm]{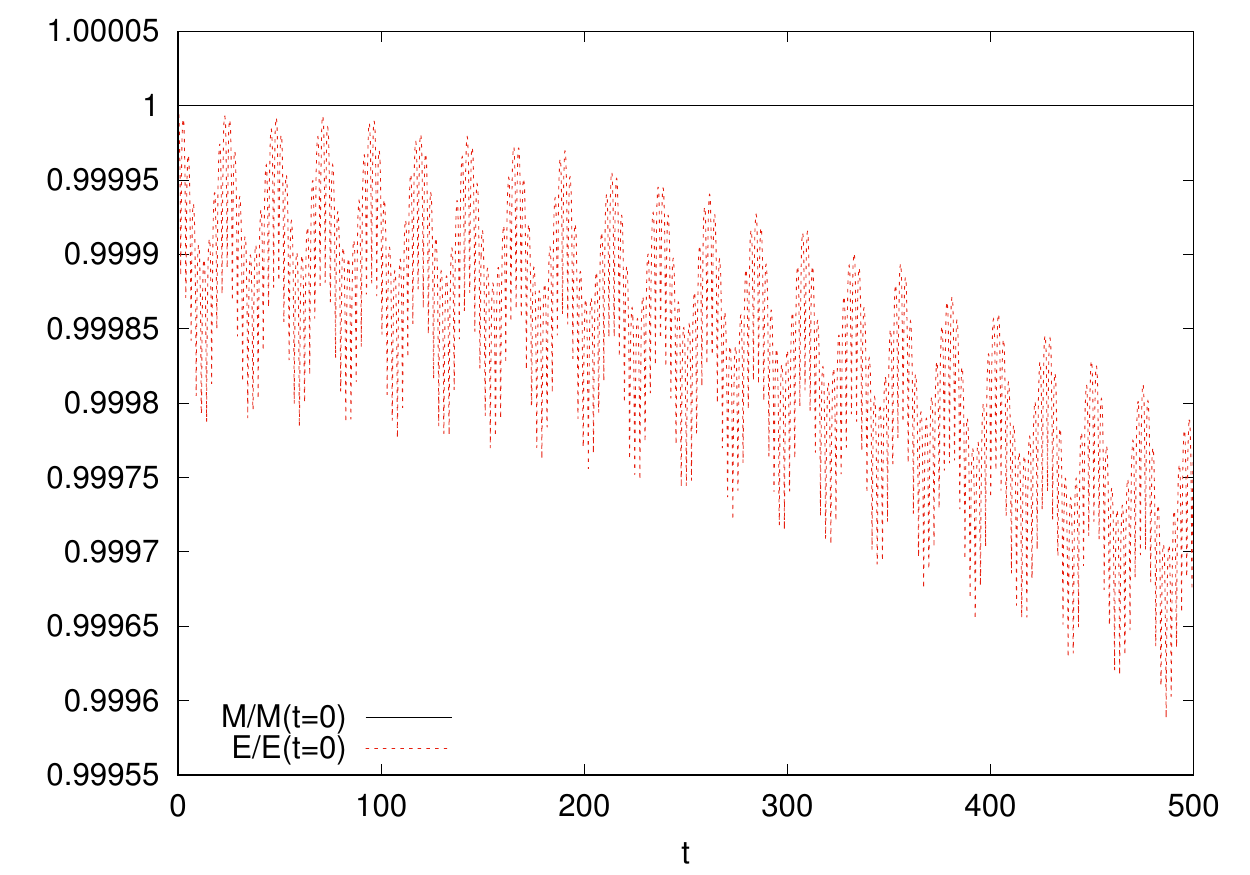}
\caption{\label{fig:boosted} Evolution of an equilibrium configuration to which an initial velocity $v_{x0}=1$ is applied in the periodic domain $[-10,10]^3$. The left panel shows snapshots of the profile $|\Psi|^2$ along the $x$ axis every $t=10$ units of time. At the right side we show the mass and total energy as functions of time, normalized with respect to their initial value.}
\end{figure}

In Figure \ref{fig:boosted} we show the density $|\Psi|^2$ along the $x-$axis at various times. The pulse travels from left to right, exits through the right end and reenters from the left as a result of the periodic domain. We also show the change of the total mass and energy with respect to its initial value, and see that both quantities are conserved with good precision, being the energy the most dissipated one, in less that $0.05\%$ during 50 crossing times.

\bibliography{BECDM}

\begin{thebibliography}{52}%
\makeatletter
\providecommand \@ifxundefined [1]{%
 \@ifx{#1\undefined}
}%
\providecommand \@ifnum [1]{%
 \ifnum #1\expandafter \@firstoftwo
 \else \expandafter \@secondoftwo
 \fi
}%
\providecommand \@ifx [1]{%
 \ifx #1\expandafter \@firstoftwo
 \else \expandafter \@secondoftwo
 \fi
}%
\providecommand \natexlab [1]{#1}%
\providecommand \enquote  [1]{``#1''}%
\providecommand \bibnamefont  [1]{#1}%
\providecommand \bibfnamefont [1]{#1}%
\providecommand \citenamefont [1]{#1}%
\providecommand \href@noop [0]{\@secondoftwo}%
\providecommand \href [0]{\begingroup \@sanitize@url \@href}%
\providecommand \@href[1]{\@@startlink{#1}\@@href}%
\providecommand \@@href[1]{\endgroup#1\@@endlink}%
\providecommand \@sanitize@url [0]{\catcode `\\12\catcode `\$12\catcode
  `\&12\catcode `\#12\catcode `\^12\catcode `\_12\catcode `\%12\relax}%
\providecommand \@@startlink[1]{}%
\providecommand \@@endlink[0]{}%
\providecommand \url  [0]{\begingroup\@sanitize@url \@url }%
\providecommand \@url [1]{\endgroup\@href {#1}{\urlprefix }}%
\providecommand \urlprefix  [0]{URL }%
\providecommand \Eprint [0]{\href }%
\providecommand \doibase [0]{http://dx.doi.org/}%
\providecommand \selectlanguage [0]{\@gobble}%
\providecommand \bibinfo  [0]{\@secondoftwo}%
\providecommand \bibfield  [0]{\@secondoftwo}%
\providecommand \translation [1]{[#1]}%
\providecommand \BibitemOpen [0]{}%
\providecommand \bibitemStop [0]{}%
\providecommand \bibitemNoStop [0]{.\EOS\space}%
\providecommand \EOS [0]{\spacefactor3000\relax}%
\providecommand \BibitemShut  [1]{\csname bibitem#1\endcsname}%
\let\auto@bib@innerbib\@empty
\bibitem [{\citenamefont {{Hu}}\ \emph {et~al.}(2000)\citenamefont {{Hu}},
  \citenamefont {{Barkana}},\ and\ \citenamefont {{Gruzinov}}}]{Hu:2000}%
  \BibitemOpen
  \bibfield  {author} {\bibinfo {author} {\bibfnamefont {W.}~\bibnamefont
  {{Hu}}}, \bibinfo {author} {\bibfnamefont {R.}~\bibnamefont {{Barkana}}}, \
  and\ \bibinfo {author} {\bibfnamefont {A.}~\bibnamefont {{Gruzinov}}},\
  }\bibfield  {title} {\enquote {\bibinfo {title} {{Fuzzy Cold Dark Matter: The
  Wave Properties of Ultralight Particles}},}\ }\href {\doibase
  10.1103/PhysRevLett.85.1158} {\bibfield  {journal} {\bibinfo  {journal}
  {Physical Review Letters}\ }\textbf {\bibinfo {volume} {85}},\ \bibinfo
  {pages} {1158--1161} (\bibinfo {year} {2000})},\ \Eprint
  {http://arxiv.org/abs/astro-ph/0003365} {astro-ph/0003365} \BibitemShut
  {NoStop}%
\bibitem [{\citenamefont {{Matos}}\ and\ \citenamefont
  {{Ure{\~n}a-L{\'o}pez}}(2000)}]{Matos-Urena:2000}%
  \BibitemOpen
  \bibfield  {author} {\bibinfo {author} {\bibfnamefont {T.}~\bibnamefont
  {{Matos}}}\ and\ \bibinfo {author} {\bibfnamefont {L.~A.}\ \bibnamefont
  {{Ure{\~n}a-L{\'o}pez}}},\ }\bibfield  {title} {\enquote {\bibinfo {title}
  {{Quintessence and scalar dark matter in the Universe}},}\ }\href {\doibase
  10.1088/0264-9381/17/13/101} {\bibfield  {journal} {\bibinfo  {journal}
  {Classical and Quantum Gravity}\ }\textbf {\bibinfo {volume} {17}},\ \bibinfo
  {pages} {L75--L81} (\bibinfo {year} {2000})},\ \Eprint
  {http://arxiv.org/abs/astro-ph/0004332} {astro-ph/0004332} \BibitemShut
  {NoStop}%
\bibitem [{\citenamefont {Sahni}\ and\ \citenamefont
  {Wang}(2000)}]{Sahni:2000}%
  \BibitemOpen
  \bibfield  {author} {\bibinfo {author} {\bibfnamefont {Varun}\ \bibnamefont
  {Sahni}}\ and\ \bibinfo {author} {\bibfnamefont {Limin}\ \bibnamefont
  {Wang}},\ }\bibfield  {title} {\enquote {\bibinfo {title} {New cosmological
  model of quintessence and dark matter},}\ }\href {\doibase
  10.1103/PhysRevD.62.103517} {\bibfield  {journal} {\bibinfo  {journal} {Phys.
  Rev. D}\ }\textbf {\bibinfo {volume} {62}},\ \bibinfo {pages} {103517}
  (\bibinfo {year} {2000})}\BibitemShut {NoStop}%
\bibitem [{\citenamefont {Hui}\ \emph {et~al.}(2017)\citenamefont {Hui},
  \citenamefont {Ostriker}, \citenamefont {Tremaine},\ and\ \citenamefont
  {Witten}}]{Hui:2016}%
  \BibitemOpen
  \bibfield  {author} {\bibinfo {author} {\bibfnamefont {Lam}\ \bibnamefont
  {Hui}}, \bibinfo {author} {\bibfnamefont {Jeremiah~P.}\ \bibnamefont
  {Ostriker}}, \bibinfo {author} {\bibfnamefont {Scott}\ \bibnamefont
  {Tremaine}}, \ and\ \bibinfo {author} {\bibfnamefont {Edward}\ \bibnamefont
  {Witten}},\ }\bibfield  {title} {\enquote {\bibinfo {title} {Ultralight
  scalars as cosmological dark matter},}\ }\href {\doibase
  10.1103/PhysRevD.95.043541} {\bibfield  {journal} {\bibinfo  {journal} {Phys.
  Rev. D}\ }\textbf {\bibinfo {volume} {95}},\ \bibinfo {pages} {043541}
  (\bibinfo {year} {2017})}\BibitemShut {NoStop}%
\bibitem [{\citenamefont {Hui}(2021)}]{Hui:2021tkt}%
  \BibitemOpen
  \bibfield  {author} {\bibinfo {author} {\bibfnamefont {Lam}\ \bibnamefont
  {Hui}},\ }\bibfield  {title} {\enquote {\bibinfo {title} {Wave dark
  matter},}\ }\href {\doibase 10.1146/annurev-astro-120920-010024} {\bibfield
  {journal} {\bibinfo  {journal} {Annual Review of Astronomy and Astrophysics}\
  }\textbf {\bibinfo {volume} {59}},\ \bibinfo {pages} {247--289} (\bibinfo
  {year} {2021})}\BibitemShut {NoStop}%
\bibitem [{\citenamefont {Niemeyer}(2020)}]{Niemeyer_2020}%
  \BibitemOpen
  \bibfield  {author} {\bibinfo {author} {\bibfnamefont {Jens~C.}\ \bibnamefont
  {Niemeyer}},\ }\bibfield  {title} {\enquote {\bibinfo {title} {Small-scale
  structure of fuzzy and axion-like dark matter},}\ }\href {\doibase
  10.1016/j.ppnp.2020.103787} {\bibfield  {journal} {\bibinfo  {journal}
  {Progress in Particle and Nuclear Physics}\ }\textbf {\bibinfo {volume}
  {113}},\ \bibinfo {pages} {103787} (\bibinfo {year} {2020})}\BibitemShut
  {NoStop}%
\bibitem [{\citenamefont {{Ferreira}}(2021)}]{ElisaFerreiraAApRev2021}%
  \BibitemOpen
  \bibfield  {author} {\bibinfo {author} {\bibfnamefont {Elisa G.~M.}\
  \bibnamefont {{Ferreira}}},\ }\bibfield  {title} {\enquote {\bibinfo {title}
  {{Ultra-light dark matter}},}\ }\href {\doibase 10.1007/s00159-021-00135-6}
  {\bibfield  {journal} {\bibinfo  {journal} {The Astronomy and Astrophysics
  Review}\ }\textbf {\bibinfo {volume} {29}},\ \bibinfo {eid} {7} (\bibinfo
  {year} {2021})},\ \Eprint {http://arxiv.org/abs/2005.03254} {arXiv:2005.03254
  [astro-ph.CO]} \BibitemShut {NoStop}%
\bibitem [{\citenamefont {Lee}(2018)}]{Lee:2017}%
  \BibitemOpen
  \bibfield  {author} {\bibinfo {author} {\bibfnamefont {Jae-Weon}\
  \bibnamefont {Lee}},\ }\bibfield  {title} {\enquote {\bibinfo {title} {{Brief
  History of Ultra-light Scalar Dark Matter Models}},}\ }\href {\doibase
  10.1051/epjconf/201816806005} {\bibfield  {journal} {\bibinfo  {journal} {EPJ
  Web Conf.}\ }\textbf {\bibinfo {volume} {168}},\ \bibinfo {pages} {06005}
  (\bibinfo {year} {2018})},\ \Eprint {http://arxiv.org/abs/1704.05057}
  {arXiv:1704.05057 [astro-ph.CO]} \BibitemShut {NoStop}%
\bibitem [{\citenamefont {Su\'arez}\ \emph {et~al.}(2014)\citenamefont
  {Su\'arez}, \citenamefont {Robles},\ and\ \citenamefont
  {Matos}}]{Suarez:2013}%
  \BibitemOpen
  \bibfield  {author} {\bibinfo {author} {\bibfnamefont {Abril}\ \bibnamefont
  {Su\'arez}}, \bibinfo {author} {\bibfnamefont {Victor~H.}\ \bibnamefont
  {Robles}}, \ and\ \bibinfo {author} {\bibfnamefont {Tonatiuh}\ \bibnamefont
  {Matos}},\ }\bibfield  {title} {\enquote {\bibinfo {title} {{A Review on the
  Scalar Field/Bose-Einstein Condensate Dark Matter Model}},}\ }\href {\doibase
  10.1007/978-3-319-02063-1_9} {\bibfield  {journal} {\bibinfo  {journal}
  {Astrophys. Space Sci. Proc.}\ }\textbf {\bibinfo {volume} {38}},\ \bibinfo
  {pages} {107--142} (\bibinfo {year} {2014})},\ \Eprint
  {http://arxiv.org/abs/1302.0903} {arXiv:1302.0903 [astro-ph.CO]} \BibitemShut
  {NoStop}%
\bibitem [{\citenamefont {{Ruffini}}\ and\ \citenamefont
  {{Bonazzola}}(1969)}]{Ruffini:1969}%
  \BibitemOpen
  \bibfield  {author} {\bibinfo {author} {\bibfnamefont {R.}~\bibnamefont
  {{Ruffini}}}\ and\ \bibinfo {author} {\bibfnamefont {S.}~\bibnamefont
  {{Bonazzola}}},\ }\bibfield  {title} {\enquote {\bibinfo {title} {Systems of
  self-gravitating particles in general relativity and the concept of an
  equation of state},}\ }\href {\doibase 10.1103/PhysRev.187.1767} {\bibfield
  {journal} {\bibinfo  {journal} {Phys. Rev.}\ }\textbf {\bibinfo {volume}
  {187}},\ \bibinfo {pages} {1767--1783} (\bibinfo {year} {1969})}\BibitemShut
  {NoStop}%
\bibitem [{\citenamefont {Guzm\'an}\ and\ \citenamefont {Ure\~na
  L\'opez}(2004)}]{GuzmanUrena2004}%
  \BibitemOpen
  \bibfield  {author} {\bibinfo {author} {\bibfnamefont {F.~S.}\ \bibnamefont
  {Guzm\'an}}\ and\ \bibinfo {author} {\bibfnamefont {L.~A.}\ \bibnamefont
  {Ure\~na L\'opez}},\ }\bibfield  {title} {\enquote {\bibinfo {title}
  {Evolution of the schr\"odinger-newton system for a self-gravitating scalar
  field},}\ }\href {\doibase 10.1103/PhysRevD.69.124033} {\bibfield  {journal}
  {\bibinfo  {journal} {Phys. Rev. D}\ }\textbf {\bibinfo {volume} {69}},\
  \bibinfo {pages} {124033} (\bibinfo {year} {2004})}\BibitemShut {NoStop}%
\bibitem [{\citenamefont {Bernal}\ and\ \citenamefont
  {Guzm\'an}(2006{\natexlab{a}})}]{BernalGuzman2006a}%
  \BibitemOpen
  \bibfield  {author} {\bibinfo {author} {\bibfnamefont {Argelia}\ \bibnamefont
  {Bernal}}\ and\ \bibinfo {author} {\bibfnamefont {F.~S.}\ \bibnamefont
  {Guzm\'an}},\ }\bibfield  {title} {\enquote {\bibinfo {title} {Scalar field
  dark matter: Head-on interaction between two structures},}\ }\href {\doibase
  10.1103/physrevd.74.103002} {\bibfield  {journal} {\bibinfo  {journal}
  {Physical Review D}\ }\textbf {\bibinfo {volume} {74}} (\bibinfo {year}
  {2006}{\natexlab{a}}),\ 10.1103/physrevd.74.103002}\BibitemShut {NoStop}%
\bibitem [{\citenamefont {{Schwabe}}\ \emph {et~al.}(2016)\citenamefont
  {{Schwabe}}, \citenamefont {{Niemeyer}},\ and\ \citenamefont
  {{Engels}}}]{Schwabe:2016}%
  \BibitemOpen
  \bibfield  {author} {\bibinfo {author} {\bibfnamefont {Bodo}\ \bibnamefont
  {{Schwabe}}}, \bibinfo {author} {\bibfnamefont {Jens~C.}\ \bibnamefont
  {{Niemeyer}}}, \ and\ \bibinfo {author} {\bibfnamefont {Jan~F.}\ \bibnamefont
  {{Engels}}},\ }\bibfield  {title} {\enquote {\bibinfo {title} {{Simulations
  of solitonic core mergers in ultralight axion dark matter cosmologies}},}\
  }\href {\doibase 10.1103/PhysRevD.94.043513} {\bibfield  {journal} {\bibinfo
  {journal} {\prd}\ }\textbf {\bibinfo {volume} {94}},\ \bibinfo {eid} {043513}
  (\bibinfo {year} {2016})},\ \Eprint {http://arxiv.org/abs/1606.05151}
  {arXiv:1606.05151 [astro-ph.CO]} \BibitemShut {NoStop}%
\bibitem [{\citenamefont {Dawoodbhoy}\ \emph {et~al.}(2021)\citenamefont
  {Dawoodbhoy}, \citenamefont {Shapiro},\ and\ \citenamefont
  {Rindler-Daller}}]{ShapiroCreTail}%
  \BibitemOpen
  \bibfield  {author} {\bibinfo {author} {\bibfnamefont {Taha}\ \bibnamefont
  {Dawoodbhoy}}, \bibinfo {author} {\bibfnamefont {Paul~R}\ \bibnamefont
  {Shapiro}}, \ and\ \bibinfo {author} {\bibfnamefont {Tanja}\ \bibnamefont
  {Rindler-Daller}},\ }\bibfield  {title} {\enquote {\bibinfo {title}
  {Core-envelope haloes in scalar field dark matter with repulsive
  self-interaction: fluid dynamics beyond the de broglie wavelength},}\ }\href
  {\doibase 10.1093/mnras/stab1859} {\bibfield  {journal} {\bibinfo  {journal}
  {Monthly Notices of the Royal Astronomical Society}\ }\textbf {\bibinfo
  {volume} {506}},\ \bibinfo {pages} {2418 -- 2444} (\bibinfo {year}
  {2021})}\BibitemShut {NoStop}%
\bibitem [{\citenamefont {Du}\ \emph {et~al.}(2018)\citenamefont {Du},
  \citenamefont {Schwabe}, \citenamefont {Niemeyer},\ and\ \citenamefont
  {B\"urger}}]{PhysRevD.97.063507}%
  \BibitemOpen
  \bibfield  {author} {\bibinfo {author} {\bibfnamefont {Xiaolong}\
  \bibnamefont {Du}}, \bibinfo {author} {\bibfnamefont {Bodo}\ \bibnamefont
  {Schwabe}}, \bibinfo {author} {\bibfnamefont {Jens~C.}\ \bibnamefont
  {Niemeyer}}, \ and\ \bibinfo {author} {\bibfnamefont {David}\ \bibnamefont
  {B\"urger}},\ }\bibfield  {title} {\enquote {\bibinfo {title} {Tidal
  disruption of fuzzy dark matter subhalo cores},}\ }\href {\doibase
  10.1103/PhysRevD.97.063507} {\bibfield  {journal} {\bibinfo  {journal} {Phys.
  Rev. D}\ }\textbf {\bibinfo {volume} {97}},\ \bibinfo {pages} {063507}
  (\bibinfo {year} {2018})}\BibitemShut {NoStop}%
\bibitem [{\citenamefont {Schive}\ \emph
  {et~al.}(2014{\natexlab{a}})\citenamefont {Schive}, \citenamefont {Chiueh},\
  and\ \citenamefont {Broadhurst}}]{Schive:2014dra}%
  \BibitemOpen
  \bibfield  {author} {\bibinfo {author} {\bibfnamefont {Hsi-Yu}\ \bibnamefont
  {Schive}}, \bibinfo {author} {\bibfnamefont {Tzihong}\ \bibnamefont
  {Chiueh}}, \ and\ \bibinfo {author} {\bibfnamefont {Tom}\ \bibnamefont
  {Broadhurst}},\ }\bibfield  {title} {\enquote {\bibinfo {title} {{Cosmic
  Structure as the Quantum Interference of a Coherent Dark Wave}},}\ }\href
  {\doibase 10.1038/nphys2996} {\bibfield  {journal} {\bibinfo  {journal}
  {Nature Phys.}\ }\textbf {\bibinfo {volume} {10}},\ \bibinfo {pages}
  {496--499} (\bibinfo {year} {2014}{\natexlab{a}})},\ \Eprint
  {http://arxiv.org/abs/1406.6586} {arXiv:1406.6586 [astro-ph.GA]} \BibitemShut
  {NoStop}%
\bibitem [{\citenamefont {Mocz}\ \emph {et~al.}(2017)\citenamefont {Mocz},
  \citenamefont {Vogelsberger}, \citenamefont {Robles}, \citenamefont {Zavala},
  \citenamefont {Boylan-Kolchin}, \citenamefont {Fialkov},\ and\ \citenamefont
  {Hernquist}}]{Mocz:2017wlg}%
  \BibitemOpen
  \bibfield  {author} {\bibinfo {author} {\bibfnamefont {Philip}\ \bibnamefont
  {Mocz}}, \bibinfo {author} {\bibfnamefont {Mark}\ \bibnamefont
  {Vogelsberger}}, \bibinfo {author} {\bibfnamefont {Victor~H.}\ \bibnamefont
  {Robles}}, \bibinfo {author} {\bibfnamefont {Jesus}\ \bibnamefont {Zavala}},
  \bibinfo {author} {\bibfnamefont {Michael}\ \bibnamefont {Boylan-Kolchin}},
  \bibinfo {author} {\bibfnamefont {Anastasia}\ \bibnamefont {Fialkov}}, \ and\
  \bibinfo {author} {\bibfnamefont {Lars}\ \bibnamefont {Hernquist}},\
  }\bibfield  {title} {\enquote {\bibinfo {title} {{Galaxy formation with BECDM
  I. Turbulence and relaxation of idealized haloes}},}\ }\href {\doibase
  10.1093/mnras/stx1887} {\bibfield  {journal} {\bibinfo  {journal} {Mon. Not.
  Roy. Astron. Soc.}\ }\textbf {\bibinfo {volume} {471}},\ \bibinfo {pages}
  {4559--4570} (\bibinfo {year} {2017})},\ \Eprint
  {http://arxiv.org/abs/1705.05845} {arXiv:1705.05845 [astro-ph.CO]}
  \BibitemShut {NoStop}%
\bibitem [{\citenamefont {Mocz}\ \emph {et~al.}(2019)\citenamefont {Mocz},
  \citenamefont {Fialkov}, \citenamefont {Vogelsberger}, \citenamefont
  {Becerra}, \citenamefont {Amin}, \citenamefont {Bose}, \citenamefont
  {Boylan-Kolchin}, \citenamefont {Chavanis}, \citenamefont {Hernquist},
  \citenamefont {Lancaster},\ and\ \citenamefont {et~al.}}]{mocz19}%
  \BibitemOpen
  \bibfield  {author} {\bibinfo {author} {\bibfnamefont {Philip}\ \bibnamefont
  {Mocz}}, \bibinfo {author} {\bibfnamefont {Anastasia}\ \bibnamefont
  {Fialkov}}, \bibinfo {author} {\bibfnamefont {Mark}\ \bibnamefont
  {Vogelsberger}}, \bibinfo {author} {\bibfnamefont {Fernando}\ \bibnamefont
  {Becerra}}, \bibinfo {author} {\bibfnamefont {Mustafa~A.}\ \bibnamefont
  {Amin}}, \bibinfo {author} {\bibfnamefont {Sownak}\ \bibnamefont {Bose}},
  \bibinfo {author} {\bibfnamefont {Michael}\ \bibnamefont {Boylan-Kolchin}},
  \bibinfo {author} {\bibfnamefont {Pierre-Henri}\ \bibnamefont {Chavanis}},
  \bibinfo {author} {\bibfnamefont {Lars}\ \bibnamefont {Hernquist}}, \bibinfo
  {author} {\bibfnamefont {Lachlan}\ \bibnamefont {Lancaster}}, \ and\ \bibinfo
  {author} {\bibnamefont {et~al.}},\ }\bibfield  {title} {\enquote {\bibinfo
  {title} {First star-forming structures in fuzzy cosmic filaments},}\ }\href
  {http://dx.doi.org/10.1103/PhysRevLett.123.141301} {\bibfield  {journal}
  {\bibinfo  {journal} {Phys. Rev. Lett.}\ }\textbf {\bibinfo {volume} {123}}
  (\bibinfo {year} {2019})}\BibitemShut {NoStop}%
\bibitem [{\citenamefont {Mocz}\ \emph {et~al.}(2020)\citenamefont {Mocz},
  \citenamefont {Fialkov}, \citenamefont {Vogelsberger}, \citenamefont
  {Becerra}, \citenamefont {Shen}, \citenamefont {Robles}, \citenamefont
  {Amin}, \citenamefont {Zavala}, \citenamefont {Boylan-Kolchin}, \citenamefont
  {Bose}, \citenamefont {Marinacci}, \citenamefont {Chavanis}, \citenamefont
  {Lancaster},\ and\ \citenamefont {Hernquist}}]{mocz19b}%
  \BibitemOpen
  \bibfield  {author} {\bibinfo {author} {\bibfnamefont {Philip}\ \bibnamefont
  {Mocz}}, \bibinfo {author} {\bibfnamefont {Anastasia}\ \bibnamefont
  {Fialkov}}, \bibinfo {author} {\bibfnamefont {Mark}\ \bibnamefont
  {Vogelsberger}}, \bibinfo {author} {\bibfnamefont {Fernando}\ \bibnamefont
  {Becerra}}, \bibinfo {author} {\bibfnamefont {Xuejian}\ \bibnamefont {Shen}},
  \bibinfo {author} {\bibfnamefont {Victor~H}\ \bibnamefont {Robles}}, \bibinfo
  {author} {\bibfnamefont {Mustafa~A}\ \bibnamefont {Amin}}, \bibinfo {author}
  {\bibfnamefont {Jesus}\ \bibnamefont {Zavala}}, \bibinfo {author}
  {\bibfnamefont {Michael}\ \bibnamefont {Boylan-Kolchin}}, \bibinfo {author}
  {\bibfnamefont {Sownak}\ \bibnamefont {Bose}}, \bibinfo {author}
  {\bibfnamefont {Federico}\ \bibnamefont {Marinacci}}, \bibinfo {author}
  {\bibfnamefont {Pierre-Henri}\ \bibnamefont {Chavanis}}, \bibinfo {author}
  {\bibfnamefont {Lachlan}\ \bibnamefont {Lancaster}}, \ and\ \bibinfo {author}
  {\bibfnamefont {Lars}\ \bibnamefont {Hernquist}},\ }\bibfield  {title}
  {\enquote {\bibinfo {title} {{Galaxy formation with BECDM II. Cosmic
  filaments and first galaxies}},}\ }\href {\doibase 10.1093/mnras/staa738}
  {\bibfield  {journal} {\bibinfo  {journal} {Mon. Not. R. Astron. Soc.}\
  }\textbf {\bibinfo {volume} {494}},\ \bibinfo {pages} {2027--2044} (\bibinfo
  {year} {2020})}\BibitemShut {NoStop}%
\bibitem [{\citenamefont {Schwabe}\ and\ \citenamefont
  {Niemeyer}(2022)}]{Gotinga2022}%
  \BibitemOpen
  \bibfield  {author} {\bibinfo {author} {\bibfnamefont {Bodo}\ \bibnamefont
  {Schwabe}}\ and\ \bibinfo {author} {\bibfnamefont {Jens~C.}\ \bibnamefont
  {Niemeyer}},\ }\bibfield  {title} {\enquote {\bibinfo {title} {Deep zoom-in
  simulation of a fuzzy dark matter galactic halo},}\ }\href {\doibase
  10.1103/PhysRevLett.128.181301} {\bibfield  {journal} {\bibinfo  {journal}
  {Phys. Rev. Lett.}\ }\textbf {\bibinfo {volume} {128}},\ \bibinfo {pages}
  {181301} (\bibinfo {year} {2022})}\BibitemShut {NoStop}%
\bibitem [{\citenamefont {Seidel}\ and\ \citenamefont
  {Suen}(1994)}]{SeidelSuenCooling}%
  \BibitemOpen
  \bibfield  {author} {\bibinfo {author} {\bibfnamefont {Edward}\ \bibnamefont
  {Seidel}}\ and\ \bibinfo {author} {\bibfnamefont {Wai-Mo}\ \bibnamefont
  {Suen}},\ }\bibfield  {title} {\enquote {\bibinfo {title} {Formation of
  solitonic stars through gravitational cooling},}\ }\href {\doibase
  10.1103/PhysRevLett.72.2516} {\bibfield  {journal} {\bibinfo  {journal}
  {Phys. Rev. Lett.}\ }\textbf {\bibinfo {volume} {72}},\ \bibinfo {pages}
  {2516--2519} (\bibinfo {year} {1994})}\BibitemShut {NoStop}%
\bibitem [{\citenamefont {Guzm\'an}\ and\ \citenamefont {Ure\~na
  L\'opez}(2006)}]{GuzmanUrena2006}%
  \BibitemOpen
  \bibfield  {author} {\bibinfo {author} {\bibfnamefont {F.~S.}\ \bibnamefont
  {Guzm\'an}}\ and\ \bibinfo {author} {\bibfnamefont {L.~Arturo}\ \bibnamefont
  {Ure\~na L\'opez}},\ }\bibfield  {title} {\enquote {\bibinfo {title}
  {Gravitational cooling of self-gravitating bose condensates},}\ }\href
  {\doibase 10.1086/504508} {\bibfield  {journal} {\bibinfo  {journal} {The
  Astrophysical Journal}\ }\textbf {\bibinfo {volume} {645}},\ \bibinfo {pages}
  {814 -- 819} (\bibinfo {year} {2006})}\BibitemShut {NoStop}%
\bibitem [{\citenamefont {{Navarro}}\ \emph {et~al.}(1997)\citenamefont
  {{Navarro}}, \citenamefont {{Frenk}},\ and\ \citenamefont {{White}}}]{NFW}%
  \BibitemOpen
  \bibfield  {author} {\bibinfo {author} {\bibfnamefont {Julio~F.}\
  \bibnamefont {{Navarro}}}, \bibinfo {author} {\bibfnamefont {Carlos~S.}\
  \bibnamefont {{Frenk}}}, \ and\ \bibinfo {author} {\bibfnamefont {Simon
  D.~M.}\ \bibnamefont {{White}}},\ }\bibfield  {title} {\enquote {\bibinfo
  {title} {{A Universal Density Profile from Hierarchical Clustering}},}\
  }\href {\doibase 10.1086/304888} {\bibfield  {journal} {\bibinfo  {journal}
  {\apj}\ }\textbf {\bibinfo {volume} {490}},\ \bibinfo {pages} {493--508}
  (\bibinfo {year} {1997})},\ \Eprint {http://arxiv.org/abs/astro-ph/9611107}
  {arXiv:astro-ph/9611107 [astro-ph]} \BibitemShut {NoStop}%
\bibitem [{\citenamefont {{Marsh}}(2016)}]{Marsh2016}%
  \BibitemOpen
  \bibfield  {author} {\bibinfo {author} {\bibfnamefont {David J.~E.}\
  \bibnamefont {{Marsh}}},\ }\bibfield  {title} {\enquote {\bibinfo {title}
  {{Axion cosmology}},}\ }\href {\doibase 10.1016/j.physrep.2016.06.005}
  {\bibfield  {journal} {\bibinfo  {journal} {Physics Reports}\ }\textbf
  {\bibinfo {volume} {643}},\ \bibinfo {pages} {1--79} (\bibinfo {year}
  {2016})},\ \Eprint {http://arxiv.org/abs/1510.07633} {arXiv:1510.07633
  [astro-ph.CO]} \BibitemShut {NoStop}%
\bibitem [{\citenamefont {{Jetzer}}(1992)}]{Jetzer1992}%
  \BibitemOpen
  \bibfield  {author} {\bibinfo {author} {\bibfnamefont {Phillippe}\
  \bibnamefont {{Jetzer}}},\ }\bibfield  {title} {\enquote {\bibinfo {title}
  {{Boson stars}},}\ }\href {\doibase 10.1016/0370-1573(92)90123-H} {\bibfield
  {journal} {\bibinfo  {journal} {Physics Reports}\ }\textbf {\bibinfo {volume}
  {220}},\ \bibinfo {pages} {163--227} (\bibinfo {year} {1992})}\BibitemShut
  {NoStop}%
\bibitem [{\citenamefont {{Membrado}}\ \emph {et~al.}(1989)\citenamefont
  {{Membrado}}, \citenamefont {{Pacheco}},\ and\ \citenamefont
  {{Sa{\~n}udo}}}]{Membrado1989}%
  \BibitemOpen
  \bibfield  {author} {\bibinfo {author} {\bibfnamefont {M.}~\bibnamefont
  {{Membrado}}}, \bibinfo {author} {\bibfnamefont {A.~F.}\ \bibnamefont
  {{Pacheco}}}, \ and\ \bibinfo {author} {\bibfnamefont {J.}~\bibnamefont
  {{Sa{\~n}udo}}},\ }\bibfield  {title} {\enquote {\bibinfo {title} {{Hartree
  solutions for the self-Yukawian boson sphere}},}\ }\href {\doibase
  10.1103/PhysRevA.39.4207} {\bibfield  {journal} {\bibinfo  {journal} {\pra}\
  }\textbf {\bibinfo {volume} {39}},\ \bibinfo {pages} {4207--4211} (\bibinfo
  {year} {1989})}\BibitemShut {NoStop}%
\bibitem [{\citenamefont {Schive}\ \emph
  {et~al.}(2014{\natexlab{b}})\citenamefont {Schive}, \citenamefont {Liao},
  \citenamefont {Woo}, \citenamefont {Wong}, \citenamefont {Chiueh},
  \citenamefont {Broadhurst},\ and\ \citenamefont {Hwang}}]{Schive:2014hza}%
  \BibitemOpen
  \bibfield  {author} {\bibinfo {author} {\bibfnamefont {Hsi-Yu}\ \bibnamefont
  {Schive}}, \bibinfo {author} {\bibfnamefont {Ming-Hsuan}\ \bibnamefont
  {Liao}}, \bibinfo {author} {\bibfnamefont {Tak-Pong}\ \bibnamefont {Woo}},
  \bibinfo {author} {\bibfnamefont {Shing-Kwong}\ \bibnamefont {Wong}},
  \bibinfo {author} {\bibfnamefont {Tzihong}\ \bibnamefont {Chiueh}}, \bibinfo
  {author} {\bibfnamefont {Tom}\ \bibnamefont {Broadhurst}}, \ and\ \bibinfo
  {author} {\bibfnamefont {W.~Y.~Pauchy}\ \bibnamefont {Hwang}},\ }\bibfield
  {title} {\enquote {\bibinfo {title} {{Understanding the Core-Halo Relation of
  Quantum Wave Dark Matter from 3D Simulations}},}\ }\href {\doibase
  10.1103/PhysRevLett.113.261302} {\bibfield  {journal} {\bibinfo  {journal}
  {Phys. Rev. Lett.}\ }\textbf {\bibinfo {volume} {113}},\ \bibinfo {pages}
  {261302} (\bibinfo {year} {2014}{\natexlab{b}})},\ \Eprint
  {http://arxiv.org/abs/1407.7762} {arXiv:1407.7762 [astro-ph.GA]} \BibitemShut
  {NoStop}%
\bibitem [{\citenamefont {{Rindler-Daller}}\ and\ \citenamefont
  {{Shapiro}}(2014)}]{RindlerDallerShapiro2014}%
  \BibitemOpen
  \bibfield  {author} {\bibinfo {author} {\bibfnamefont {Tanja}\ \bibnamefont
  {{Rindler-Daller}}}\ and\ \bibinfo {author} {\bibfnamefont {Paul~R.}\
  \bibnamefont {{Shapiro}}},\ }\bibfield  {title} {\enquote {\bibinfo {title}
  {{Complex Scalar Field Dark Matter on Galactic Scales}},}\ }\href {\doibase
  10.1142/S021773231430002X} {\bibfield  {journal} {\bibinfo  {journal} {Modern
  Physics Letters A}\ }\textbf {\bibinfo {volume} {29}},\ \bibinfo {eid}
  {1430002} (\bibinfo {year} {2014})},\ \Eprint
  {http://arxiv.org/abs/1312.1734} {arXiv:1312.1734 [astro-ph.CO]} \BibitemShut
  {NoStop}%
\bibitem [{\citenamefont {{Dawoodbhoy}}\ \emph {et~al.}(2021)\citenamefont
  {{Dawoodbhoy}}, \citenamefont {{Shapiro}},\ and\ \citenamefont
  {{Rindler-Daller}}}]{DawoodbhoyShapiroRindlerDaller2021}%
  \BibitemOpen
  \bibfield  {author} {\bibinfo {author} {\bibfnamefont {Taha}\ \bibnamefont
  {{Dawoodbhoy}}}, \bibinfo {author} {\bibfnamefont {Paul~R.}\ \bibnamefont
  {{Shapiro}}}, \ and\ \bibinfo {author} {\bibfnamefont {Tanja}\ \bibnamefont
  {{Rindler-Daller}}},\ }\bibfield  {title} {\enquote {\bibinfo {title}
  {{Core-envelope haloes in scalar field dark matter with repulsive
  self-interaction: fluid dynamics beyond the de Broglie wavelength}},}\ }\href
  {\doibase 10.1093/mnras/stab1859} {\bibfield  {journal} {\bibinfo  {journal}
  {Monthly Notices of the Royal Astronomical Society}\ }\textbf {\bibinfo
  {volume} {506}},\ \bibinfo {pages} {2418--2444} (\bibinfo {year} {2021})},\
  \Eprint {http://arxiv.org/abs/2104.07043} {arXiv:2104.07043 [astro-ph.CO]}
  \BibitemShut {NoStop}%
\bibitem [{\citenamefont {{Shapiro}}\ \emph {et~al.}(2022)\citenamefont
  {{Shapiro}}, \citenamefont {{Dawoodbhoy}},\ and\ \citenamefont
  {{Rindler-Daller}}}]{ShapiroDawoodbhoyRindlerDaller2022}%
  \BibitemOpen
  \bibfield  {author} {\bibinfo {author} {\bibfnamefont {Paul~R.}\ \bibnamefont
  {{Shapiro}}}, \bibinfo {author} {\bibfnamefont {Taha}\ \bibnamefont
  {{Dawoodbhoy}}}, \ and\ \bibinfo {author} {\bibfnamefont {Tanja}\
  \bibnamefont {{Rindler-Daller}}},\ }\bibfield  {title} {\enquote {\bibinfo
  {title} {{Cosmological structure formation in scalar field dark matter with
  repulsive self-interaction: the incredible shrinking Jeans mass}},}\ }\href
  {\doibase 10.1093/mnras/stab2884} {\bibfield  {journal} {\bibinfo  {journal}
  {Monthly Notices of the Royal Astronomical Society}\ }\textbf {\bibinfo
  {volume} {509}},\ \bibinfo {pages} {145--173} (\bibinfo {year} {2022})},\
  \Eprint {http://arxiv.org/abs/2106.13244} {arXiv:2106.13244 [astro-ph.CO]}
  \BibitemShut {NoStop}%
\bibitem [{\citenamefont {{Widrow}}\ and\ \citenamefont
  {{Kaiser}}(1993)}]{WidrowKaiser1993}%
  \BibitemOpen
  \bibfield  {author} {\bibinfo {author} {\bibfnamefont {Lawrence~M.}\
  \bibnamefont {{Widrow}}}\ and\ \bibinfo {author} {\bibfnamefont {Nick}\
  \bibnamefont {{Kaiser}}},\ }\bibfield  {title} {\enquote {\bibinfo {title}
  {{Using the Schroedinger Equation to Simulate Collisionless Matter}},}\
  }\href {\doibase 10.1086/187073} {\bibfield  {journal} {\bibinfo  {journal}
  {The Astrophysical journal Letters}\ }\textbf {\bibinfo {volume} {416}},\
  \bibinfo {pages} {L71} (\bibinfo {year} {1993})}\BibitemShut {NoStop}%
\bibitem [{\citenamefont {{Mocz}}\ \emph {et~al.}(2018)\citenamefont {{Mocz}},
  \citenamefont {{Lancaster}}, \citenamefont {{Fialkov}}, \citenamefont
  {{Becerra}},\ and\ \citenamefont
  {{Chavanis}}}]{MoczFialkovBecerraChavanis2018}%
  \BibitemOpen
  \bibfield  {author} {\bibinfo {author} {\bibfnamefont {Philip}\ \bibnamefont
  {{Mocz}}}, \bibinfo {author} {\bibfnamefont {Lachlan}\ \bibnamefont
  {{Lancaster}}}, \bibinfo {author} {\bibfnamefont {Anastasia}\ \bibnamefont
  {{Fialkov}}}, \bibinfo {author} {\bibfnamefont {Fernando}\ \bibnamefont
  {{Becerra}}}, \ and\ \bibinfo {author} {\bibfnamefont {Pierre-Henri}\
  \bibnamefont {{Chavanis}}},\ }\bibfield  {title} {\enquote {\bibinfo {title}
  {{Schr{\"o}dinger-Poisson-Vlasov-Poisson correspondence}},}\ }\href {\doibase
  10.1103/PhysRevD.97.083519} {\bibfield  {journal} {\bibinfo  {journal}
  {\prd}\ }\textbf {\bibinfo {volume} {97}},\ \bibinfo {eid} {083519} (\bibinfo
  {year} {2018})},\ \Eprint {http://arxiv.org/abs/1801.03507} {arXiv:1801.03507
  [astro-ph.CO]} \BibitemShut {NoStop}%
\bibitem [{\citenamefont {{Madelung}}(1927)}]{Madelung1927}%
  \BibitemOpen
  \bibfield  {author} {\bibinfo {author} {\bibfnamefont {E.}~\bibnamefont
  {{Madelung}}},\ }\bibfield  {title} {\enquote {\bibinfo {title}
  {{Quantentheorie in hydrodynamischer Form}},}\ }\href {\doibase
  10.1007/BF01400372} {\bibfield  {journal} {\bibinfo  {journal} {Zeitschrift
  fur Physik}\ }\textbf {\bibinfo {volume} {40}},\ \bibinfo {pages} {322--326}
  (\bibinfo {year} {1927})}\BibitemShut {NoStop}%
\bibitem [{\citenamefont {Bohm}(1952{\natexlab{a}})}]{Bohm1952a}%
  \BibitemOpen
  \bibfield  {author} {\bibinfo {author} {\bibfnamefont {David}\ \bibnamefont
  {Bohm}},\ }\bibfield  {title} {\enquote {\bibinfo {title} {A suggested
  interpretation of the quantum theory in terms of "hidden" variables. i},}\
  }\href {\doibase 10.1103/PhysRev.85.166} {\bibfield  {journal} {\bibinfo
  {journal} {Phys. Rev.}\ }\textbf {\bibinfo {volume} {85}},\ \bibinfo {pages}
  {166--179} (\bibinfo {year} {1952}{\natexlab{a}})}\BibitemShut {NoStop}%
\bibitem [{\citenamefont {Bohm}(1952{\natexlab{b}})}]{Bohm1952b}%
  \BibitemOpen
  \bibfield  {author} {\bibinfo {author} {\bibfnamefont {David}\ \bibnamefont
  {Bohm}},\ }\bibfield  {title} {\enquote {\bibinfo {title} {A suggested
  interpretation of the quantum theory in terms of "hidden" variables. ii},}\
  }\href {\doibase 10.1103/PhysRev.85.180} {\bibfield  {journal} {\bibinfo
  {journal} {Phys. Rev.}\ }\textbf {\bibinfo {volume} {85}},\ \bibinfo {pages}
  {180--193} (\bibinfo {year} {1952}{\natexlab{b}})}\BibitemShut {NoStop}%
\bibitem [{\citenamefont {{Alvarez-R{\'\i}os}}\ and\ \citenamefont
  {{Guzm{\'a}n}}(2022)}]{SPvsMadelung}%
  \BibitemOpen
  \bibfield  {author} {\bibinfo {author} {\bibfnamefont {Iv{\'a}n}\
  \bibnamefont {{Alvarez-R{\'\i}os}}}\ and\ \bibinfo {author} {\bibfnamefont
  {Francisco~S.}\ \bibnamefont {{Guzm{\'a}n}}},\ }\bibfield  {title} {\enquote
  {\bibinfo {title} {{Construction and Evolution of Equilibrium Configurations
  of the Schr{\"o}dinger{\textendash}Poisson System in the Madelung Frame}},}\
  }\href {\doibase 10.3390/universe8080432} {\bibfield  {journal} {\bibinfo
  {journal} {Universe}\ }\textbf {\bibinfo {volume} {8}},\ \bibinfo {pages}
  {432} (\bibinfo {year} {2022})},\ \Eprint {http://arxiv.org/abs/2210.15608}
  {arXiv:2210.15608 [gr-qc]} \BibitemShut {NoStop}%
\bibitem [{\citenamefont {{Takabayasi}}(1954)}]{Takabayasi1954}%
  \BibitemOpen
  \bibfield  {author} {\bibinfo {author} {\bibfnamefont {T.}~\bibnamefont
  {{Takabayasi}}},\ }\bibfield  {title} {\enquote {\bibinfo {title} {{The
  Formulation of Quantum Mechanics in terms of Ensemble in Phase Space}},}\
  }\href {\doibase 10.1143/PTP.11.341} {\bibfield  {journal} {\bibinfo
  {journal} {Progress of Theoretical Physics}\ }\textbf {\bibinfo {volume}
  {11}},\ \bibinfo {pages} {341--373} (\bibinfo {year} {1954})}\BibitemShut
  {NoStop}%
\bibitem [{\citenamefont {Wigner}(1932)}]{Wigner1932}%
  \BibitemOpen
  \bibfield  {author} {\bibinfo {author} {\bibfnamefont {E.}~\bibnamefont
  {Wigner}},\ }\bibfield  {title} {\enquote {\bibinfo {title} {On the quantum
  correction for thermodynamic equilibrium},}\ }\href {\doibase
  10.1103/PhysRev.40.749} {\bibfield  {journal} {\bibinfo  {journal} {Phys.
  Rev.}\ }\textbf {\bibinfo {volume} {40}},\ \bibinfo {pages} {749--759}
  (\bibinfo {year} {1932})}\BibitemShut {NoStop}%
\bibitem [{\citenamefont {Skodje}\ \emph {et~al.}(1989)\citenamefont {Skodje},
  \citenamefont {Rohrs},\ and\ \citenamefont {VanBuskirk}}]{Skodje1989}%
  \BibitemOpen
  \bibfield  {author} {\bibinfo {author} {\bibfnamefont {Rex~T.}\ \bibnamefont
  {Skodje}}, \bibinfo {author} {\bibfnamefont {Henry~W.}\ \bibnamefont
  {Rohrs}}, \ and\ \bibinfo {author} {\bibfnamefont {James}\ \bibnamefont
  {VanBuskirk}},\ }\bibfield  {title} {\enquote {\bibinfo {title} {Flux
  analysis, the correspondence principle, and the structure of quantum phase
  space},}\ }\href {\doibase 10.1103/PhysRevA.40.2894} {\bibfield  {journal}
  {\bibinfo  {journal} {Phys. Rev. A}\ }\textbf {\bibinfo {volume} {40}},\
  \bibinfo {pages} {2894--2916} (\bibinfo {year} {1989})}\BibitemShut {NoStop}%
\bibitem [{\citenamefont {HUSIMI}(1940)}]{Husimi1940}%
  \BibitemOpen
  \bibfield  {author} {\bibinfo {author} {\bibfnamefont {K.}~\bibnamefont
  {HUSIMI}},\ }\bibfield  {title} {\enquote {\bibinfo {title} {Some formal
  properties of the density matrix},}\ }\href {\doibase
  10.11429/ppmsj1919.22.4_264} {\bibfield  {journal} {\bibinfo  {journal}
  {Proceedings of the Physico-Mathematical Society of Japan. 3rd Series}\
  }\textbf {\bibinfo {volume} {22}},\ \bibinfo {pages} {264--314} (\bibinfo
  {year} {1940})}\BibitemShut {NoStop}%
\bibitem [{\citenamefont {{Ahn}}\ and\ \citenamefont
  {{Shapiro}}(2005)}]{AhnShapiro2005}%
  \BibitemOpen
  \bibfield  {author} {\bibinfo {author} {\bibfnamefont {Kyungjin}\
  \bibnamefont {{Ahn}}}\ and\ \bibinfo {author} {\bibfnamefont {Paul~R.}\
  \bibnamefont {{Shapiro}}},\ }\bibfield  {title} {\enquote {\bibinfo {title}
  {{Formation and evolution of self-interacting dark matter haloes}},}\ }\href
  {\doibase 10.1111/j.1365-2966.2005.09492.x} {\bibfield  {journal} {\bibinfo
  {journal} {Monthly Notices of the Royal Astronomical Society}\ }\textbf
  {\bibinfo {volume} {363}},\ \bibinfo {pages} {1092--1110} (\bibinfo {year}
  {2005})},\ \Eprint {http://arxiv.org/abs/astro-ph/0412169}
  {arXiv:astro-ph/0412169 [astro-ph]} \BibitemShut {NoStop}%
\bibitem [{\citenamefont {{Alvarez}}\ \emph {et~al.}(2003)\citenamefont
  {{Alvarez}}, \citenamefont {{Ahn}},\ and\ \citenamefont
  {{Shapiro}}}]{AlvarezAhnShapiro2003}%
  \BibitemOpen
  \bibfield  {author} {\bibinfo {author} {\bibfnamefont {M.~A.}\ \bibnamefont
  {{Alvarez}}}, \bibinfo {author} {\bibfnamefont {K.}~\bibnamefont {{Ahn}}}, \
  and\ \bibinfo {author} {\bibfnamefont {P.~R.}\ \bibnamefont {{Shapiro}}},\
  }\bibfield  {title} {\enquote {\bibinfo {title} {{Density Profiles of Dark
  Halos from their Mass Accretion Histories}},}\ }in\ \href@noop {} {\emph
  {\bibinfo {booktitle} {Revista Mexicana de Astronomia y Astrofisica
  Conference Series}}},\ \bibinfo {series} {Revista Mexicana de Astronomia y
  Astrofisica Conference Series}, Vol.~\bibinfo {volume} {18},\ \bibinfo
  {editor} {edited by\ \bibinfo {editor} {\bibfnamefont {M.}~\bibnamefont
  {{Reyes-Ruiz}}}\ and\ \bibinfo {editor} {\bibfnamefont {E.}~\bibnamefont
  {{V{\'a}zquez-Semadeni}}}}\ (\bibinfo {year} {2003})\ pp.\ \bibinfo {pages}
  {4--7},\ \Eprint {http://arxiv.org/abs/astro-ph/0302336}
  {arXiv:astro-ph/0302336 [astro-ph]} \BibitemShut {NoStop}%
\bibitem [{\citenamefont {{Shapiro}}\ \emph {et~al.}(2004)\citenamefont
  {{Shapiro}}, \citenamefont {{Iliev}}, \citenamefont {{Martel}}, \citenamefont
  {{Ahn}},\ and\ \citenamefont {{Alvarez}}}]{Shapiroetal2004}%
  \BibitemOpen
  \bibfield  {author} {\bibinfo {author} {\bibfnamefont {Paul~R.}\ \bibnamefont
  {{Shapiro}}}, \bibinfo {author} {\bibfnamefont {Ilian~T.}\ \bibnamefont
  {{Iliev}}}, \bibinfo {author} {\bibfnamefont {Hugo}\ \bibnamefont
  {{Martel}}}, \bibinfo {author} {\bibfnamefont {Kyungjin}\ \bibnamefont
  {{Ahn}}}, \ and\ \bibinfo {author} {\bibfnamefont {Marcelo~A.}\ \bibnamefont
  {{Alvarez}}},\ }\bibfield  {title} {\enquote {\bibinfo {title} {{The
  Equilibrium Structure of CDM Halos}},}\ }\href@noop {} {\bibfield  {journal}
  {\bibinfo  {journal} {arXiv e-prints}\ ,\ \bibinfo {eid} {astro-ph/0409173}}
  (\bibinfo {year} {2004})},\ \Eprint {http://arxiv.org/abs/astro-ph/0409173}
  {arXiv:astro-ph/0409173 [astro-ph]} \BibitemShut {NoStop}%
\bibitem [{\citenamefont {{Shapiro}}\ \emph {et~al.}(2006)\citenamefont
  {{Shapiro}}, \citenamefont {{Ahn}}, \citenamefont {{Alvarez}}, \citenamefont
  {{Iliev}},\ and\ \citenamefont {{Martel}}}]{Shapiro2006}%
  \BibitemOpen
  \bibfield  {author} {\bibinfo {author} {\bibfnamefont {P.~R.}\ \bibnamefont
  {{Shapiro}}}, \bibinfo {author} {\bibfnamefont {K.}~\bibnamefont {{Ahn}}},
  \bibinfo {author} {\bibfnamefont {M.}~\bibnamefont {{Alvarez}}}, \bibinfo
  {author} {\bibfnamefont {I.~T.}\ \bibnamefont {{Iliev}}}, \ and\ \bibinfo
  {author} {\bibfnamefont {H.}~\bibnamefont {{Martel}}},\ }\bibfield  {title}
  {\enquote {\bibinfo {title} {{Understanding the Equilibrium Structure of CDM
  Halos}},}\ }in\ \href {\doibase 10.1051/eas:2006036} {\emph {\bibinfo
  {booktitle} {EAS Publications Series}}},\ \bibinfo {series} {EAS Publications
  Series}, Vol.~\bibinfo {volume} {20},\ \bibinfo {editor} {edited by\ \bibinfo
  {editor} {\bibfnamefont {Gary~A.}\ \bibnamefont {{Mamon}}}, \bibinfo {editor}
  {\bibfnamefont {Francoise}\ \bibnamefont {{Combes}}}, \bibinfo {editor}
  {\bibfnamefont {Cedric}\ \bibnamefont {{Deffayet}}}, \ and\ \bibinfo {editor}
  {\bibfnamefont {Bernard}\ \bibnamefont {{Fort}}}}\ (\bibinfo {year} {2006})\
  pp.\ \bibinfo {pages} {5--10},\ \Eprint
  {http://arxiv.org/abs/astro-ph/0510146} {arXiv:astro-ph/0510146 [astro-ph]}
  \BibitemShut {NoStop}%
\bibitem [{\citenamefont {Chavanis}(2022)}]{Chavanis22}%
  \BibitemOpen
  \bibfield  {author} {\bibinfo {author} {\bibfnamefont {Pierre-Henri}\
  \bibnamefont {Chavanis}},\ }\bibfield  {title} {\enquote {\bibinfo {title}
  {{A heuristic wave equation parameterizing BEC dark matter halos with a
  quantum core and an isothermal atmospher}},}\ }\href {\doibase
  10.1140/epjb/s10051-022-00299-9} {\bibfield  {journal} {\bibinfo  {journal}
  {The European Physical Journal B}\ }\textbf {\bibinfo {volume} {95}},\
  \bibinfo {pages} {48} (\bibinfo {year} {2022})}\BibitemShut {NoStop}%
\bibitem [{\citenamefont {Guzm\'an}\ \emph {et~al.}(2014)\citenamefont
  {Guzm\'an}, \citenamefont {Lora-Clavijo}, \citenamefont
  {Gonz\'alez-Avil\'es},\ and\ \citenamefont {Rivera-Paleo}}]{Nkode3d}%
  \BibitemOpen
  \bibfield  {author} {\bibinfo {author} {\bibfnamefont {F.~S.}\ \bibnamefont
  {Guzm\'an}}, \bibinfo {author} {\bibfnamefont {F.~D.}\ \bibnamefont
  {Lora-Clavijo}}, \bibinfo {author} {\bibfnamefont {J.~J.}\ \bibnamefont
  {Gonz\'alez-Avil\'es}}, \ and\ \bibinfo {author} {\bibfnamefont {F.~J.}\
  \bibnamefont {Rivera-Paleo}},\ }\bibfield  {title} {\enquote {\bibinfo
  {title} {Rotation curves of rotating galactic bose-einstein condensate dark
  matter halos},}\ }\href {\doibase 10.1103/PhysRevD.89.063507} {\bibfield
  {journal} {\bibinfo  {journal} {Phys. Rev. D}\ }\textbf {\bibinfo {volume}
  {89}},\ \bibinfo {pages} {063507} (\bibinfo {year} {2014})}\BibitemShut
  {NoStop}%
\bibitem [{\citenamefont {Alvarez-Rios}\ and\ \citenamefont
  {Guzman}(2022)}]{AlvarezGuzman2022}%
  \BibitemOpen
  \bibfield  {author} {\bibinfo {author} {\bibfnamefont {I.}~\bibnamefont
  {Alvarez-Rios}}\ and\ \bibinfo {author} {\bibfnamefont {F.~S.}\ \bibnamefont
  {Guzman}},\ }\bibfield  {title} {\enquote {\bibinfo {title} {Exploration of
  simple scenarios involving fuzzy dark matter cores and gas at local
  scales},}\ }\href {https://arxiv.org/abs/2207.00062} {\  (\bibinfo {year}
  {2022})},\ \Eprint {http://arxiv.org/abs/2207.00062} {arXiv:2207.00062}
  \BibitemShut {NoStop}%
\bibitem [{\citenamefont {Guzm\'an}(2019)}]{Guzman2019}%
  \BibitemOpen
  \bibfield  {author} {\bibinfo {author} {\bibfnamefont {F.~S.}\ \bibnamefont
  {Guzm\'an}},\ }\bibfield  {title} {\enquote {\bibinfo {title} {Oscillation
  modes of ultralight bec dark matter cores},}\ }\href {\doibase
  10.1103/physrevd.99.083513} {\bibfield  {journal} {\bibinfo  {journal}
  {Physical Review D}\ }\textbf {\bibinfo {volume} {99}} (\bibinfo {year}
  {2019}),\ 10.1103/physrevd.99.083513}\BibitemShut {NoStop}%
\bibitem [{\citenamefont {Guzm\'an}\ \emph {et~al.}(2021)\citenamefont
  {Guzm\'an}, \citenamefont {Alvarez-R\'ios},\ and\ \citenamefont
  {Gonz\'alez}}]{GuzmanAlvarezGonzalez2021}%
  \BibitemOpen
  \bibfield  {author} {\bibinfo {author} {\bibfnamefont {F.~S.}\ \bibnamefont
  {Guzm\'an}}, \bibinfo {author} {\bibfnamefont {I.}~\bibnamefont
  {Alvarez-R\'ios}}, \ and\ \bibinfo {author} {\bibfnamefont {J.~A.}\
  \bibnamefont {Gonz\'alez}},\ }\bibfield  {title} {\enquote {\bibinfo {title}
  {Merger of galactic cores made of ultralight bosonic dark matter},}\ }\href
  {\doibase 10.31349/revmexfis.67.75} {\bibfield  {journal} {\bibinfo
  {journal} {Rev. Mex. Fis.}\ }\textbf {\bibinfo {volume} {67}},\ \bibinfo
  {pages} {75--83} (\bibinfo {year} {2021})}\BibitemShut {NoStop}%
\bibitem [{\citenamefont {Li}\ \emph {et~al.}(2021)\citenamefont {Li},
  \citenamefont {Hui},\ and\ \citenamefont {Yavetz}}]{Li_2021}%
  \BibitemOpen
  \bibfield  {author} {\bibinfo {author} {\bibfnamefont {Xinyu}\ \bibnamefont
  {Li}}, \bibinfo {author} {\bibfnamefont {Lam}\ \bibnamefont {Hui}}, \ and\
  \bibinfo {author} {\bibfnamefont {Tomer~D.}\ \bibnamefont {Yavetz}},\
  }\bibfield  {title} {\enquote {\bibinfo {title} {Oscillations and random walk
  of the soliton core in a fuzzy dark matter halo},}\ }\href {\doibase
  10.1103/physrevd.103.023508} {\bibfield  {journal} {\bibinfo  {journal}
  {Physical Review D}\ }\textbf {\bibinfo {volume} {103}} (\bibinfo {year}
  {2021}),\ 10.1103/physrevd.103.023508}\BibitemShut {NoStop}%
\bibitem [{\citenamefont {{Liu}}\ \emph {et~al.}(2023)\citenamefont {{Liu}},
  \citenamefont {{Proukakis}},\ and\ \citenamefont
  {{Rigopoulos}}}]{KangLiu2023}%
  \BibitemOpen
  \bibfield  {author} {\bibinfo {author} {\bibfnamefont {I.~Kang}\ \bibnamefont
  {{Liu}}}, \bibinfo {author} {\bibfnamefont {Nick~P.}\ \bibnamefont
  {{Proukakis}}}, \ and\ \bibinfo {author} {\bibfnamefont {Gerasimos}\
  \bibnamefont {{Rigopoulos}}},\ }\bibfield  {title} {\enquote {\bibinfo
  {title} {{Coherent and incoherent structures in fuzzy dark matter halos}},}\
  }\href {\doibase 10.1093/mnras/stad591} {\bibfield  {journal} {\bibinfo
  {journal} {Monthly Notices of the Royal Astronomical Society}\ } (\bibinfo
  {year} {2023}),\ 10.1093/mnras/stad591},\ \Eprint
  {http://arxiv.org/abs/2211.02565} {arXiv:2211.02565 [astro-ph.CO]}
  \BibitemShut {NoStop}%
\bibitem [{\citenamefont {Bernal}\ and\ \citenamefont
  {Guzm\'an}(2006{\natexlab{b}})}]{BernalGuzman2006b}%
  \BibitemOpen
  \bibfield  {author} {\bibinfo {author} {\bibfnamefont {Argelia}\ \bibnamefont
  {Bernal}}\ and\ \bibinfo {author} {\bibfnamefont {F.~S.}\ \bibnamefont
  {Guzm\'an}},\ }\bibfield  {title} {\enquote {\bibinfo {title} {Scalar field
  dark matter: Nonspherical collapse and late-time behavior},}\ }\href
  {\doibase 10.1103/physrevd.74.063504} {\bibfield  {journal} {\bibinfo
  {journal} {Physical Review D}\ }\textbf {\bibinfo {volume} {74}} (\bibinfo
  {year} {2006}{\natexlab{b}}),\ 10.1103/physrevd.74.063504}\BibitemShut
  {NoStop}%
\end{thebibliography}%

\end{document}